\documentclass[useAMS,usenatbib]{mn2e}
\usepackage{graphicx}
\usepackage{epsfig}
\usepackage{rotating}
\usepackage{pdflscape}
\usepackage{longtable}

\newcommand{\HI}{H\,{\sc i}}
\newcommand{\HII}{H\,{\sc ii}}
\newcommand{\SII}{[S\,{\sc ii}]}

\newcommand{\OIII}{[O\,{\sc iii}]}
\newcommand{\OII}{[O\,{\sc ii}]}
\newcommand{\OI}{[O\,{\sc i}]}

\newcommand{\NII}{[N\,{\sc ii}]}

\newcommand{\HeII}{He\,{\sc ii}}
\newcommand{\HeI}{He\,{\sc i}}

\newcommand{\CaII}{Ca\,{\sc ii}}
\newcommand{\MnII}{Mn\,{\sc ii}}
\newcommand{\CrII}{Cr\,{\sc ii}}
\newcommand{\EuII}{Eu\,{\sc ii}}
\newcommand{\SrII}{Sr\,{\sc ii}}
\newcommand{\SiII}{Si\,{\sc ii}}
\newcommand{\SiIII}{Si\,{\sc iii}}
\newcommand{\SiIV}{Si\,{\sc iv}}
\newcommand{\CIII}{C\,{\sc iii}}
\newcommand{\CII}{C\,{\sc ii}}
\newcommand{\MgII}{Mg\,{\sc ii}}
\newcommand{\MgIII}{Mg\,{\sc iii}}
\newcommand{\FeII}{[Fe\,{\sc ii}]}
\def\p0{\phantom{0}}

\title[Emission-line stars discovered in the LMC]{Emission-line stars discovered in the UKST H$\alpha$ survey of the Large Magellanic Cloud; Part 1: Hot stars}
\author[Warren A. Reid and Quentin A. Parker]{Warren A. Reid$^{1,2}$\thanks{E-mail:
warren.reid@mq.edu.au; war@aao.gov.au (WR); } and Quentin A.
Parker$^{1,2,3}$\footnotemark[1]\thanks{E-mail: quentin.parker@mq.edu.au (QAP)}\\
$^{1}$Department of Physics, Macquarie University, Sydney, NSW 2109, Australia\\
$^2$Macquarie University Research Centre in Astronomy, Astrophysics \& Astrophotonics, \\
Macquarie University, Sydney, NSW 2109, Australia\\
$^{3}$Australian Astronomical Observatory, PO Box 296, Epping, NSW 1710
Australia}
\begin{document}

\date{Accepted 2012 June 7. Received 2012 June 5; in original form 2012 February 1}

\pagerange{\pageref{firstpage}--\pageref{lastpage}} \pubyear{2012}

\maketitle

\label{firstpage}

\begin{abstract}
We present new, accurate positions, spectral classifications, radial and rotational velocities, H$\alpha$ fluxes, equivalent widths and B,V,I,R magnitudes for 579 hot emission-line
stars (classes B0 - F9) in the Large Magellanic Cloud (LMC) which include 469 new discoveries. Candidate emission line stars were
discovered using a deep, high resolution H$\alpha$ map of the
central 25 deg$^{2}$ of the LMC obtained by median stacking a dozen
2 hour H$\alpha$ exposures taken with the UK Schmidt Telescope
(UKST). Spectroscopic follow-up observations on the Anglo-Australian Telescope (AAT), the UKST,
the Very Large Telescope (VLT), the South African Astronomical Observatory (SAAO) 1.9m and the 2.3m telescope at Siding Spring Observatory have established the identity of these faint sources down
to magnitude $R_{\textrm{equiv}}\sim$23 for H$\alpha$
($4.5\times10^{-17}\textrm{ergs~cm}^{-2}~\textrm{s}^{-1}~$\AA$^{-1}$).

Confirmed emission-line stars have been assigned an underlying spectral classification through cross-correlation against 131 absorption line template spectra covering the range O1 to F8. We confirm 111 previously identified emission line stars and 64 previously known variable stars with spectral types hotter than F8. The majority of hot stars identified (518 stars or 89\%) are class B. Of all the hot emission-line stars in classes B-F, 130 or 22\% are type B[e], characterised by the presence of forbidden emission lines such as \SII, \NII~and \OII. We report on the physical location of these stars with reference to possible contamination from ambient \HII~emission. Only 13 of the emission-line stars require additional high resolution spectroscopic observations in order to assign a spectroscopic classification. They have nonetheless been added to the catalogue.

Along with flux calibration of the H$\alpha$ emission we provide the first H$\alpha$ luminosity function for selected sub-samples after correction for any possible nebula or ambient contamination. We find a moderate correlation between the intensity of H$\alpha$ emission and the V magnitude of the central star based on SuperCOSMOS magnitudes and the Optical Gravitational Lensing Experiment (OGLE-II) photometry where possible. Cool stars from classes G-S, with and without strong H$\alpha$ emission, will be the focus of part 2 in this series.
\end{abstract}


\section{Introduction}


The Large Magellanic Cloud (LMC) is a unique laboratory in which to
study the peculiar characteristics of massive and luminous emission-line stars. At a known distance of $\sim$50 kpc (see Reid \& Parker 2010 and references therein) to all
LMC members, modest inclination angle to the line of sight ($\sim$21deg) and with relatively low interstellar extinction (Rv = 3.41 $\pm$ 0.06; Gordon et al. 2003), apparent
brightness is a good indicator of absolute luminosity to within a
few tenths of a magnitude.

We take advantage of these benefits as we identify and begin basic analysis of emission-line stars in the LMC. The most prominent observational feature
of the emission-line stellar group is the presence of the H$\alpha$
line. The presence of this emission feature has been widely used as an identifier in the many
previous searches for emission-line stars in the LMC (eg. Feast et al.
1960; Henize 1956; Lindsay 1963,1974; Bohannan \& Epps 1974; Grebel
1997; Keller et al. 1999; Grebel \& Chu 2000; Keller et al 2000;
Olsen et al. 2001). None of these surveys went particularly deep. More recently, the OGLE II database has prevailed as the main tool used to uncover emission-line star candidates (Sabogal et al. 2005).


The UKST H$\alpha$ survey of the central 25deg$^{2}$ of the LMC has changed this situation. It was adjunct to the successful Southern Galactic Plane H$\alpha$ survey (Parker et al. 2005) and has revealed large numbers of various
emission objects. In addition to revealing 460 new planetary nebulae
within the survey region which were confirmed spectroscopically (Reid \& Parker, 2006a,b), spectroscopic followup and
careful analysis has revealed 579 hot emission-line stars with spectral classes B-F out of a total sample of 1,062 emission-line stars of all spectral types uncovered. Only 111 of these were previously known or identified while 469 are newly discovered.
The majority are Be, B[e], Bpe and HAeBe stars but two are Luminous Blue Variable (LBV) candidates. 
Identifying these objects will assist
our understanding of the main sequence evolution of massive stars. We have also identified 6 new and 33 previously known Wolf-Rayet stars, which are not included in this number but will be the special focus of a
follow-up paper.

 Be stars
are known to be variables which undergo active and quiescent stages
(Telting 2000; Bjorkman et al. 2002). A single epoch survey
could miss many of these stars if they were undergoing a quiescent
stage. This problem has already been demonstrated by several follow-up
investigations (Hummel et al. 1999; Keller et al. 1999; Wisniewski
and Bjorkman 2006) which were unable to identify all of the
previously identified Be stars in the Magellanic Clouds and in the Galaxy. In addition, these same follow-up studies revealed previously unidentified Be stars. Our H$\alpha$
survey, utilising 12 H$\alpha$ exposures taken over a three year
period has largely alleviated such problems and revealed a large number of
emission-line stars in the survey region to a magnitude of $R_{\textrm{equiv}}\sim$22
for H$\alpha$.


In order to study the Balmer emission we have measured the Equivalent Width (EW) and Full Width Half Maximum (FWHM) of the H$\alpha$ emission-lines. In addition, we include H$\alpha$ fluxes from medium resolution spectroscopy of 575 (99.3\%) of the detected emission-line stars within the survey area. Our follow-up spectroscopy was conducted from November 2004 to February 2005 on a variety of telescopes, allowing us to re-observe several known variable stars and detect minor changes in spectral characteristics. All but 2 candidate emission-line stars found in the H$\alpha$ survey had some degree of H$\alpha$ emission detectable in their spectrum at the time of observation. After describing flux calibration (section~\ref{section3}), we explain the method used to assign a spectral classification and luminosity class to each star using cross-correlation against well-established templates (section~\ref{section4}). Section~\ref{section6} describes our method for deriving the rotational velocities and section~\ref{section7} outlines a simple method for correcting or at least estimating nebula contribution in the spectrum. Section~\ref{section8} details our routine for assigning accurate positions to each star.

In section~\ref{section9} we describe the method used for measuring the radial velocity of each star. Velocities accurate to $\sim$4 km s$^{-1}$ have been found for 572
emission-line stars using both the weighted emission-line and
cross-correlation techniques on our higher dispersion spectroscopic data. These velocities can be used to search
for kinematical substructures in the LMC disk, create a 3D kinematic
map of the LMC for comparison with the \HI~disk, assist studies of age-metallicity dispersion
and distribution, potentially find stellar associations and streams, and
compare medium to old age populations such as planetary nebulae within the
LMC (Reid \& Parker 2006b).

 In section~\ref{section10} we show the projected distribution of emission-line stars and late-type stars across the survey field of the LMC. In section~\ref{section11} we measure the intensity of the H$\alpha$ emission considering ambient sky and any nebula contamination in order to create the first luminosity function for these stars in the LMC. Then, in section 13 we assess the emission by comparing BVI photometry from SuperCOSMOS and OGLE-II data where available. We discuss the stellar photometry, its reliability and problems associated with variability. In section~\ref{section12} we briefly discuss the variability already found in many of the candidate emission-line stars. The full catalogue of emission-line stars is described in section~\ref{section13} and presented in the appendix. Individual spectra and H$\alpha$ images will be available (2nd half 2012) on a dedicated web page hosted by the Astronomy Department at Macquarie University.



\section{Background to hot emission-line stars}

The origin of emission-lines in hot stars such as Be stars is not well understood. Such emission-line stars are found near the main sequence of luminosity classes V to III
exhibiting Balmer emission (Jaschek et al. 1981, Frew \& Parker 2010).
Various mechanisms have been proposed to explain how gaseous
circumstellar disks may form around Be stars (see Porter \&
Rivinius, 2003 for a review). Struve (1931) was the first to
speculate that Be stars exhibit rapid rotation. Recent theoretical
studies suggest that classical Be stars may be rotating close to
their critical velocity (Townsend et al. 2004) and exhibiting a strange form of variability (Kaler 1997). Other models of
circumstellar disk formation include the wind-compressed model
(Bjorkman \& Cassinelli 1993), pulsations arising from the stellar
photosphere (Rivinius et al. 2001) and the magnetically torqued and
wind compressed disk model (Cassinelli et al. 2002). While variables such as Cepheids and Miras are known
to pulsate radially, many stars also pulsate non-radially, producing
subtle magnitude variations and changing the shape of absorption
lines. It has been suggested that these oscillations, common on the
O and B main sequence, may be powerful enough to drive the winds
which produce the Be phenomenon (Kaler 1997).

In the case of pre
main sequence (PMS) T Tauri stars, the origin of the emission lines
is understood in terms of the magnetospheric accretion model, where
the emission lines originate from magnetospheric accretion columns
(Uchida \& Shibata 1985; K\"{o}nigl 1991; Hartmann et al. 1994;
Muzerolle et al. 1998, 2001). With the detection of magnetic fields
in a few Herbig AeBe stars (Hubrig et al. 2004; Wade et al. 2005),
the magnetospheric accretion model was successfully applied to these objects
(Muzerolle et al. 2004). However, the mechanism for triggering the
accretion is still not known.

B[e] stars have all the characteristics of Be stars but they additionally include forbidden emission lines in their spectra. Although lines such as \OIII$\lambda$5007 are suggestive of planetary nebulae (PNe), the presence of Fe emission and He absorption in the strong blue continuum clearly separate B[e] stars from PNe.

The evolutionary sequence of these stars is still not well known. Nor is the non-spherically symmetric circumstellar environment which is responsible for the B[e] phenomenon. Strong variability often reported from these objects has been explained by outbursts and shell phases (Hutsemékers 1985; Andrillat \& Houziaux 1991).

Related stars such as Herbig Ae/Be (HAeBe) stars, first discussed by Herbig (1960) are
found above the main sequence on the HR diagram and are believed to
be making their way toward it along radiative tracks as first
postulated by Henyey et al (1955). Along with T Tauri stars, they
share the characteristic of being associated with a nebula and
infra-red emission indicating the presence of circumstellar dust
(Hillenbrand et al. 1992; Lada \& Adams, 1992). What immediately
separates them from T Tauri stars is their larger mass of between
2M$_{\odot}$ and 10M$_{\odot}$. In order to separate HAeBe stars
from B[e] supergiants, Waters and Waelkens (1998) included the
condition that HAeBe stars should be of luminosity classes V to III.

As well as the Balmer lines, other optical emission-lines often
observed in HAeBe emission-line stars include HeI
($\lambda$5876\AA~and $\lambda$6678\AA), OI ($\lambda$7774\AA~and
$\lambda$8446\AA) and the \CaII~triplet ($\lambda$8498\AA,
$\lambda$8542\AA and $\lambda$8662\AA) (Herbig 1960; Hamann 1994;
B\"{o}hm \& Catala 1994; B\"{o}hm \& Hirth 1997; Corcoran \& Ray 1998; Viera
et al. 2003; Acke et al. 2005). We do not attempt to separate HAeBe stars from B[e] stars since many HAeBe and B[e] stars are spectroscopically indistinguishable.

\section{Optical observations}
\label{section1}

\subsection{The H$\alpha$ survey}

Over a period of three years, from 1997, a series of repeated
narrow-band H$\alpha$ and matching broad-band short red (SR)
exposures of the central LMC field were taken in order to
produce a deep H$\alpha$ and SR image with a 1 magnitude depth
gain over a single image frame. The twelve
highest quality and well-matched UK Schmidt Telescope 2-hour
H$\alpha$ exposures and six 15-minute equivalent SR-band exposures
were selected. From these exposures, deep, homogeneous, narrow-band
H$\alpha$ and matching broad-band SR maps of the
entire central 25 deg$^{2}$ region of the LMC were constructed.

The full aperture H$\alpha$ filter used for this survey was effectively the world's largest monolithic
interference filter to be used in astronomy (Parker \&
Bland-Hawthorn 1998). The choice of central wavelength ($\lambda$6590\AA) and
bandpass (70 \AA~FWHM) work effectively in the UKST's fast f/2.48
converging beam meaning the H$\alpha$ line remains within the filter band-pass for velocities up to 400\,km\,s$^{-1}$. Peak
filter transmission is $>$85\%. The fields for the survey were
exposed on non-standard, overlapping 4-degree centres due to the circular
aperture of the H$\alpha$ filter. These overlapped fields enabled
full contiguous coverage of the entire LMC/SMC region in H$\alpha$ despite the circular
aperture.

The successful implementation of high resolution, panchromatic
Tech-Pan film on the UKST, coupled with its peak sensitivity at
H$\alpha$, was a prime motivation for the survey. Tech-pan film
was an ideal wide-field photographic detector for use with
an H$\alpha$ filter. The resulting images produced were unequalled in
terms of their combined resolution, sensitivity and LMC coverage. Further
details of the properties of Tech-Pan can be found in Parker \&
Malin (1999).

The SuperCOSMOS plate-measuring machine at the Royal Observatory
Edinburgh (Hambly et al. 2001) was used to scan, co-add and pixel
match these selected exposures creating 10$\mu$m (0.67~arcsec) pixel data
which extends 1.35 (H$\alpha$) and 1 (SR) magnitude deeper than individual exposures,
achieving the full canonical Poissonian depth gain, e.g.
Bland-Hawthorn, Shopbell \& Malin (1993). This gives a depth
$\sim$21.5 for the SR images and $R_{\textrm{equiv}}\sim$22 for H$\alpha$
($4.5\times10^{-17}\textrm{ergs~cm}^{-2}~\textrm{s}^{-1}~$\AA$^{-1}$) which is at
least 1 magnitude deeper than the best wide-field narrow-band LMC
images previously available. An accurate world co-ordinate system
was applied to yield sub-arcsec astrometry (see section~\ref{section8}), essential for success of
the spectroscopic follow-up observations.

 \begin{figure}
\begin{minipage}[b]{0.355\linewidth}
\centering
\includegraphics[scale=0.75]{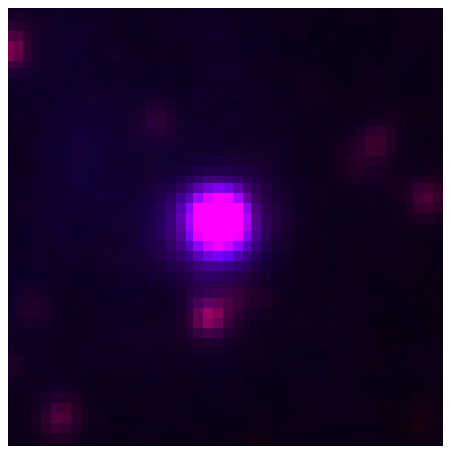}
\end{minipage}
\hspace{0.1cm}
\begin{minipage}[b]{0.65\linewidth}
\centering
\includegraphics[scale=0.36]{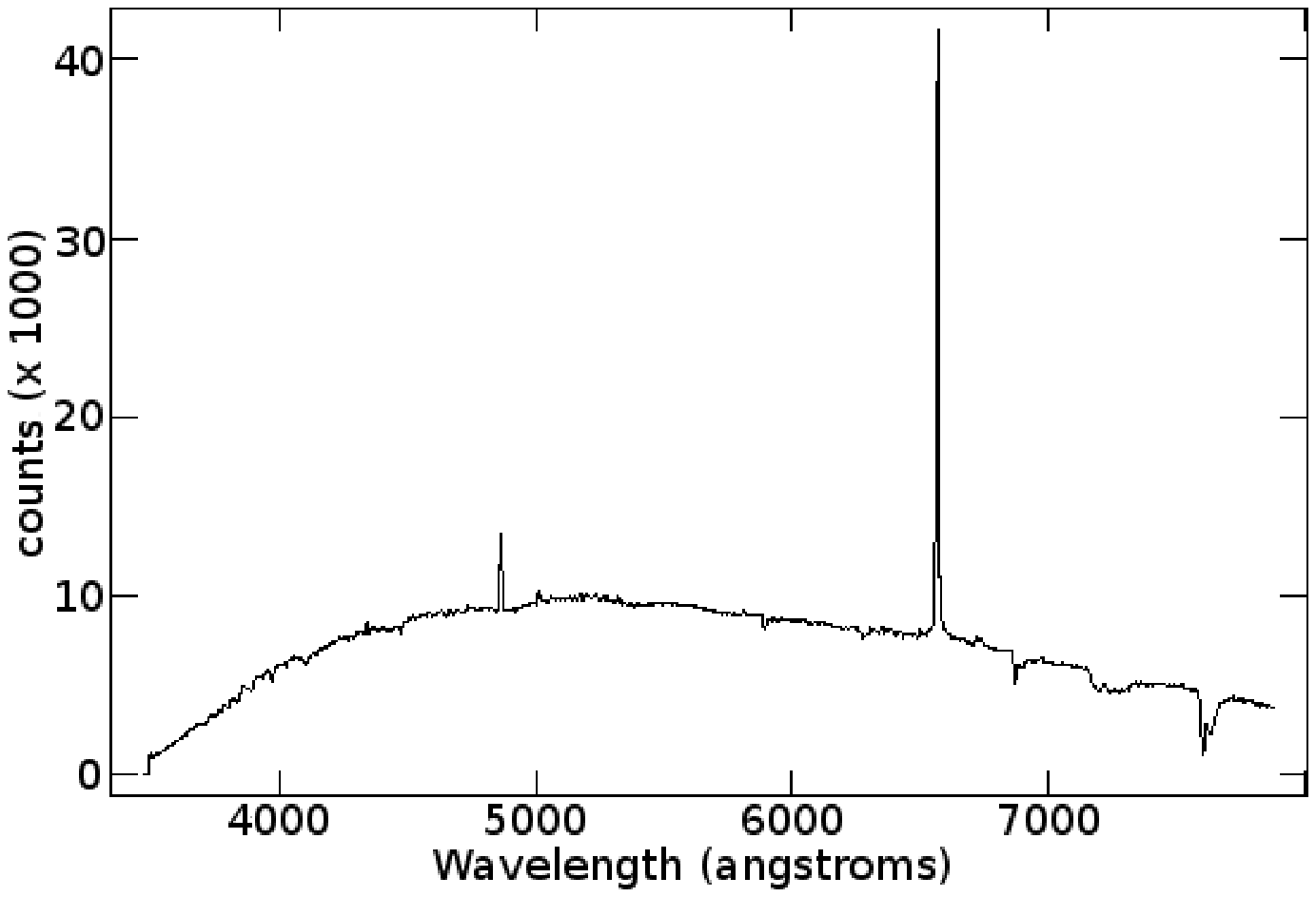}
\end{minipage}
\caption{H$\alpha$/R 30 $\times$ 30 arcsec image and 2dF low resolution spectrum of RPs255 also known as BE474 (Bohannan \& Epps, 1974) and as L333 (Lindsay, 1963). M$_{H\alpha}$=16.37. Compact H$\alpha$ emission 9.6 arcsec dia is largely due to PSF. North is upwards.}
\label{Figure1}
\begin{minipage}[b]{0.355\linewidth}
\centering
\includegraphics[scale=0.57]{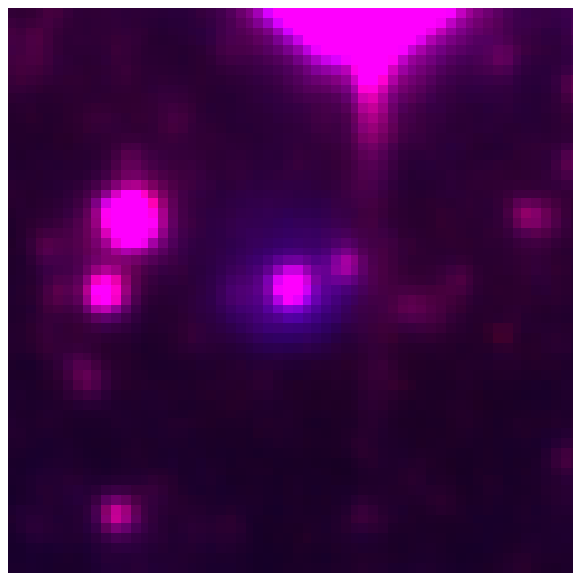}
\end{minipage}
\hspace{0.1cm}
\begin{minipage}[b]{0.65\linewidth}
\centering
\includegraphics[scale=0.36]{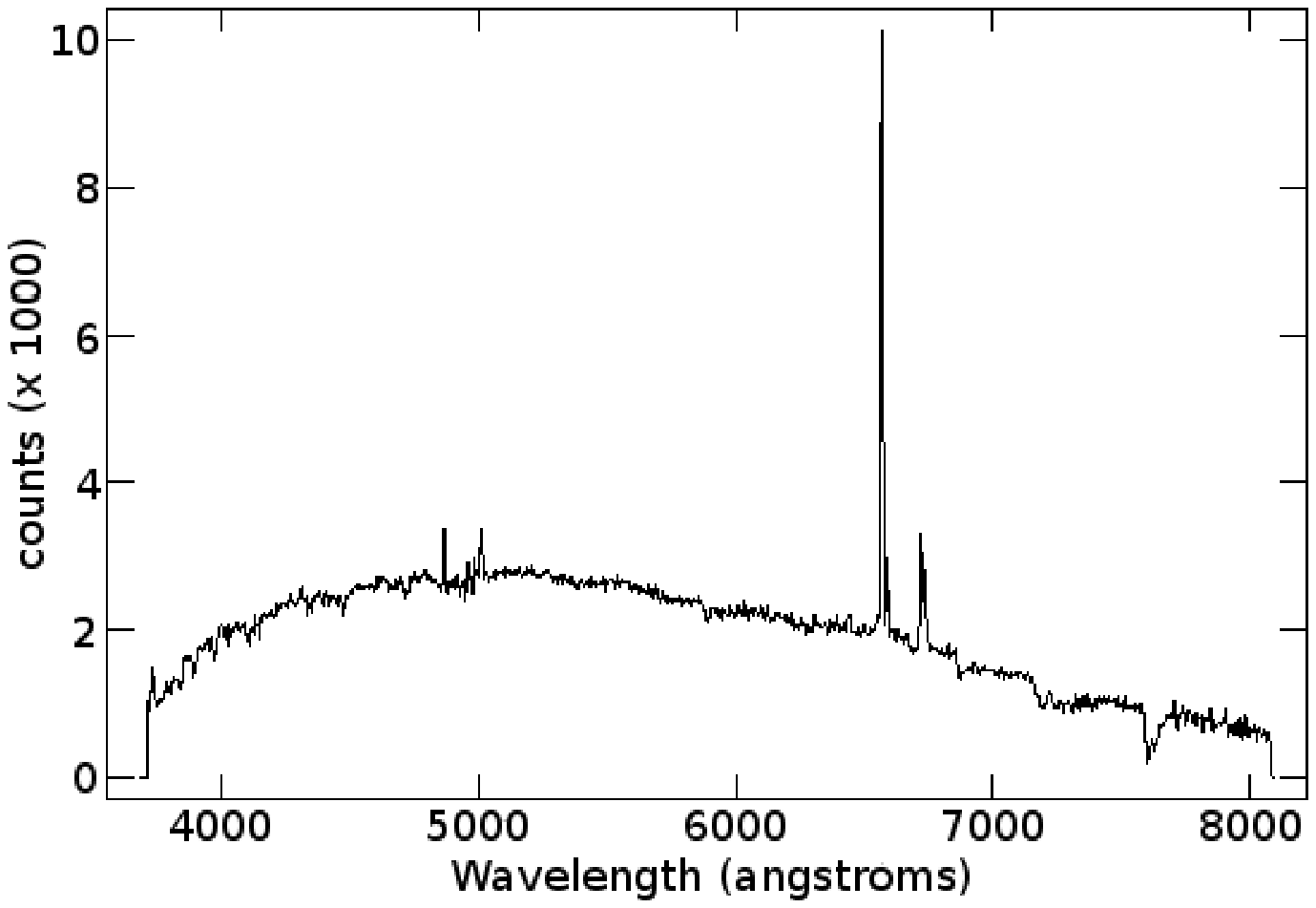}
\end{minipage}
\caption{Same as above for newly discovered emission-line star RPs256. M$_{H\alpha}$=19.34. Forbidden lines lead to B[e] classification.}
\label{Figure2}
\begin{minipage}[b]{0.355\linewidth}
\centering
\includegraphics[scale=0.57]{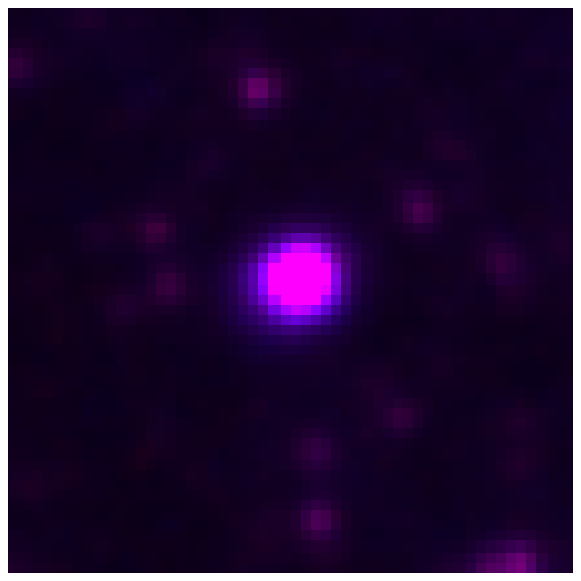}
\end{minipage}
\hspace{0.1cm}
\begin{minipage}[b]{0.65\linewidth}
\centering
\includegraphics[scale=0.36]{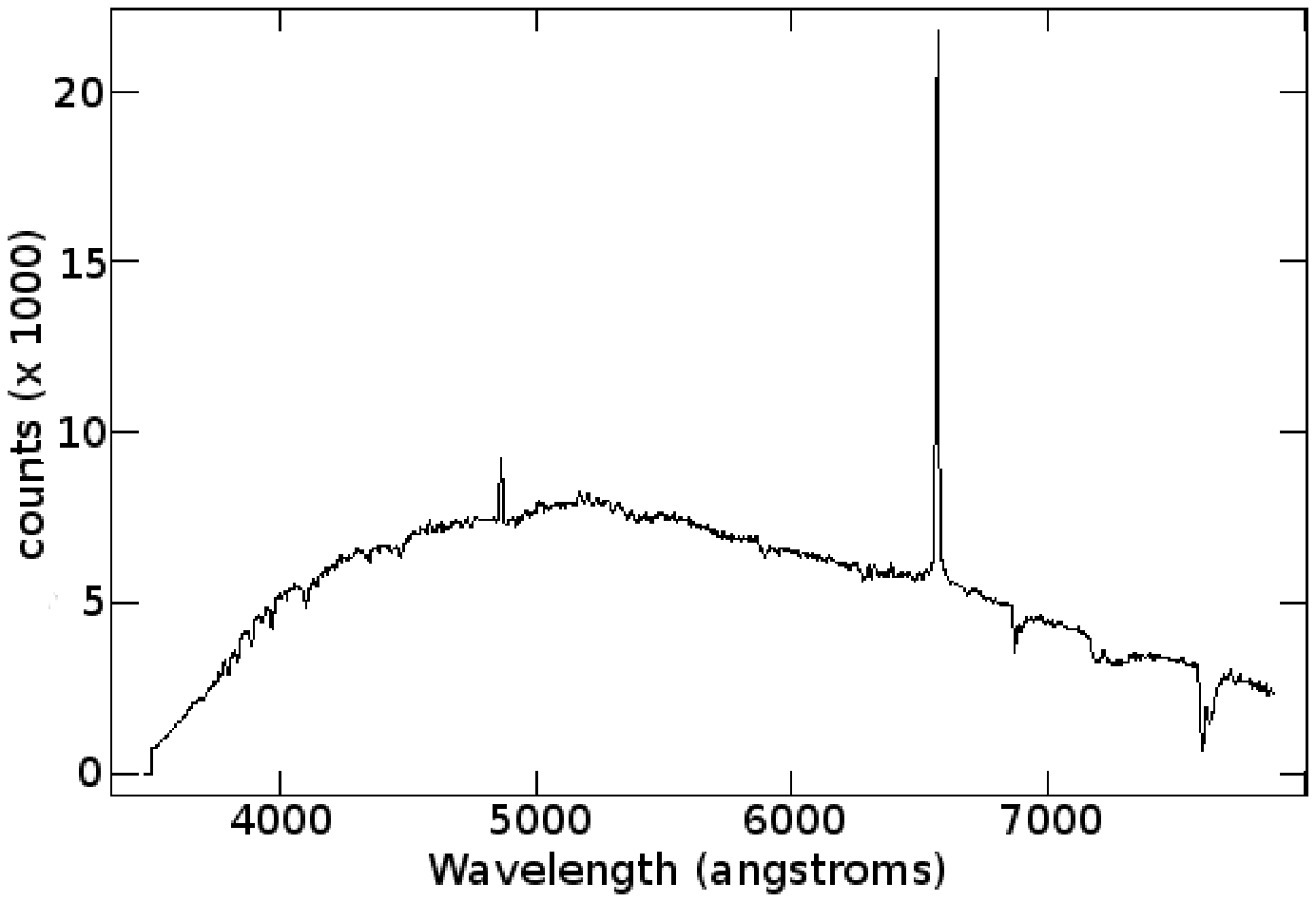}
\end{minipage}
\caption{Same as above for RPs285 also known as BE411 (Bohannan \& Epps, 1974). M$_{H\alpha}$=16.92.}
\label{Figure3}
\begin{minipage}[b]{0.355\linewidth}
\centering
\includegraphics[scale=0.57]{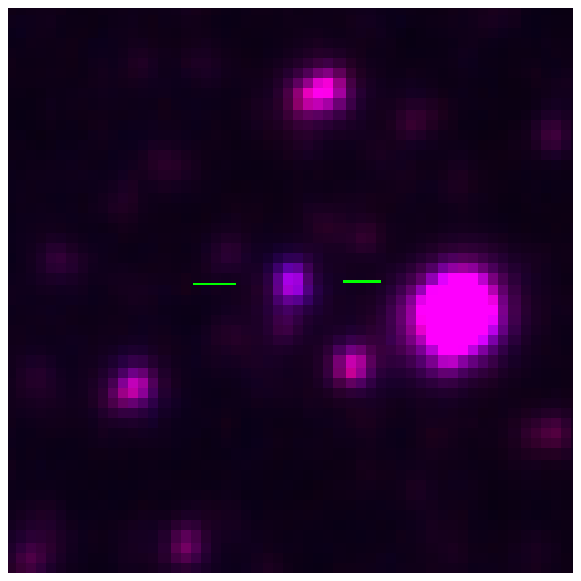}
\end{minipage}
\hspace{0.1cm}
\begin{minipage}[b]{0.65\linewidth}
\centering
\includegraphics[scale=0.36]{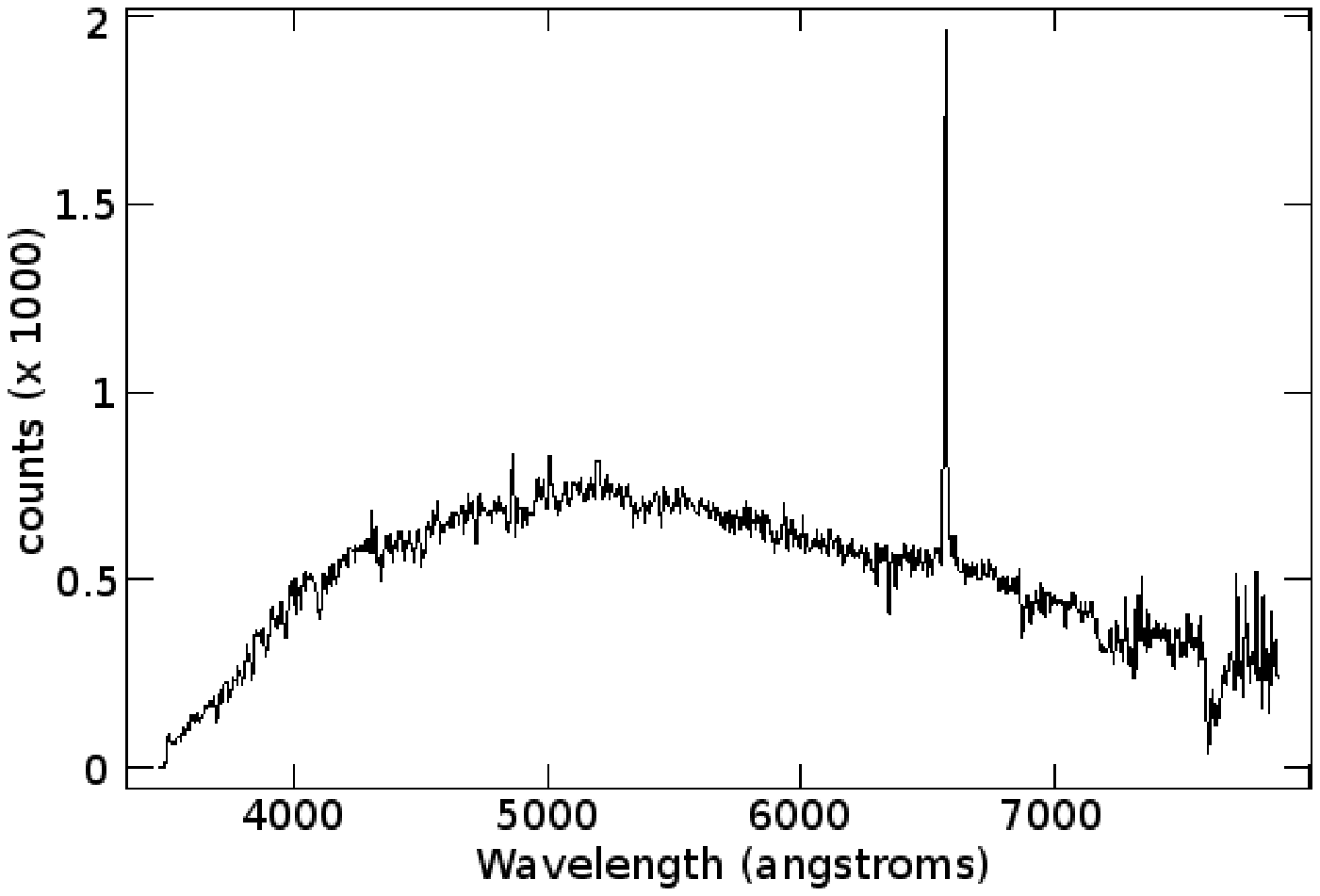}
\end{minipage}
\caption{Same as above for RPs286 also known as BE426 (Bohannan \& Epps, 1974). M$_{H\alpha}$=19.37. Only 2.4 arcsec dia on the image including minor PSF contribution.}
\label{Figure4}
\begin{minipage}[b]{0.355\linewidth}
\centering
\includegraphics[scale=0.57]{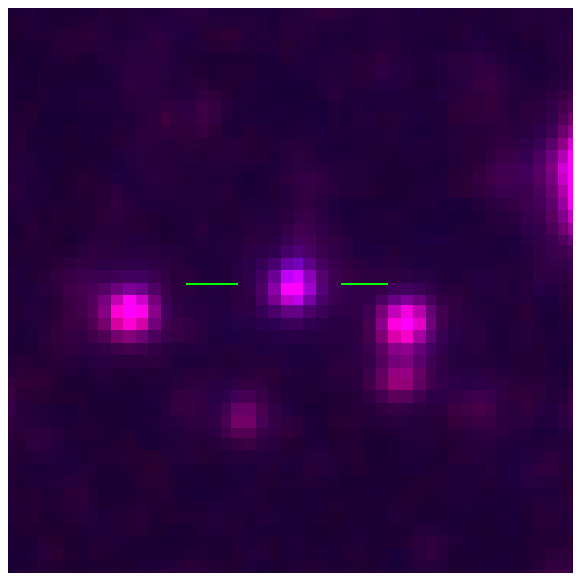}
\end{minipage}
\hspace{0.1cm}
\begin{minipage}[b]{0.65\linewidth}
\centering
\includegraphics[scale=0.36]{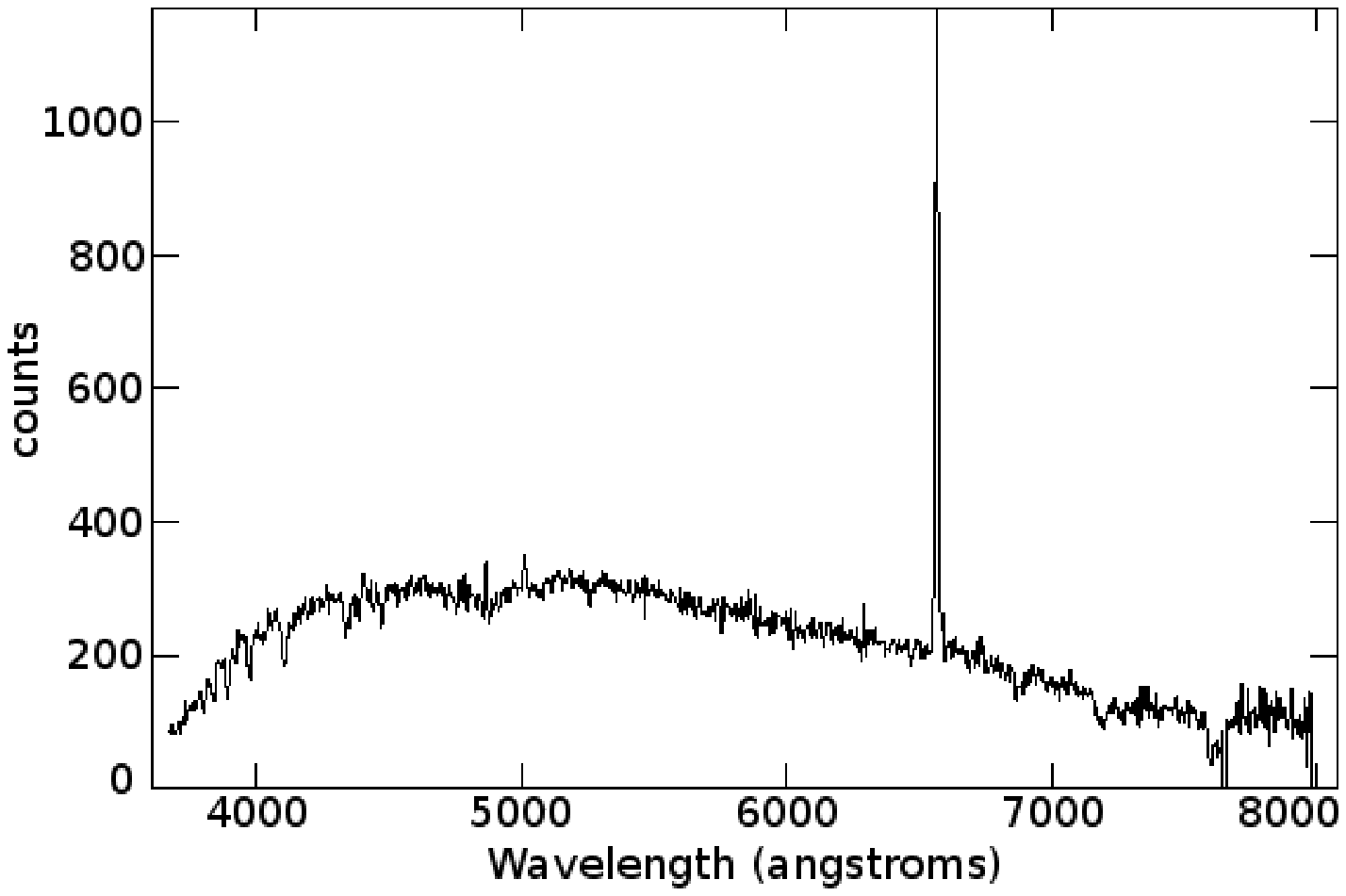}
\end{minipage}
\caption{Same as above for newly discovered emission-line star RPs338. M$_{H\alpha}$=20.19. Only 2.8 arcsec dia. on image.}
\label{Figure5}
\end{figure}

\label{subsection1.1}

\subsection{Emission-line star discovery technique and criteria}

The deep UKST H$\alpha$ survey of the LMC was originally undertaken in order to uncover multiple compact emission sources. Our successful search for extremely faint PNe (Reid \& Parker 2006a,b) is proof of its worth. It soon became clear, however, that the depth and resolution of the maps allowed us to also search for low luminosity stellar sources which exhibit detectable emission-lines. Since our aim was to uncover faint sources, we largely ignored extremely bright stars, whether or not they exhibited H$\alpha$ emission. Many of the better known, bright emission-line stars in the LMC will therefore not appear in this work. What we have included in our survey is a large sample of emission-line star candidates that comply with the expected luminosity of brightest to faintest LMC PNe (M$_{B}$ 13 - 24).

Candidate emission-line stars were found using an adaptation of a
technique available within {\scriptsize KARMA}, first reported in Reid \& Parker (2005). The SR images were
assigned a false red colour and merged with the H$\alpha$ narrow-band images
assigned a blue colour. Careful selection of software parameters allowed the
intensity of the matched H$\alpha$ and SR {\small .fits} images to be
perfectly balanced allowing only peculiarities of one or other
pass-band to be observed and measured. Using this technique, normal continuum stars appear uniformly pinkish in colour. Emission objects such as
\HII~regions and PNe are strongly coloured
blue. The broader point spread function (PSF) of the H$\alpha$ line in emission-line stars creates a
faint blue aura around the star, allowing them to be easily detected. Figures~\ref{Figure1} to 5 show a small 30 $\times$~30 arcsec area of the stacked
SR and H$\alpha$ maps featuring Be stars, RPs255, RPs256, RPs285, RPs286 and RPs338\footnote{RPs refers to \textbf{R}eid \textbf{P}arker \textbf{s}tar}~respectively at the centre together with their confirmatory 2dF spectrum. Spectroscopic confirmation shows us that the wider and more diffuse halo seen surrounding examples such as RPs256 (Figure~\ref{Figure2}) strongly indicates the presence of forbidden lines in the spectrum, leading to its classification as a B[e] star.

Even with a narrow band H$\alpha$ filter, the presence of faint Balmer lines in LMC emission-line stars can be very difficult to detect. Although it can be quite easy to miss such faint sources, the colouring and merging of
the maps makes detection straightforward,
preventing objects above a certain EW threshold from being overlooked and allowing the full depth gain
of the maps to be utilized.

\label{subssection1.2}

\section{Spectroscopic confirmation of candidate emission-line stars}

Having used the stacked H$\alpha$ and SR maps to catalogue over 2,000 emission sources, a large follow-up spectroscopic programme was undertaken in order to identify and classify each source. The most effective and efficient way to follow-up such a large number of objects was to use wide-field multi-object spectroscopic (MOS) systems such as 2dF on the Anglo-Australian Telescope (AAT), 6dF on the UK Schmidt Telescope (UKST) and FLAMES on the Very Large Telescope (VLT). Bright, extended emission objects were selected to be observed individually using long-slit spectroscopic systems on the South African Astronomical Observatory (SAAO) 1.9m telescope and the Mount Stromlo and Siding Spring Observatory (MSSSO) 2.3m telescope.

In Table~\ref{table 1} we summarise details regarding the
spectroscopic follow-up observations. The field names are observation identifications or
object names in the case of the 2.3m observations. Each of these multi-fibre observations have different
central coordinates. The first three 2dF fields, with prefix ST.., are service time runs. Classical observations using 2dF on the AAT provided 15 pointings of 1 degree radius labeled A to O. FLAMES observations on the VLT provided 9 field pointings with an 11 arcminute radius. The FLAMES observations were centred on several
of the densest areas on the LMC main bar. The three fields observed with 6dF on the UKST were repeated, subsequently with a different set of stars and extended objects, maximising use of the wide 6 arcsec fibres.
\label{section2}

\subsection{2dF observations}

A five night observing run on the AAT using 2dF (Lewis et al.
2002) was undertaken in December 2004 to spectroscopically confirm LMC emission candidates. The identification of peculiarities associated with H$\alpha$ excess in various object types (see Reid \& Parker 2006a for more details) indicated that we could expect our candidates to be a mixture of PNe, compact \HII~regions, and emission-line stars such as Be, Ae, WRs, T Tauri, M giants, carbon stars and a number of symbiotics. 2dF was an ideal choice of
instrument for the spectroscopic follow-up of large numbers of
candidate emission objects due to its unique ability to simultaneously observe
400 targets (including objects, fiducial stars and sky positions) with 2 arcsec fibres over a wide 2 degree diameter field area. The large corrector lens incorporates an atmospheric dispersion compensator, which is essential for wide wavelength coverage using small diameter fibres.

\begin{table*}
\caption{Observing logs for LMC Emission-line object follow-up. In some instances the same object has been observed multiple times at different resolutions and in overlapping fields.}
\begin{tabular}{|l|c|c|c|c|c|c|c|c|c|c|}
  \hline \hline
   &   &   &    &   &    &   &    &   \\
  Field Name & Telesc. & Date & Grating  &  Dispersion &Central  & Coverage &
   T$_{exp}$ & N$_{exp}$ & N$_{obj}$  \\
     &   &   &  Dispenser &  \AA/pixel & $\lambda$ (\AA) & $\lambda$ (\AA) & s  &
       &     \\
   \hline
  2dF-ST1 &  AAT & 26 Nov-03 & 300B & 4.299 & 5841 & 3650 - 7960 & 1500 & 2 & 131 \\
  2dF-ST2 &  AAT & 26 Nov-03 & 300B &  4.299 & 5841 &3650 - 7960 & 1500 & 2 & 80 \\
  2dF-ST3 &  AAT  & 15 March-03 &  300B  &  4.299  &  5852 & 3660 - 7970 &  1800  &
  2  &  81  \\
     a1550,061-213      &   1.9m    &   09-13 Nov-04   &   300 &   5   &   5800  & 3850 - 7738  &   800 &   2   &   11  \\
a1550,214-324      &   1.9m    &   11-15 Nov-04   &   1200    &   1   &   6563  & 6000 - 7120  &   1000    &   2   &   10  \\
FLAMES 1-9   &  VLT   &  5-7 Dec-04 &  LR2   &  0.339  &  4272 & 3960 - 4567 &  1000    &   3  & 420  \\
FLAMES 1-9  &   VLT   &  5-7 Dec-04  & LR3   &  0.339  &  4797 & 4500 - 5077 &  1000    &   3  & 420  \\
FLAMES 1-9   &  VLT   &  5-7 Dec-04  & LR6   &  0.339  &  6822 & 6438 - 7172 &  1000    &   3 &  420  \\
2dF A-O   &     AAT &   13-16 Dec-04   &   300B    &   4.3    &   5852 & 3660 - 7970 &   1200    &   3   &   3603 \\
2dF-1200R A-O &     AAT &   17-18 Dec-04   &   1200R   &   1.105   &   6793 & 6220 - 7340 &   1200    &   2   &   3303 \\
RPs   &   2.3m    &   07-18 Jan-05   &   600R+B  &   2.2 &   4600  & 3600 - 5570 &   900 &   2   &   56   \\
RPs   &   2.3m    &   07-18 Jan-05   &   600R+B  &   2.2 &   6563  & 5515 - 7520 &   900 &   2   &   56   \\
6dF 1-3  &   UKST  &  3-5 Feb-05   &    425R    &   0.62    &    6750 & 5318 - 7576 &     600    &   3 &  573   \\
6dF 1-3  &   UKST  &  3-5 Feb-05   &    580V    &  0.62     &    4750 & 3948 - 5600 &     600   &   3 &  573   \\
    \hline
\end{tabular}
\label{table 1}
\end{table*}

The observations provided $\sim$4,000 spectra. Individual exposure times were mostly 1200s using the 300B grating
with a central wavelength of $\lambda$5852\AA~and wavelength range
$\lambda$3600-8000\AA~at a dispersion of 4.30\AA/pixel. These low-resolution
observations, at 9.0\AA~FWHM, were the primary means of
object identification and were used in cross-correlation to provide spectral classifications. All fields were then re-observed using the higher resolution 1200R grating to gain our radial velocities with wavelength range $\lambda$6220-7340\AA.
\label{subsection2.1}

\subsection{ESO VLT FLAMES observations}

Our data includes additional spectroscopic observations in dense regions of the LMC main
bar, undertaken using the multi-object fibre spectroscopic
system, FLAMES (Pasquini 2002) on the VLT UT2 over three nights in
December 2004. The OzPoz positioner on FLAMES was used to position the 130 available fibres with an accuracy of better
than 0.1 arcsec. Gratings LR2 and
LR3 allowed us to cover the most important optical diagnostic lines
for emission-line stars in the blue including \OIII\,$\lambda$4363, \HeII\,$\lambda$4686,
H$\beta$ and \OIII\,$\lambda$4959 \& $\lambda$5007 in emission and absorption lines such as \HeI$\lambda$4471, $\lambda$4387, $\lambda$4144, $\lambda$4121, $\lambda$4026, $\lambda$4009 and $\lambda$3820. Grating LR6 covered the
H$\alpha$, \NII\,$\lambda$6548 and $\lambda$6583~lines as well as the \SII\,$\lambda$6716 \&
$\lambda$6731~lines. Using these low resolution gratings allowed us to both identify and classify objects and observe micro-structures such as self-absorption within the Balmer emission lines. The observed FLAMES 25 arcmin diameter fields, containing a total of 420 objects, overlapped with 2dF fields, providing a continuous coverage of the main bar region.
\label{subsection2.2}

\subsection{6dF observations}

A 3 night observing run was also undertaken on the 3-5th February 2005 using the 6dF 150 fibre MOS system on the UKST. Each of these observations covered an impressive 6 degree diameter field on the sky and allowed us to observe candidates that were missed in 2dF observations due to crowding. The separate 580V and 425R gratings provided continuous coverage across the optical range from $\lambda$3700\AA~to $\lambda$7550\AA~for 573 objects observed. A proportion close to 50\% were re-observations of objects previously covered using 2dF, providing additional object confirmation. The wider 6 arcsec fibres on 6dF, compared to the 2 arcsec fibres on 2dF, meant that we had to re-examine the location of each object in order to avoid observing close stellar sources with that instrument. On the other hand, the 6 arcsec fibre meant that it was an excellent choice of instrument for extended sources such as large PNe with post AGB halos and compact \HII~regions.

\label{subsection2.3}

\subsection{Long-slit observations}

Long-slit spectra were obtained using the 1.9m telescope at the South African Astronomical Observatory in November 2004 and 2.3m telescope at Mount Stromlo and Siding Spring Observatory (MSSSO) in January 2005. Both of these observing runs not only provided spectra for object confirmation and classification but assisted our flux calibration for fibre-based observations. Individually, the 1.9m telescope provided both low dispersion spectra for object identification and higher resolution spectra for radial velocities. Light fed to the double-beam spectrograph on the MSSSO 2.3m telescope was split by a dichroic and sent to red and blue optimised detectors. The resulting medium resolution red and blue spectra also provided spectroscopic confirmation of individual objects that were missed during multispec-observations due to overcrowding on field plates.

\label{subsection2.4}

\subsection{Reduction of spectra}

The 2dF data were reduced using the sophisticated {\scriptsize 2dFDR} reduction software provided
by the Australian Astronomical Observatory (AAO) specifically for the reduction of 2dF multi-fibre
spectra. The software performed the standard reduction procedures of
bias and dark subtraction, flat fielding, sky subtraction, tram-line mapping to the
fibre locations on the CCD, fibre extraction, arc line identification,
wavelength calibration and fibre throughput calibration as well as providing a user interface with
several options, specific to 2dF multi-fibre reductions. Specific bias frames are not required as the software simply makes use of the underscan/overscan bias strips on each CCD exposure.

The {\scriptsize FIT} method
of fibre extraction was used as it simultaneously fits Gaussians to the spectrum being extracted and
to the two either side of it, allowing the amount of overlap at each
point along the spectrum to be evaluated. This method also minimises
contamination between fibres and was applied to all the reductions.

To perform the sky subtraction, the data was first corrected for the
relative fibre throughput, based on a throughput map derived from
about 15 dedicated sky fibres which were carefully selected across the 2dF field to avoid stars and ambient emission. The relative intensities of the skylines in the object data frame were used to determine the relative fibre
throughput. This method saves time, as no off-set sky observations
were required.

Cosmic rays were rejected either automatically during the process of
combining two or more observations on the same field setup. This method was used because under certain
circumstances the spatial profile is sometimes sensitive to the
spectral structure of the data and it can mistake a strong emission-line for a cosmic ray.

Raw data from 6dF on the UKST was reduced using a tailored 6dF variant of the same ({\scriptsize 2dFDR}) data reduction software. A specific input file informs the software that 6dF data is to be reduced. Like 2dF, a separate file relating to the specific grating must be used. Again, cosmic rays were rejected automatically during the process of
combining two or more observations of the same field.

VLT FLAMES data were reduced using the pipeline system provided by ESO through the `{\scriptsize GASGANO}' Java-based data file organiser developed and maintained by ESO. This graphic interface identifies the input file types, produces a master bias, flat, and dark frame, then reduces and combines the science frames.

The 2.3m and 1.9m telescope spectra were reduced using the standard long-slit {\scriptsize IRAF} tasks
{\scriptsize IMRED, SPECRED} and {\scriptsize CCDRED} and {\scriptsize FIGARO}'s task
{\scriptsize BCLEAN}. Cosmic rays were rejected when combining frames. One-dimensional spectra were created
and the background sky was subtracted. Final flux calibration used
the standard stars LTT7987, LTT9239, LTT2415 and LTT9491.

\label{subsection2.5}

\section{Flux Calibration of the 2dF Fibre Spectra}

The large proportion of objects observed with 2dF means that a reliable flux calibration of the LMC stellar emission-lines was required in order to compare stellar spectra from different 2dF fields, to make meaningful comparisons between fibre spectroscopy and long-slit observations of individual objects, and to create a luminosity function.

Altogether, 18 overlapping 2dF fields, 9 FLAMES fields and 6 6DF multi-object fields were observed in order to cover the entire central 25deg$^{2}$ survey region of the LMC. To calibrate the resulting data counts, we used PNe with low continuum levels and well determined fluxes gained from HST observations (see Reid \& Parker 2006a, 2006b). These objects were deliberately included and observed on each field plate for use as flux calibrators for each individual field.

The process involved matching each spectral line on each field plate from each {\scriptsize CCD} camera
to raw PN fluxes gained from HST exposures. The individual H$\beta$ and H$\alpha$~2dF line intensities for known PNe observed on each {\scriptsize CCD} and each field plate exposure were
 plotted against HST-gained published fluxes for the same lines (see Figure 2 in Reid \& Parker, 2010).

 The agreement of flux-calibrated PNe from each spectrograph/field plate combination was considered robust enough (within 0.2 dex) to allow calibration to all the H$\beta$~and H$\alpha$~emission-lines for other emission objects observed in the same field.
 In each case, a line of best fit was derived and the underlying linear equation extracted. This
 equation became the calibrator for each emission-line in each object where the {\scriptsize CCD} and individual 2dF field
 plate exposure was the same. Full details including a discussion on the reliability of the method are presented in Reid \& Parker (2010).

\label{section3}

\section{Spectral Classification}

Spectral classification of all the emission-line stars was undertaken to assist in various studies such as the distribution of emission by stellar population, the estimation of central star temperatures, creation of H-R diagrams and improving our understanding of Balmer emission in stars of varying temperatures. We touch on some of these issues later in this paper.

\label{section4}

\subsection{Method of classification}

\begin{figure}
\begin{center}
  \includegraphics[width=0.44\textwidth]{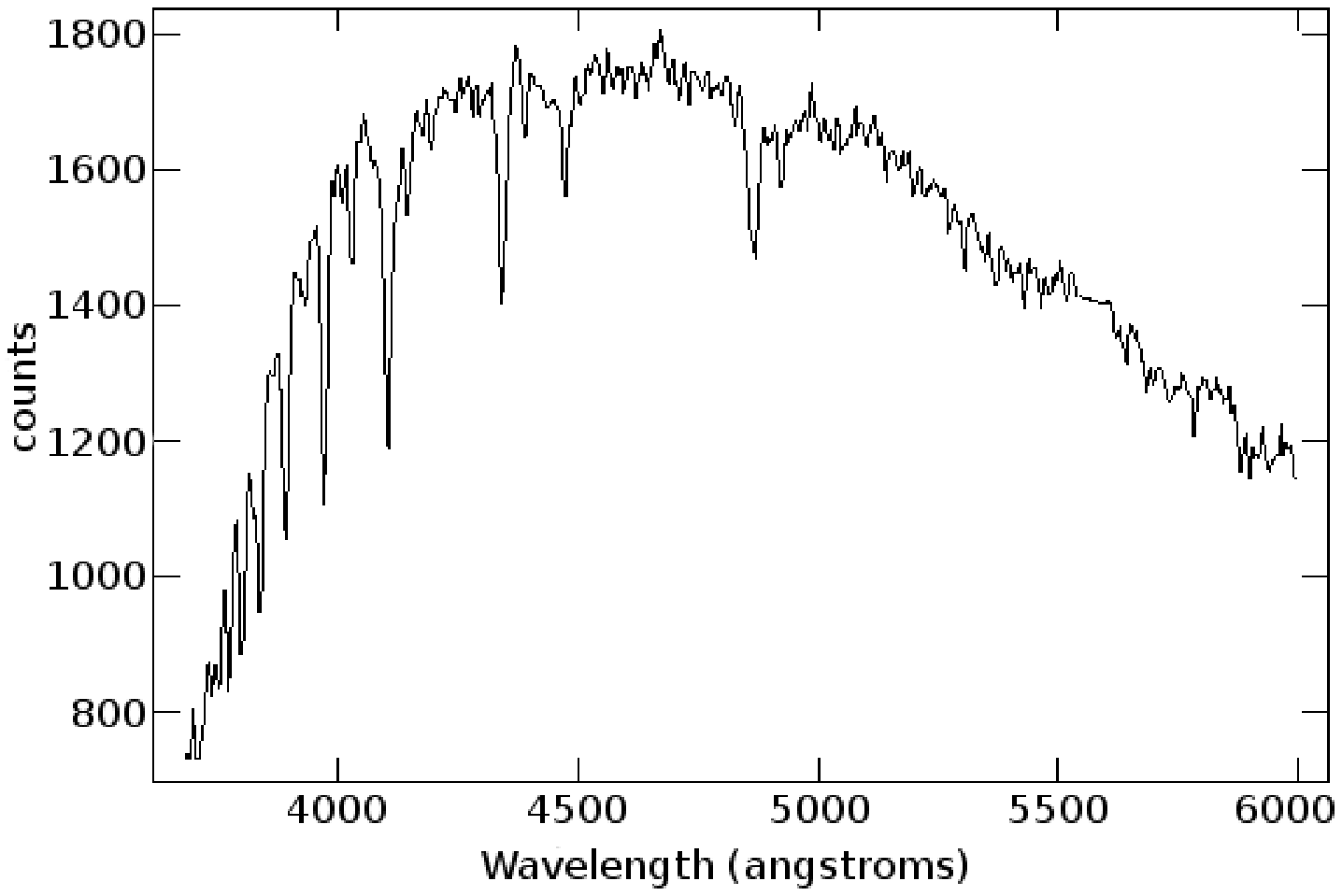}\\
  \includegraphics[width=0.45\textwidth,height=66mm]{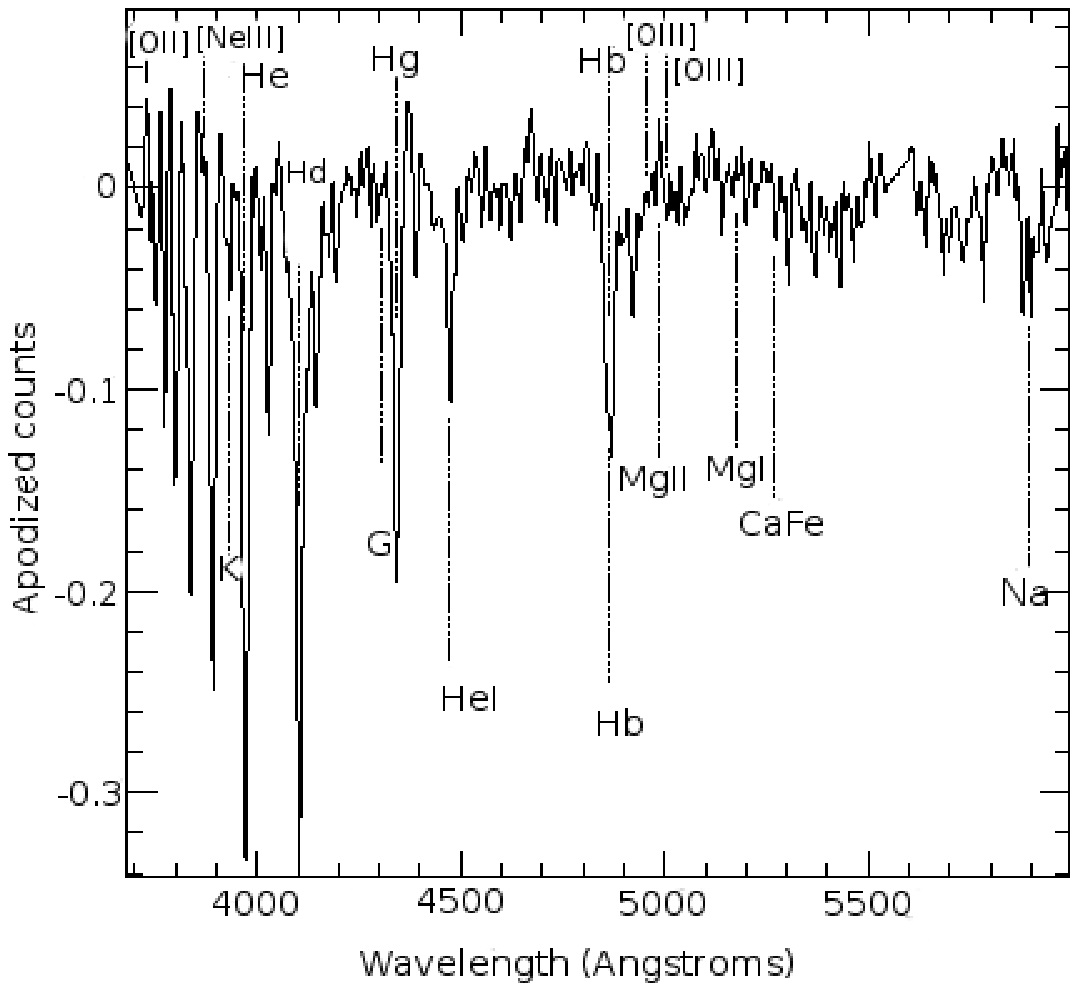}\\
  \includegraphics[width=0.45\textwidth]{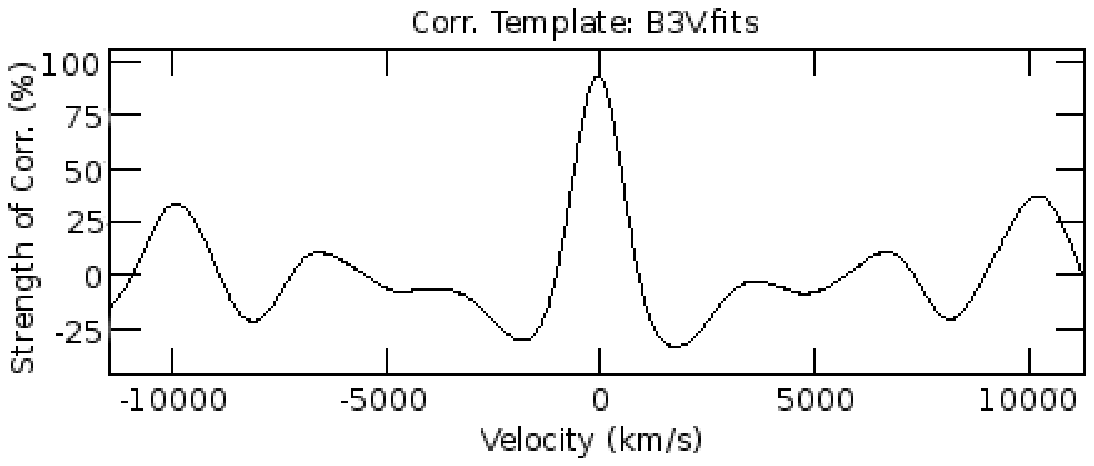}\\
  \caption{The top image shows the blue end of the RPs1326 optical spectrum prior to the removal of emission-lines and continuum. The centre frame shows the apodized and continuum subtracted spectrum, created within XCSAO and used in cross-correlation to match the best fitting template. The lower frame shows the strength of the resulting correlation, represented by the central gaussian curve, once the task has found the best-fitting template.}
  \label{Figure6}
  \end{center}
  \end{figure}
\begin{figure}
\begin{center}
  \includegraphics[width=0.45\textwidth]{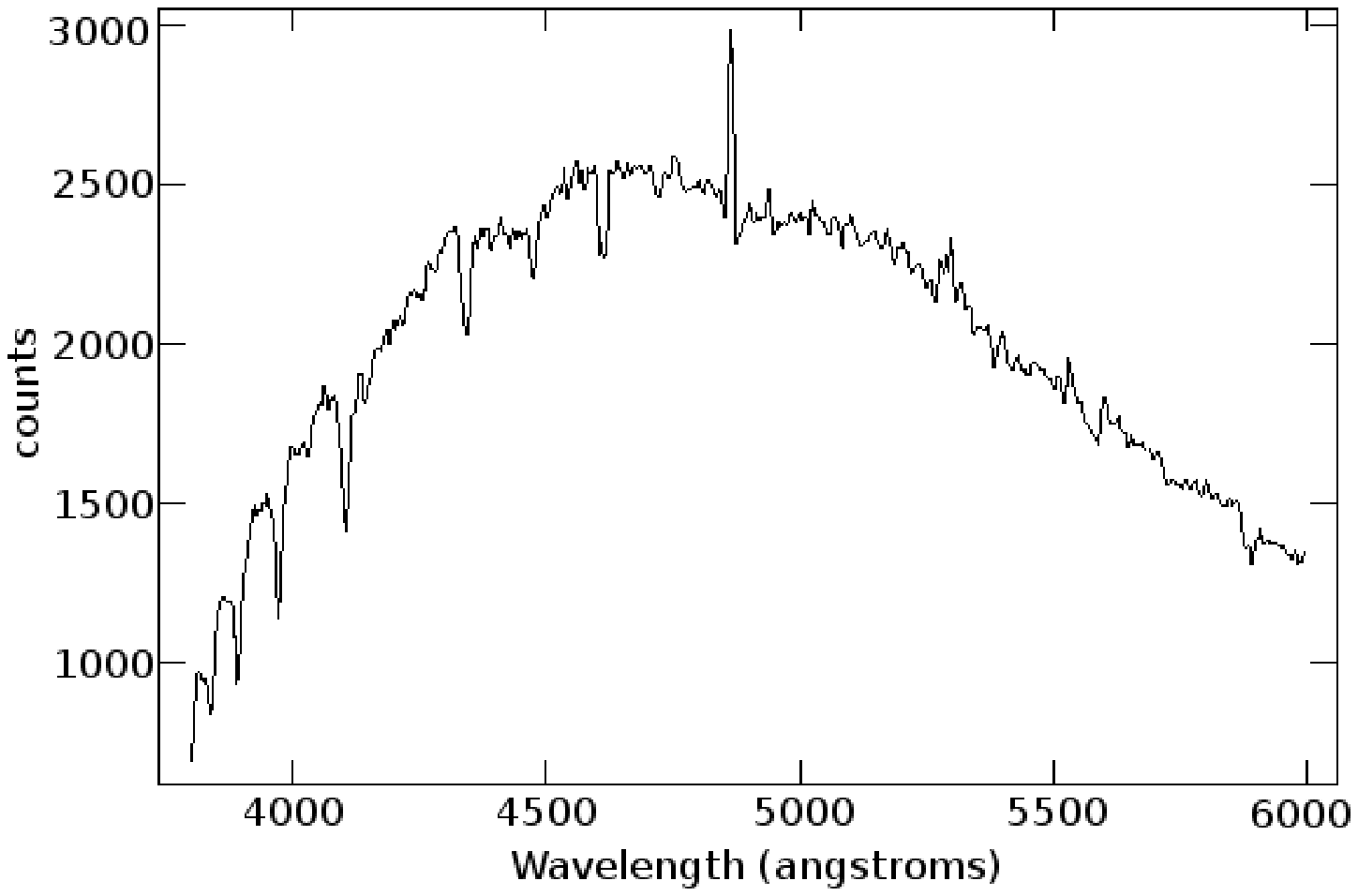}\\
  \includegraphics[width=0.45\textwidth,height=66mm]{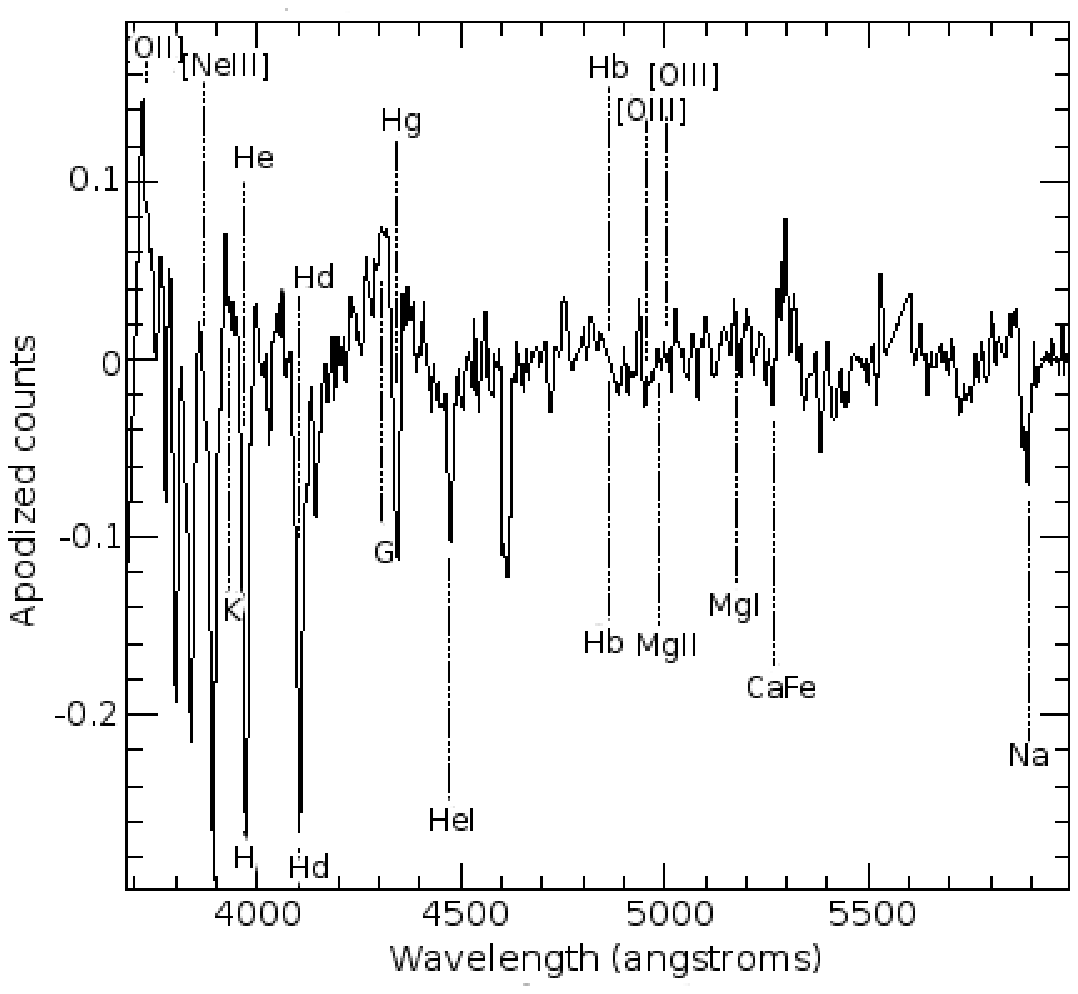}\\
  \includegraphics[width=0.45\textwidth]{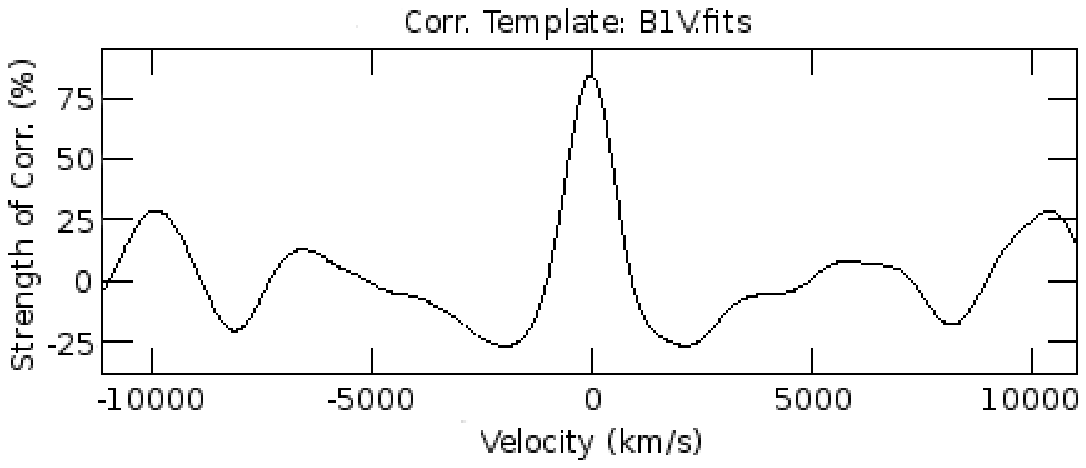}\\
  \caption{The same as Figure~\ref{Figure6} but showing the blue end of the RPs1262, a B1V optical spectrum with top: the original blue end including emission-lines and continuum prior to their removal, middle: the continuum-subtracted and apodised spectrum with detected lines identified automatically by the software, bottom: the correlation. }
  \label{Figure7}
  \end{center}
  \end{figure}
  \begin{figure}
\begin{center}
  \includegraphics[width=0.45\textwidth]{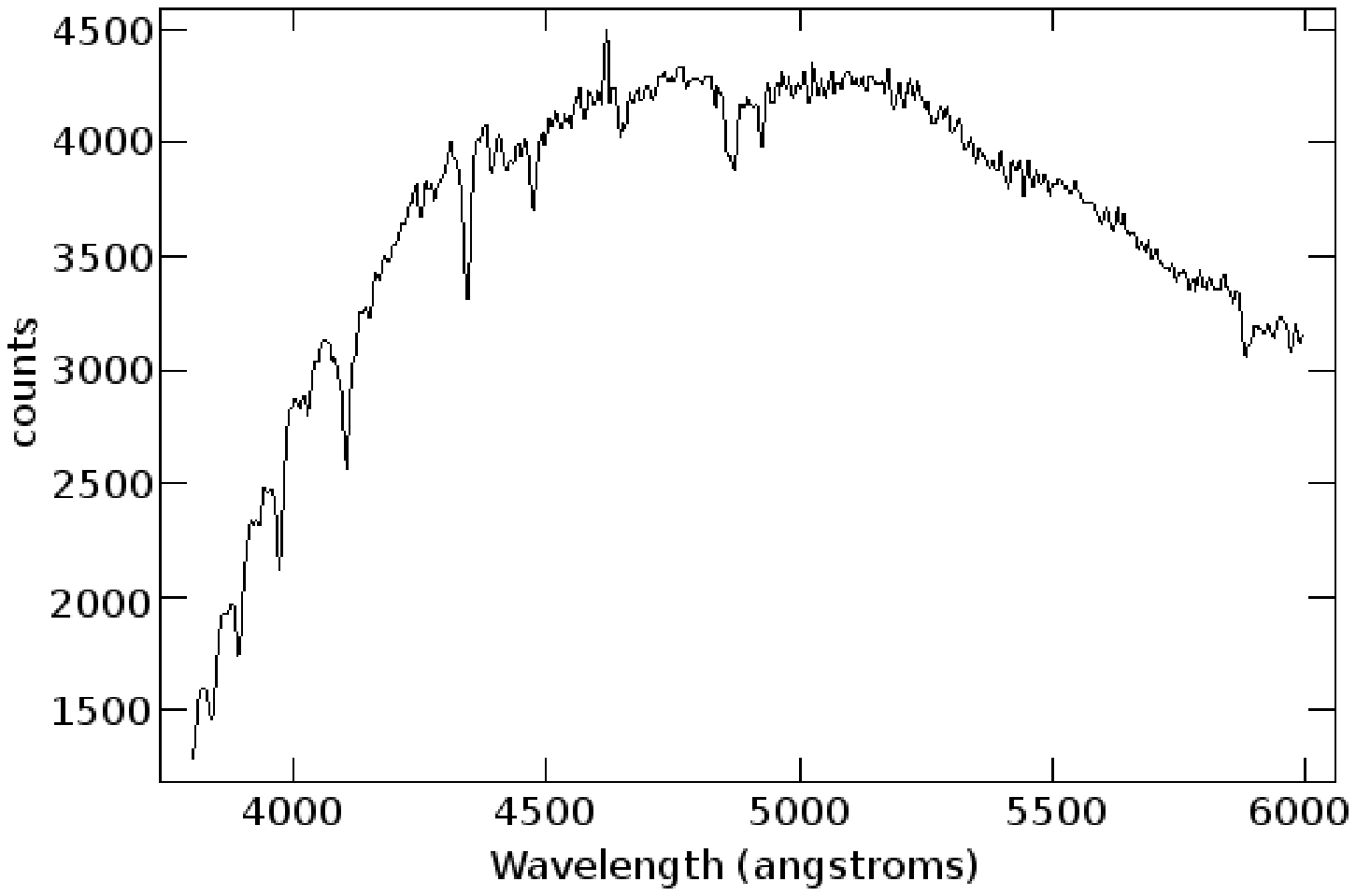}\\
  \includegraphics[width=0.45\textwidth,height=66mm]{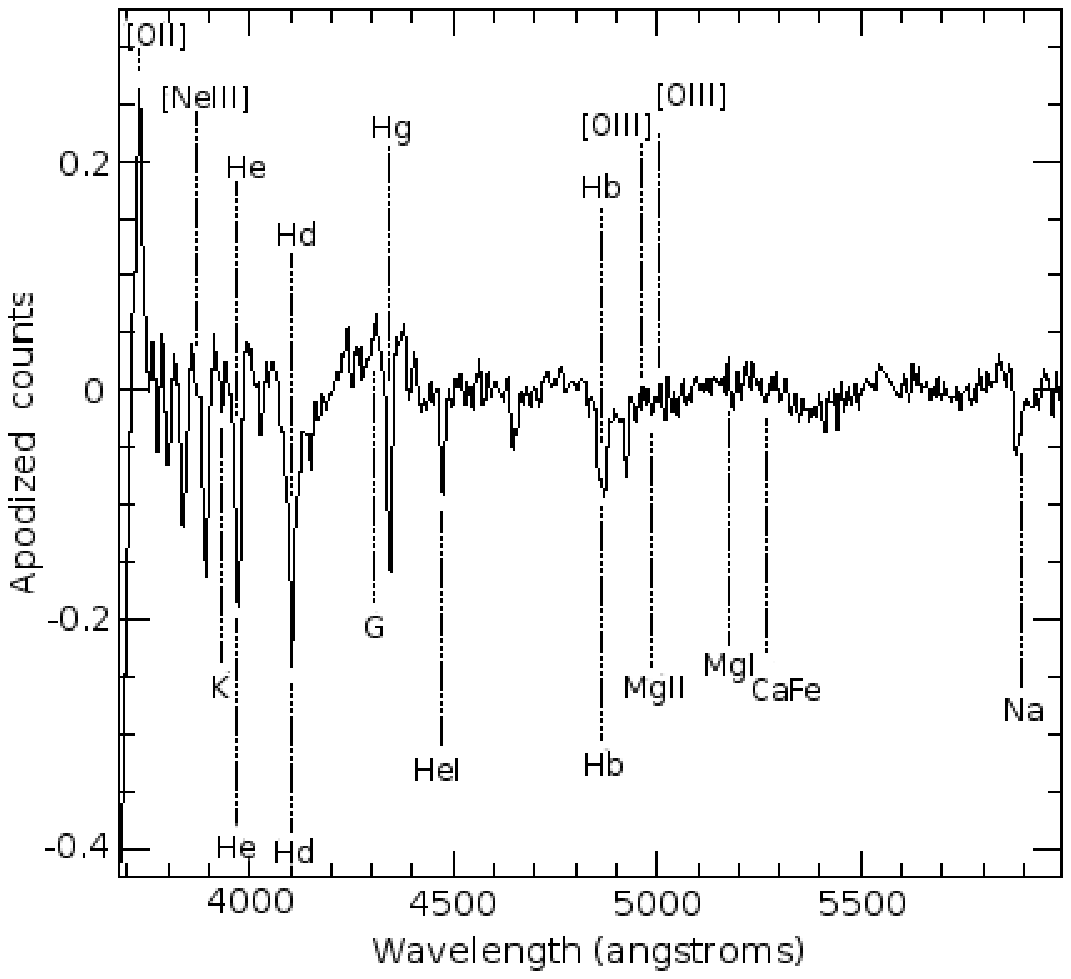}\\
  \includegraphics[width=0.45\textwidth]{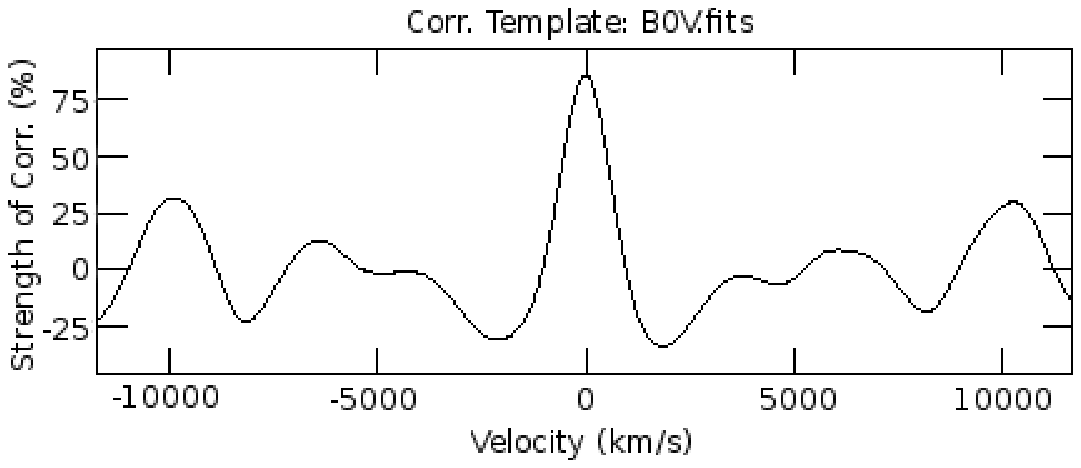}\\
  \caption{The same as Figure~\ref{Figure6} but showing the blue end of RPs1367, a B0V optical spectrum with top: the original blue end including emission-lines and continuum prior to their removal, middle: the continuum-subtracted and apodized spectrum, bottom: the correlation. }
  \label{Figure8}
  \end{center}
  \end{figure}

To assign a spectral
classification, it is neccessary to measure the strengths and widths of
various absorption features which depict specific stellar
temperatures and surface gravities, independent of any associated emission characteristics. To assist this process we used standard stellar spectra supplemented by 10 LMC emission-line stars from our sample with recognised spectral classifications as templates. The spectral standards were based on observations available
from Jacoby et al (1984), Turnshek et al (1985), Silva and Cornell
(1992), Pickles (1998) and Le Borgne et al.
(2003).

The classification of emission-line stars is complex and
often problematic due to their variability and atmospheric activity.
The strength and profile of the Balmer lines in emission only lend moderate assistance to classification, although the equivalent width of H$\gamma$ can be a good indicator of spectral type and luminosity in main sequence stars (Underhill \& Doazan 1982). In the spectra of young stars such
as T Tauri stars, photospheric absorption lines can be filled in or
disguised by UV radiation from accretion hotspots (Hartigan et al.
1995; Gullbring et al. 1998), making classification difficult.
Further complication arises from active Post Main Sequence (PMS) stars where most
spectral lines are in emission (Cohen \& Kuhi, 1979; Hern\'{a}ndez
et al., 2004). These types are usually denoted as `continuum stars'
since it is virtually impossible to accurately assign a spectral
type.

Due to the large number of emission-lines stars to be classified in
this survey, as a first step, a cross-correlation routine was employed. Although the
{\small IRAF} {\small XCSAO} task was originally developed in order
to cross correlate galactic spectra against templates and gain
redshifts (Tonry and Davis, 1979), it works equally as well
as a spectral classification tool with stellar templates. The task identifies the closest spectrum in
terms of line strengths and widths found and then returns a velocity
along with the name of the best matching template through cross-correlation based on fourier transformations.

  In order to produce the most accurate result, emission-lines (mainly the Balmer series and any residual telluric sky lines which can effect the cross-correlation) were removed. The continuum was then removed using the {\small IRAF} {\small CONTINUUM} task in order to cross-correlate the absorption lines alone. This negated the influence of the continuum where it was either stronger or weaker than the best matching spectrum in the templates, which were also continuum subtracted. Apodization within {\small XCSAO} uses a cosine bell to attenuate data on the ends of the spectrum, reducing high wave number fourier components that would be produced by abrupt cutoffs at the ends of the spectra, effectively smoothing out the continuum across the full length of the spectrum. Examples are shown in figures~\ref{Figure6},~\ref{Figure7}~and~\ref{Figure8} using only the blue end of the optical spectrum which contains the main diagnostic lines for spectral classification. The same is not true for late-type G to SC stars. For these stars, removal of emission-lines is still necessary but the overall shape of the spectrum becomes increasingly important with decreasing spectral type. By late K types it was already necessary to match the continuum with the templates using the wider optical spectrum ($\lambda$3700 to $\lambda$8000).


 Having run the above-mentioned tasks, the raw spectra, including the emission lines, were then inspected and measured. B-type stars are strongly characterised by He and Balmer emission-lines. HeI lines show a very broad intensity maximum by B2 and B3. The intensity of Balmer lines remains almost constant for Supergiant stars from B0 Ia through to A0 Ia but strengthens in late B giants. For main sequence stars, however, the Balmer lines strengthen from B0V to A0V. A more precise spectral type can therefore be confirmed by defining the ionisation temperature of Si and He supplemented by \CII~and \CIII. The main luminosity criteria are summarised in Table 2. In early B and late A main sequence through to F stars, line ratios are much easier to use for identification due to the larger number of lines available. The luminosity class can be tested by assessing the wings of the Balmer lines, which widen from classes I to V.
 \begin{table*}
\caption{The most important lines examined to assist in follow-up spectral classification after cross-correlation. These include the ratios and equivalent widths of the Balmer lines and the ratios of classification lines in the $\lambda$3500-4800 region of B-type stars. It should be noted that the ratios shown in columns 2-5 more or less depend on the luminosity. For example, HeI is only weakly visible in A0 supergiants.}
 \begin{tabular}{|l|c|c|c|c|}
   \hline
   Class & Ratios \& EW  &                     Ratios     &      Ratios (later types) &  Ratios (latest types)         \\
   \hline\hline
   \textbf{Supergiants} & & & &   \\
   B0 Ia-B2 Ia & H$\alpha$/H$\beta$/H$\gamma$ & \HeII\,4542/\HeI\,4471  & \SiIII\,4552/\SiIV\,4089 & \CIII\,4068/\OII\,4076    \\
     &    &    \HeII\,4200/\HeI\,4144    &         &  \CII\,4267/\HeI\,4121     \\
   B2 Ia-B5 Ia & H$\alpha$/H$\beta$/H$\gamma$  &  \SiIII\,4552/\SiIV\,4089 & \CIII\,4068/\OII\,4076  & \CII\,4267/\HeI\,4121  \\
   B5 Ia-A0 Ia & H$\alpha$/H$\beta$  & \SiII\,4128,31/\HeI\,4121 & \SiII\,4128,31/HeI\,4026 & \SiII\,3856, 63/HeI\,3820, 4026   \\
   \textbf{Giants} & & & &  \\
   O5 III-B0 III & H$\alpha$/H$\beta$/H$\gamma$  & \HeII\,4200, 4542/\HeI\,4471  &  \SiIV\,4089/H$\delta$  & \HeI\,4388/H$\gamma$    \\
   B0 III-B5 III  & H$\alpha$/H$\beta$/H$\gamma$  & \SiIV\,4089/\SiIII\,4553  &  \CIII\,4647-51/\HeI\,4388 & \MgII\,4481/\HeI\,4471  \\
   B5 III-A0 III  & H$\alpha$/H$\beta$  &  \MgIII\,4481/HeI\,4471  &  \SiII\,4128,31/\HeI\,4144,4026  &  \\
   \textbf{Main Sequence} & & & &  \\
   O4 V-B0 V & H$\alpha$/H$\beta$/H$\gamma$  & \HeII\,4542/\HeI\,4471  &  \HeII\,4686/\HeI\,4922 & \SiIV\,4089/\HeI\,4144    \\
   B0 V-B5 V & H$\alpha$/H$\beta$/H$\gamma$  & \SiIV\,4089,4116/\HeI\,4121 & \HeII\,4686/\HeI\,4713  &  \CIII\,4068-70/\HeI\,4009  \\
     &    &    &   &   \CIII\,4647-51/\HeI\,4713   \\
   B5 V - A0 V  &H$\alpha$/H$\beta$  & \SiII\,4128,31/\HeI\,4144,4026  &  \MgII\,4481/\HeI\,4471*  &  \CII\,4267/\MgII\,4481   \\
   \hline
 \end{tabular}\label{table 2}\\
 $^{*}$ The \HeI$\lambda$4471 line is all but gone in main sequence stars by B8\,V.
 \end{table*}

 By applying these criteria, we re-classified 40 Be stars which were automatically classified as luminosity class I supergiants in the cross-correlation routine. Most of these were re-identified as giants or subgiants. Fast rotation of the Be stars causes the Balmer lines to broaden thereby matching spectra to supergiant stellar templates. This effect was countered by examining each spectrum with reference to the ratios as shown in Table~\ref{table 2}.

 \label{subsection4.1}

\subsection{Results of spectral classification}

Although this paper is presenting the hot emission-line stars, it is important to note that the UKST H$\alpha$~survey also uncovered a large number of cooler G to SC stars which either emit strongly or are bright at H$\alpha$. These late stellar objects, which will be the subject of a second paper of this series, are listed briefly here in order to compare detection rates.

The majority of emission stars found have been classed as Be, [Be] (V\,-\,III) stars and M (III) giant stars. The letter `e' indicates that, at the very least, the first member of the Balmer series (H$\alpha$) is in emission. Although we identified 13 supergiant B stars with H$\alpha$ emission, these types are not generally known as Be stars, a classification reserved for luminosity classes V, IV and III.  Table~\ref{table 3} provides a quick breakdown of the various emission-line stars found in the central 25deg$^{2}$ LMC survey. Of these stars, 64 Be stars are previously known variable stars.

\begin{figure}
\begin{center}
  \includegraphics[width=0.5\textwidth]{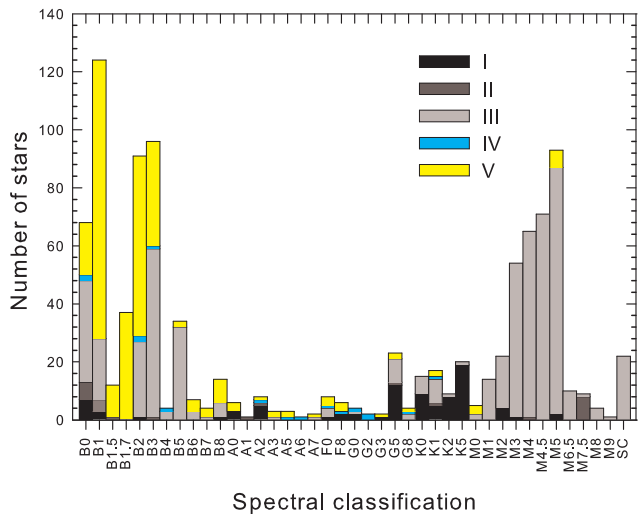}\\
  \caption{Distribution of all types of emission-line stars found within our survey region of the LMC according to number and spectral classification. The number of stars in each class is subdivided in order to express the luminosity class of the star according to the Morgan-Keenan system. It is clear that there are more main sequence stars, bright in H$\alpha$ emission, within the B spectral class. Class SC may be divided into 4 $\times$ SC4/9 stars, 1 $\times$ S4/2, 1 $\times$ S4/6 and 16 carbon stars, 13 of which are C6.  }
  \label{Figure9}
  \end{center}
  \end{figure}

Figure~\ref{Figure9} shows the spectral classification of the identified B to K emission-line stars in our survey.
The number of stars found is subdivided by luminosity class according to the Morgan-Keenan system (Morgan et al. 1943)
where the width of absorption lines are a measure of the size of the star and thus the total luminosity. As per the standard
convention, class I are supergiants, class III are giants and class V are main sequence stars. It is clear that the largest
number of emission-line stars found belong to class B and, of those, the supergiants are mainly found at B0. These supermassive
stars again dominate our detections from classes G5 to K5. The largest spectral class of Be stars represented in our sample
are those on the main sequence.

\begin{table}
\caption{Classification of stellar emission sources for the whole catalogue. The numbers provided in column 4 are not additional but represent the number of stars previously known as variable.}
\begin{tabular}{|l|c|c|c|}
  \hline
  Object Type & Previously  & Newly & known  \\
          &     known      &    discovered  & variable  \\
  \hline\hline
  O stars  &     &  1     &  \\
  Be stars & 82 &  306 & 55 \\
  B[e] stars &  23  &  107 & 9 \\
  Ae stars  &  3  &  29  &    \\
  F  stars   & 3  &   25  &   \\
  G stars &  & 29 &  \\
  K stars  &    & 49 &  \\
  M stars & 86 & 315 & 33 \\
  WR stars & 33 & 6 &  \\
  Carbon stars &  & 16 & \\
  CVs     &   4  &   &  4  \\
  Eclipsing Binaries  &   3  &   & 3  \\
  LBVs    &     &  2   &    \\
  Bp stars   &   2   &  5  &    \\
  AGB stars   &   4   &   &     \\
  Symbiotic stars   &   & 18  &   \\
  hot stars without id. &  & 14 &  \\
  cool stars without id. &    & 7  &   \\
  \hline
\end{tabular}
\label{table 3}
\end{table}

\label{subsection4.2}

\subsection{Types of emission-line stars found}

Of the 468 newly discovered emssion-line stars, we identified 107 B[e] stars that exhibit forbidden emission-lines. They were found in spectral types B0-B9. The most common forbidden emission-lines found in the B[e] stars were \FeII$\lambda$4244,4287,4415,5273,7155, \OI$\lambda$6300,6363, \NII$\lambda$5755,6548,6584, \SII$\lambda$4068,6717,6730, \OII$\lambda$7320, 7330, and \OIII$\lambda$4959,5007, the most frequent being \FeII~and \OI. The ionisation potentials of the last two, less than 25eV, place them lower than the ion energies found in planetary nebulae.

We have also identified early B-type stars with anomalies (weak or strong) in carbon, nitrogen and usually oxygen. These were first labelled CNO stars by Jaschek \& Jaschek (1967). The stars with anomalies in their heavier elements are called Bp stars, where `p' designates `peculiar'. We have identified 5 Bp candidates. They are particularly enhanced in Si-$\lambda$4200, \MnII, \CrII, \EuII~and \SrII.


As the cores of intermediate mass stars (M$_{\ast}$ = 1-8M$_{\odot}$)
become too depleted in hydrogen for fusion reactions, they leave the
main sequence to ascend the Red Giant Branch (RGB) and Asymptotic
Giant Branch (AGB).  At this point, the stars are seen as Miras or OH/IR
stars with maser activity (Winckel, 2003). Although these stars will
become the central stars of planetary nebulae, they are not yet hot
enough to ionise a potential vast halo of expelled material. Nevertheless, the dense, complex atmospheric matter, including possible extended circumstellar envelopes, is ionised sufficiently to be detected in H$\alpha$ and \NII.

The second largest group of stars uncovered in this survey are the M giants. Due to their cooler temperature, these stars
have an spectral energy distribution (SED) that peaks towards the red end of the spectrum. They often exhibit strong excess H$\alpha$ emission originating from the chromosphere which strengthens
with increasing spectral type or decreasing luminosity. For this reason the H$\alpha$ line cannot be used as a classification
criteria and was removed prior to cross-correlation. 

Late-type M giants feature TiO and VO bands which strengthen with decreasing temperature. They also feature Mg $\lambda$5167,5173,5184 until M4III and M6.5V as well as Na\,I $\lambda$5890,5896 although the latter can be overwhelmed by TiO absorption in stars later than M2III.

Our survey uncovered 401 M giant stars with emission, 315 of which are newly identified. Of the 86 previously known M giants, 33 have been found to have variable luminosity. These M giants, together with a number of G and K emission-line stars will be the subject of the next paper in this series.

\subsection{Observed emission line profiles}

The emission line profiles can represent a combination of instrumental broadening, small absorption features which are often broadened by rotation originating from the photosphere of the star, and the emission-line profile produced by the star's circumstellar envelope. Both emission and absorption lines may include kinematic and non-kinematic broadening from effects such as radiative transfer and Thomson scattering which affect the envelope (eg. Hanuschik, 1989). Absorption lines are generally less affected by such effects leaving emission lines to provide important information about the rotation and physical conditions affecting the star and it's circumstellar envelope.

\begin{figure}
\begin{minipage}[b]{0.455\linewidth}
\centering
\includegraphics[scale=0.295]{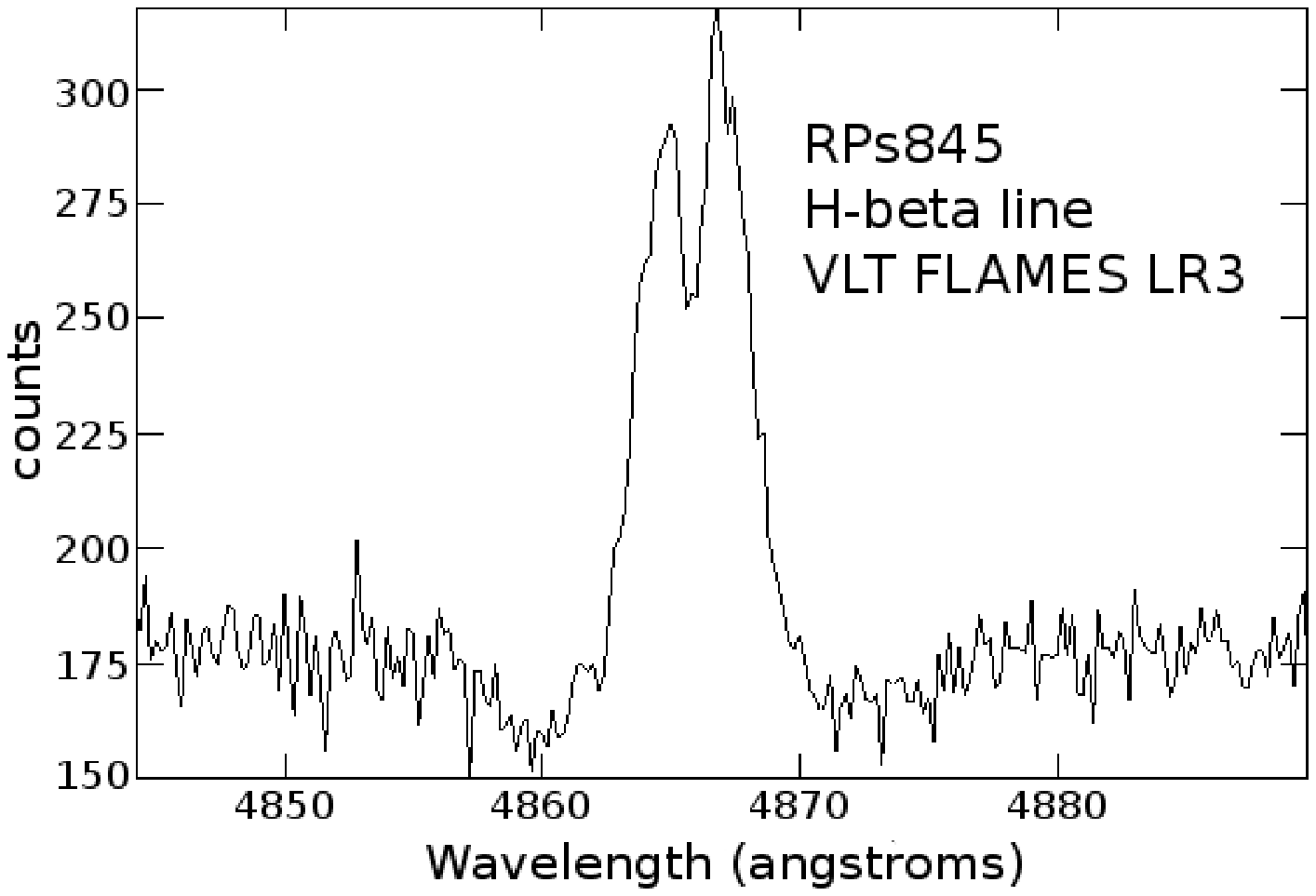}
\end{minipage}
\hspace{0.2cm}
\begin{minipage}[b]{0.455\linewidth}
\centering
\includegraphics[scale=0.295]{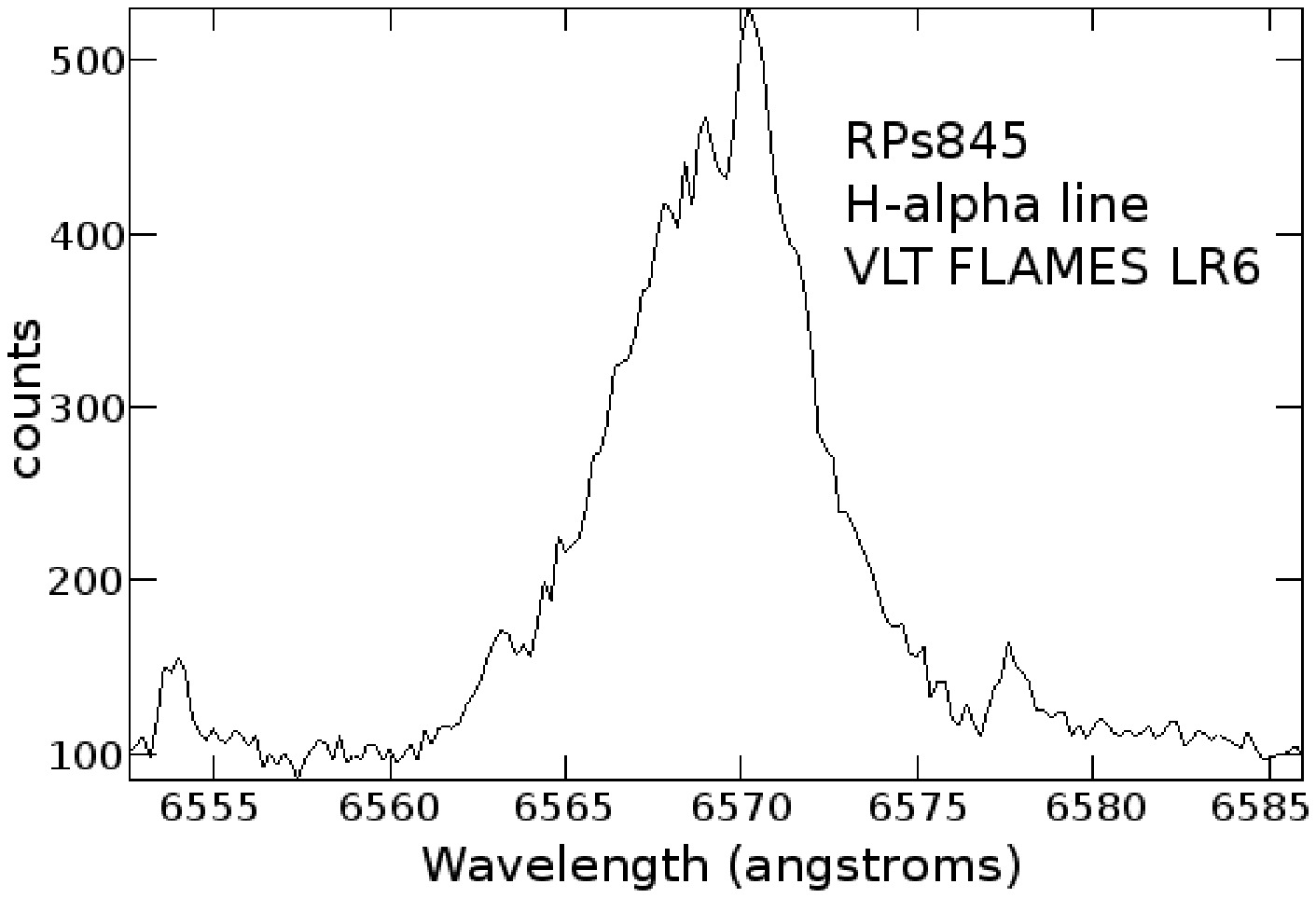}
\end{minipage}
\caption{An example of Balmer line splitting where we can see fine structure components. These profiles often feature three or more emission peaks and minute detailed features extending down to the continuum. The example shown is RPs845 with H$\beta$ left (FLAMES LR3 grating) and H$\alpha$ right (FLAMES LR6 grating). The absorption wings of the H$\beta$ line are also
greatly broadened by the Stark effect indicating that these are main sequence stars where the
gravity and electron pressure is large.}
\label{Figure10}
\end{figure}
\begin{figure}
\begin{minipage}[b]{0.455\linewidth}
\centering
\includegraphics[scale=0.29]{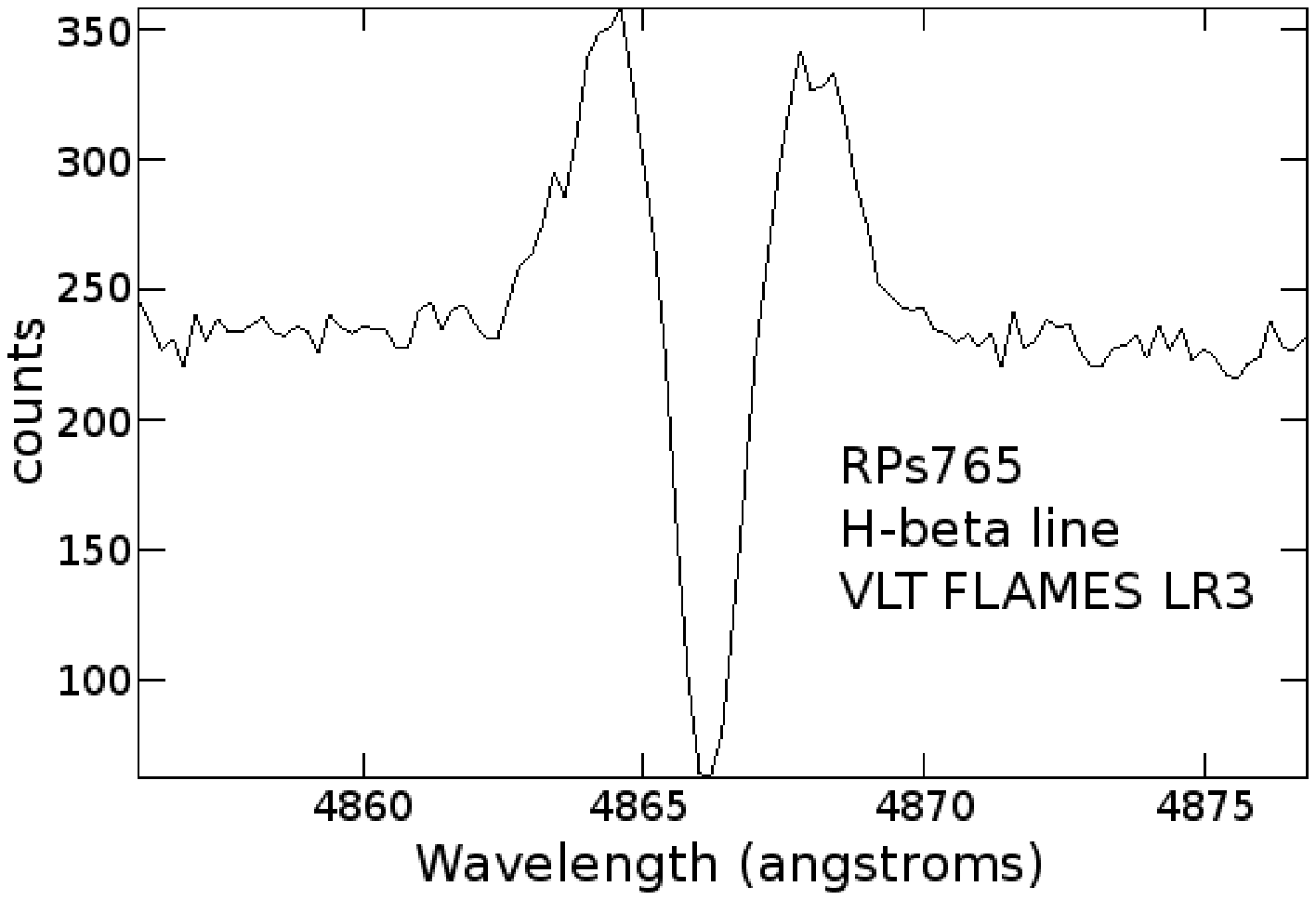}
\end{minipage}
\hspace{0.2cm}
\begin{minipage}[b]{0.455\linewidth}
\centering
\includegraphics[scale=0.29]{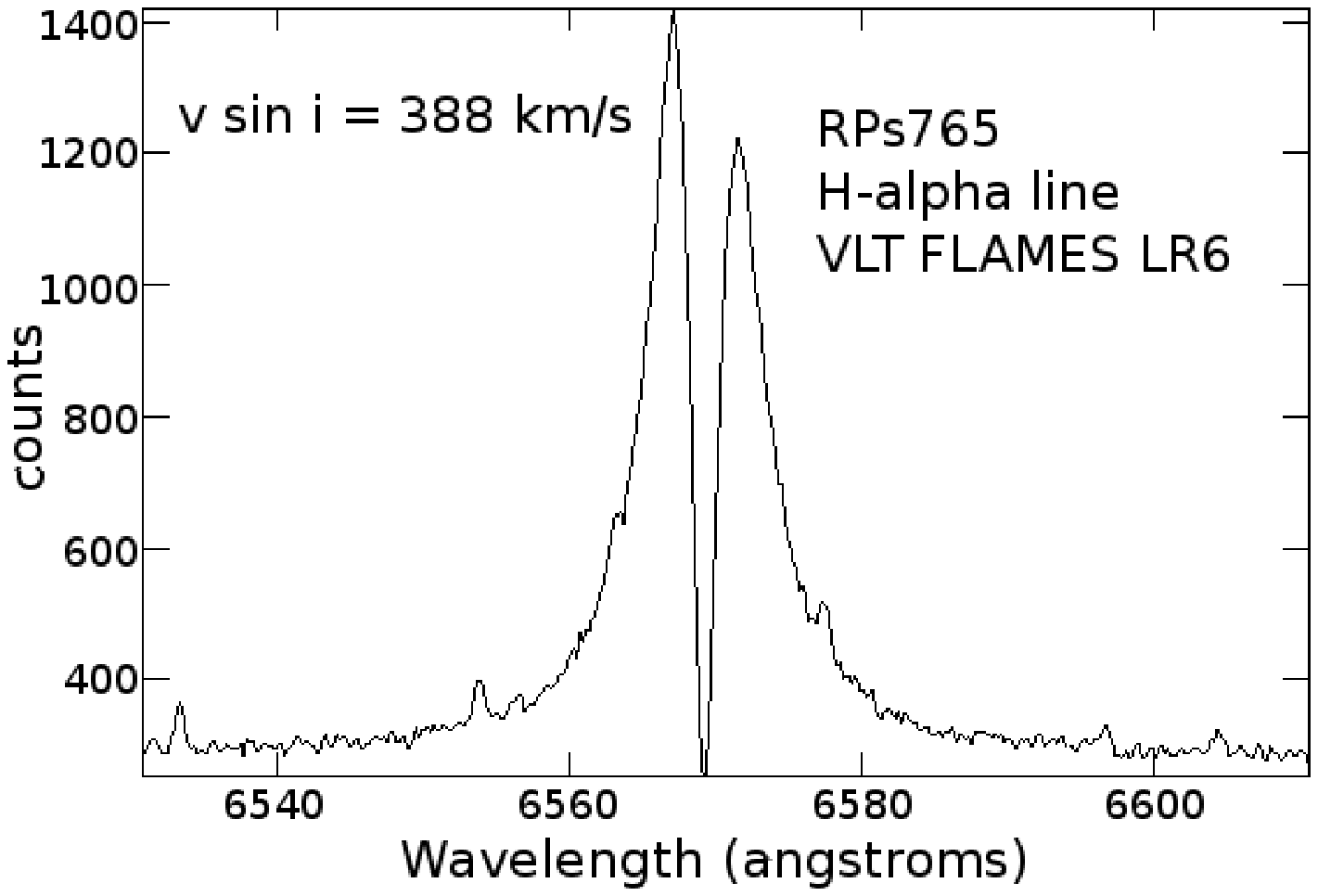}
\end{minipage}
\caption{An example of Balmer line splitting, typical of a `shell star' or more correctly, a star going through a shell phase, where the central absorption on the H$\alpha$ line extends below the stellar continuum. The example shown is RPs765 with H$\beta$
left (FLAMES LR3 grating) and H$\alpha$ right (FLAMES LR6 grating). }
\label{Figure11}
\end{figure}
\begin{figure}
\begin{minipage}[b]{0.455\linewidth}
\centering
\includegraphics[scale=0.29]{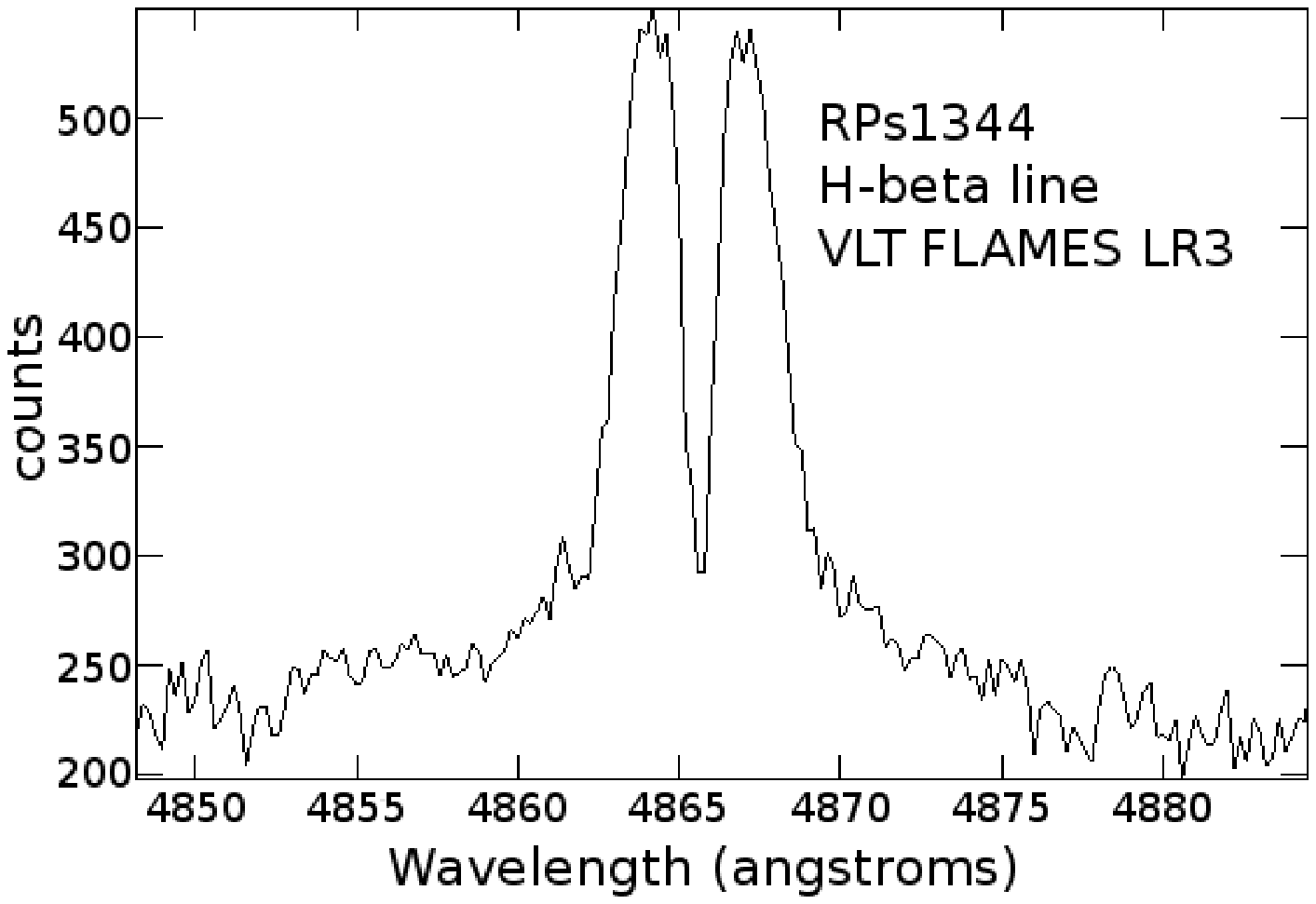}
\end{minipage}
\hspace{0.2cm}
\begin{minipage}[b]{0.455\linewidth}
\centering
\includegraphics[scale=0.29]{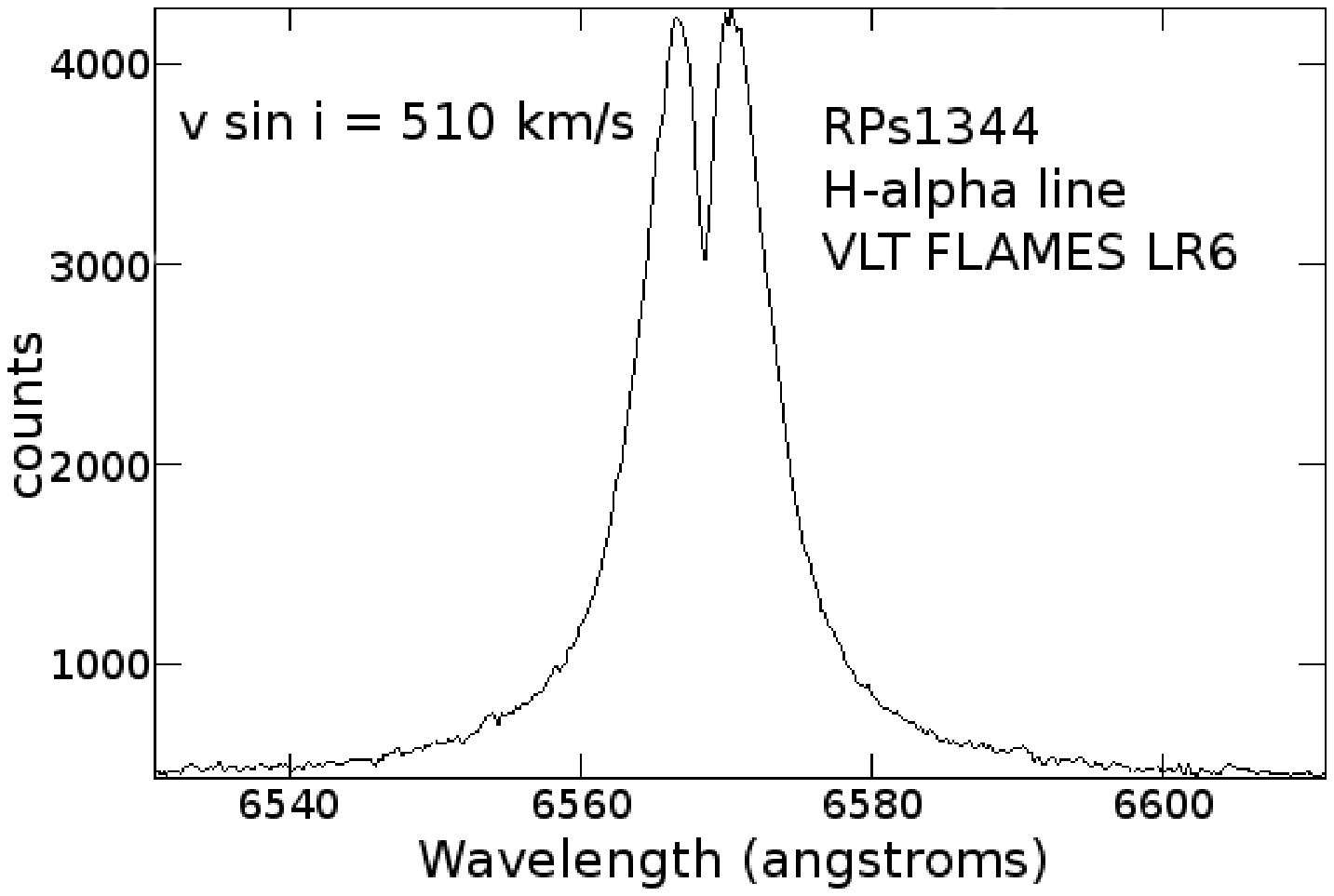}
\end{minipage}
\caption{An example of Balmer line splitting in the high velocity circumstellar emission of RPs1344 with H$\beta$
left (FLAMES LR3 grating) and H$\alpha$ right (FLAMES LR6 grating). The narrow profile of the shell absorption
line indicates its origin closer to the outer, slowly rotating parts of the shell. 
}
\label{Figure12}
\end{figure}

The Balmer emission lines demonstrate the most diverse range of profiles. Profile variations are believed to be dependent on the observer's angle of inclination to the star's pole. According to the model of Struve (1931), shell profiles occur where the star is viewed equatorially ($\textit{i}$ = 90\,deg), double peaked profiles occur at mid-inclination angles and singly peaked profiles occur by viewing towards the pole ($\textit{i}$ = 0\,deg). The measurement of accurate inclination angles, however, is complicated by other influences on the emission profile such as temperature, density and rotational velocity (Underhill \& Doazan, 1982; Quirrenbach et al. 1997; Miroshnichenko et al. 2001).

We present some representative examples of Balmer emission profiles using our VLT observations. Several of the Be stars in our VLT-observed sample show some shallow double reversal more or less central to the H$\alpha$ line. Some stars also have emission profiles with three emission peaks. It is these fine structure components (see Figure~\ref{Figure10})~that are known to show the greatest variability, down to the order of hours (Hubert \& Floquet, 1998). Stars whose emission lines have sharp, very deep absorption cores such as the example shown in Figure~\ref{Figure11}~have come to be known as \textit{shell} stars. The intrinsic variability of Be stars, however, has proven that over time these stars can lose and regain these shell characteristics (Underhill \& Doazan, 1982). We therefore refer to them as going through a `shell phase' at the time of our observation. Following the convention proposed by Hanuschik et al. (1996), we formally identify a shell star where the central absorption extends below the stellar continuum.

Further to this definition, we add that this only applies to absorption on the H$\alpha$ line. The H$\beta$ line is more dramatically affected by the atomic absorption since the reversal feature is not dependent or correlated to the strength of any individual Balmer emission line. For example, a medium absorption of H$\alpha$ resulting in a small reversal feature will correspondingly extend very deeply into the H$\beta$ emissive flux (see Figure~\ref{Figure12}).

\begin{figure}
\begin{minipage}[b]{0.46\linewidth}
\centering
\includegraphics[scale=0.295]{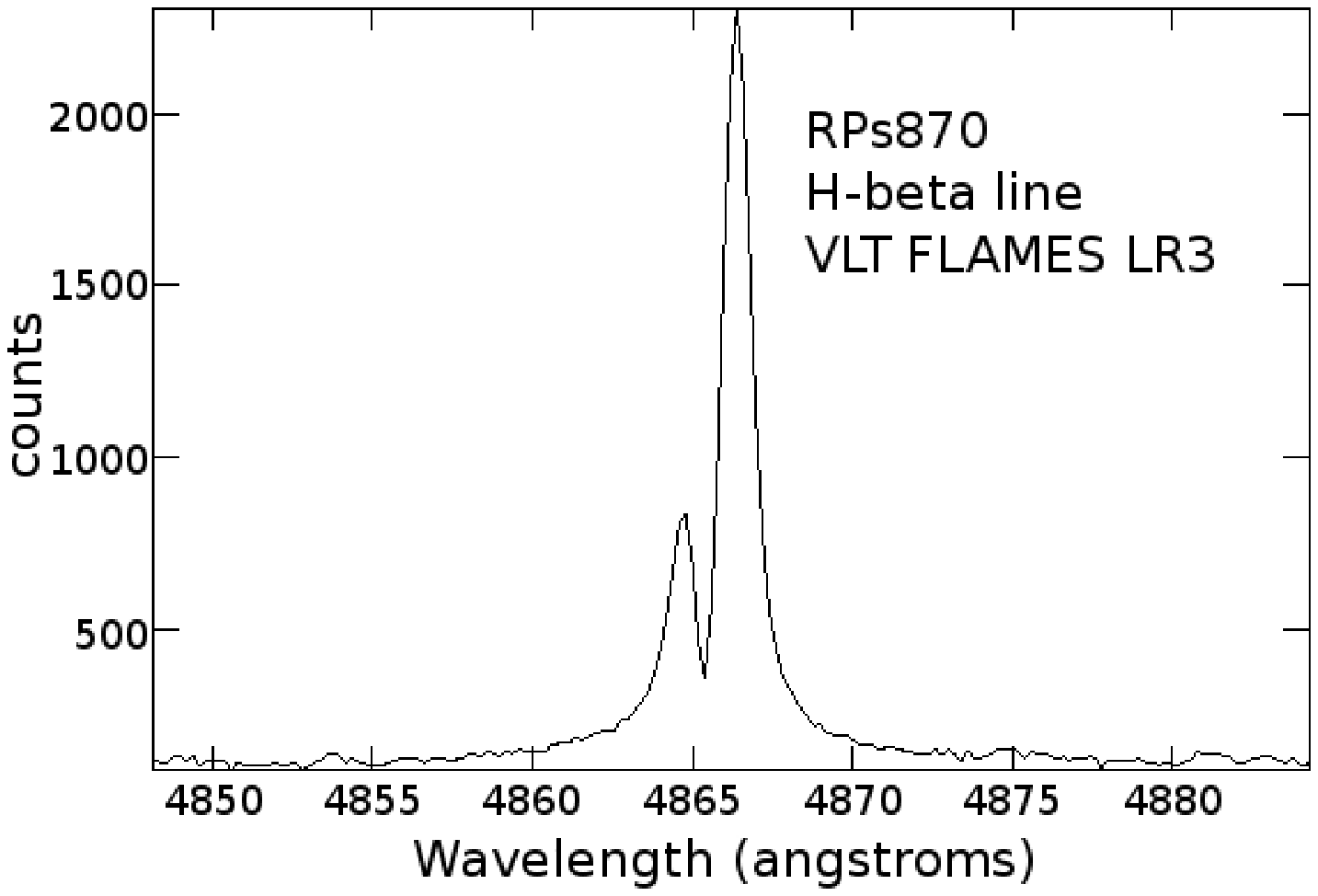}
\end{minipage}
\hspace{0.2cm}
\begin{minipage}[b]{0.46\linewidth}
\centering
\includegraphics[scale=0.295]{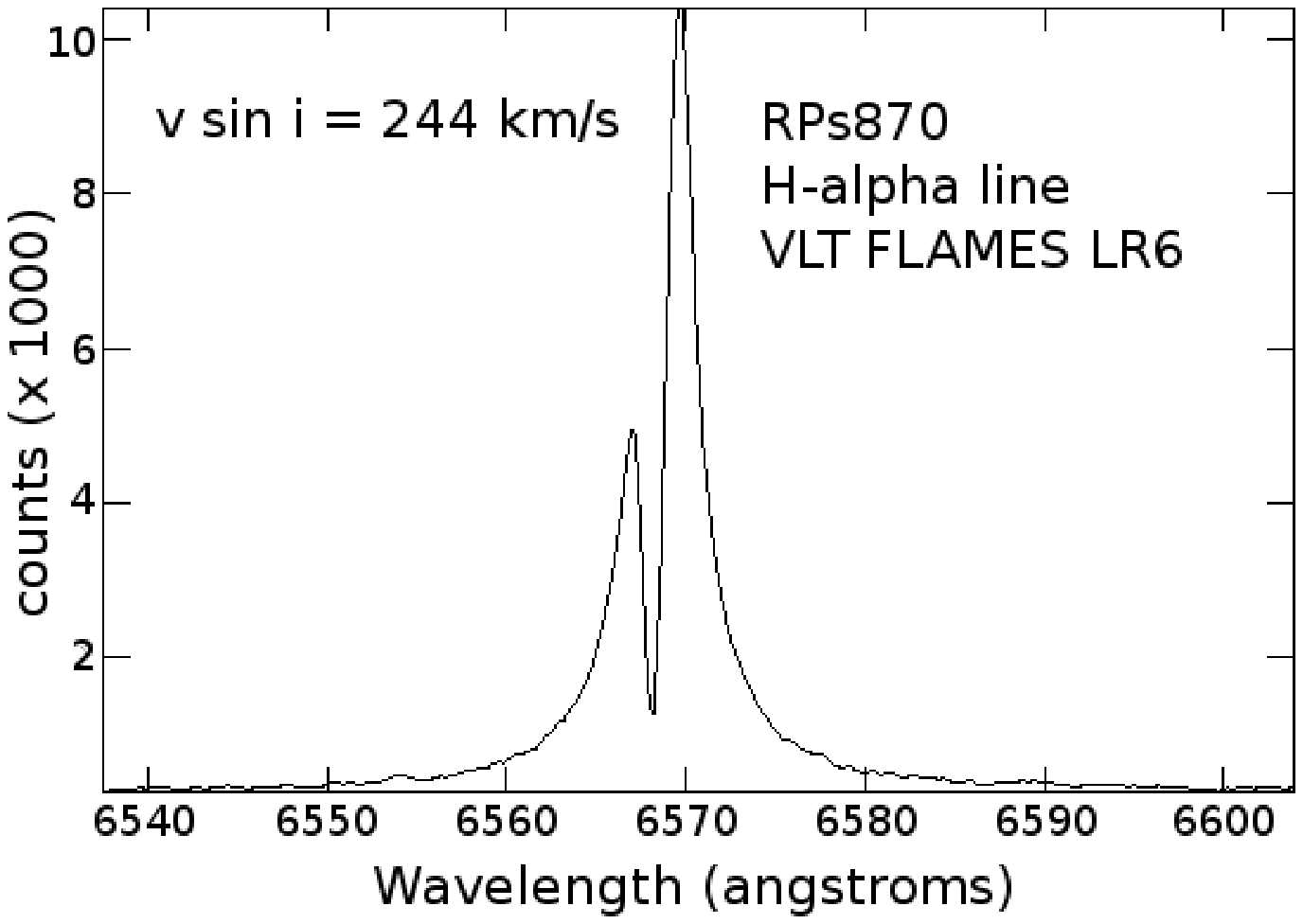}
\end{minipage}
\caption{RPs870 is an example of Balmer line splitting which appears to the left of centre
with H$\beta$ left (FLAMES LR3 grating) and H$\alpha$ right (FLAMES LR6 grating). The peak R$>$V affects both hydrogen emission lines and arises from one-armed density waves in the circumstellar disk. }
\label{Figure13}
\end{figure}
\begin{figure}
\begin{minipage}[b]{0.46\linewidth}
\centering
\includegraphics[scale=0.295]{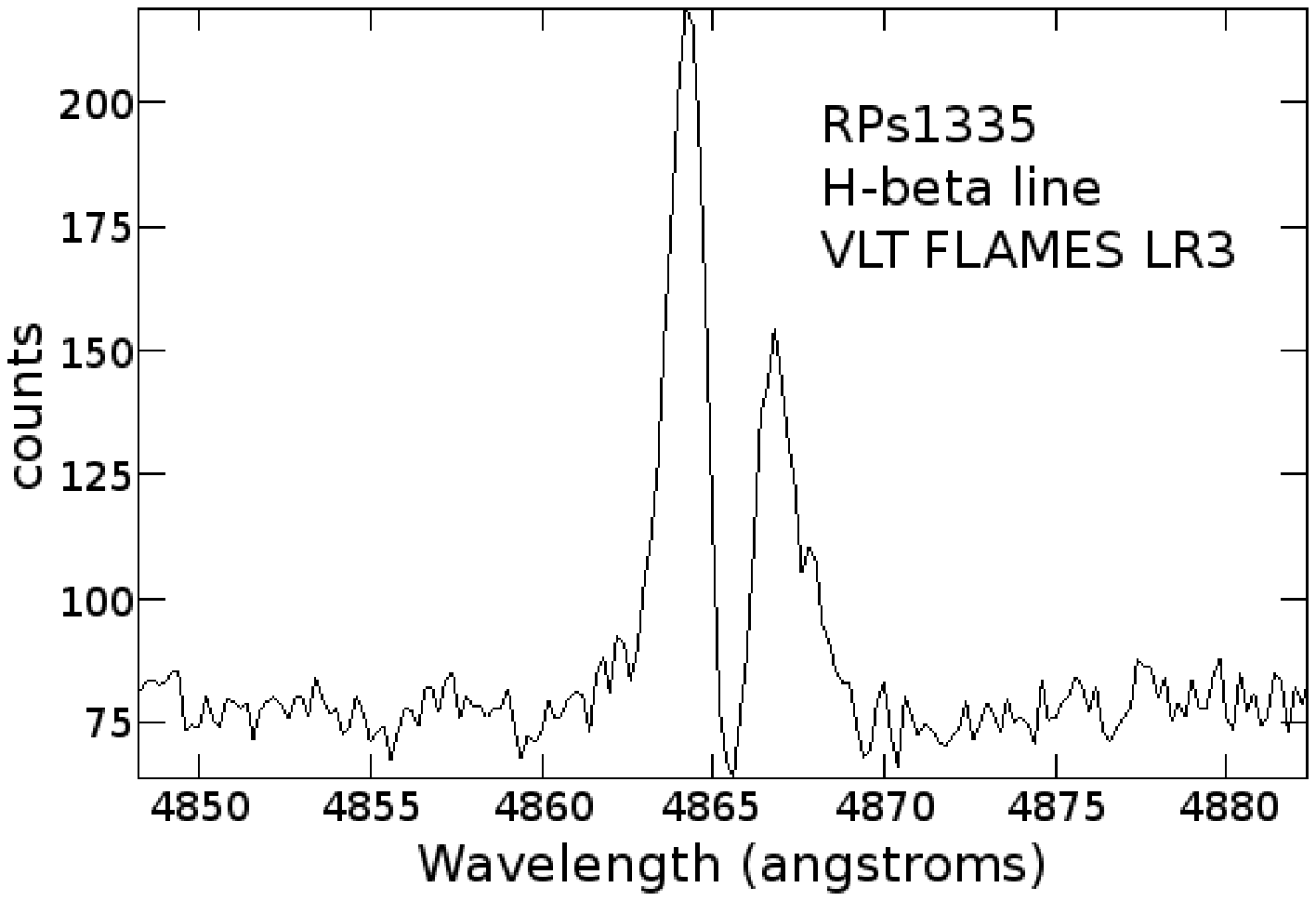}
\end{minipage}
\hspace{0.2cm}
\begin{minipage}[b]{0.46\linewidth}
\centering
\includegraphics[scale=0.295]{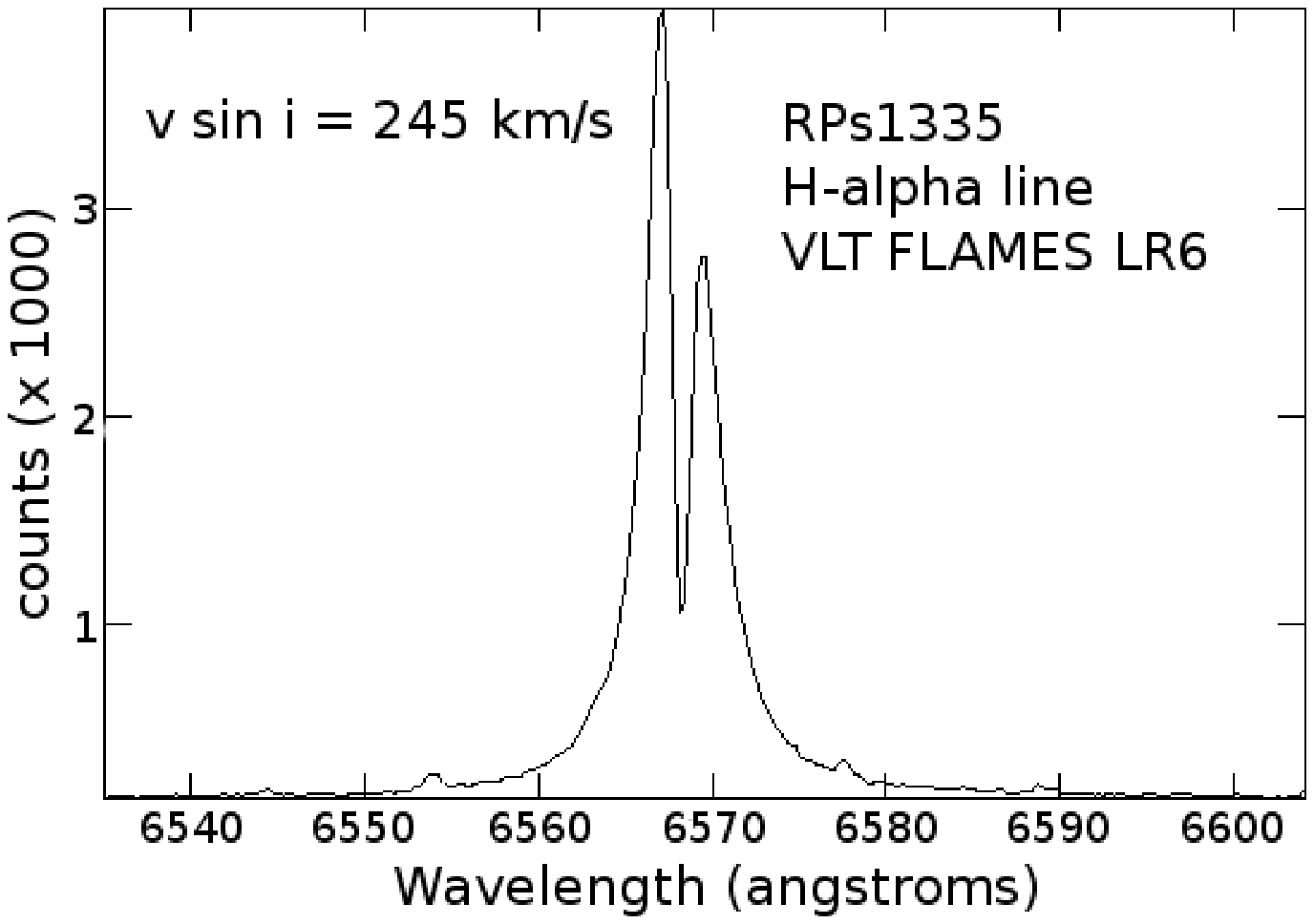}
\end{minipage}
\caption{RPs1335 is an example of Balmer line splitting which appears to the right of centre
with H$\beta$ left (FLAMES LR3 grating) and H$\alpha$ right (FLAMES LR6 grating). In this case V$>$R.
}
\label{Figure14}
\end{figure}
\begin{figure}
\begin{center}
  \includegraphics[width=0.40\textwidth]{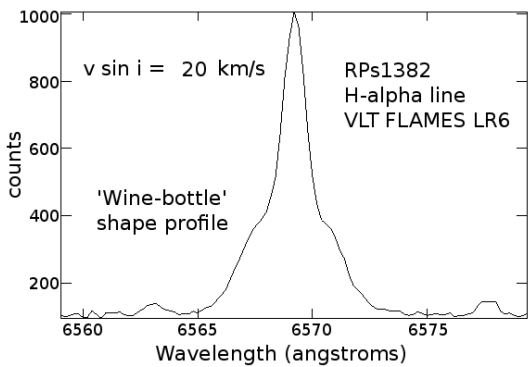}\\
  \caption{RPs1382 is an example of a `wine-bottle' profile which may occur by viewing the rotating star close to pole-on ($\textit{i}$ = 0\,deg angle). The low rotation velocity is a direct result this viewing angle and is measured using the central profile of the H$\alpha$ line (see section~\ref{section6}). The broadening is circumstellar.}
  \label{Figure15}
  \end{center}
  \end{figure}

  Asymmetry is a sub-feature found in a small percentage of Be star profiles. This is currently thought to arise from one-armed density waves in the circumstellar disk, also known as the global disk oscillation model (Silaj et al. 2010). In Figure~\ref{Figure13}~we show asymmetry where the reversal is left of centre while Figure~\ref{Figure14}~shows reversal to the right of centre, affecting both H$\beta$ (left example) and H$\alpha$ (right example) Balmer lines the same way. The resulting emission peak on the left is known as the Violet (V) component and the emission peak on the right is known as the Red (R) component. These asymmetries are also seen in single emission-lines and are probably the result of minor or isolated density waves.

\begin{figure}
\begin{minipage}[b]{0.46\linewidth}
\centering
\includegraphics[scale=0.295]{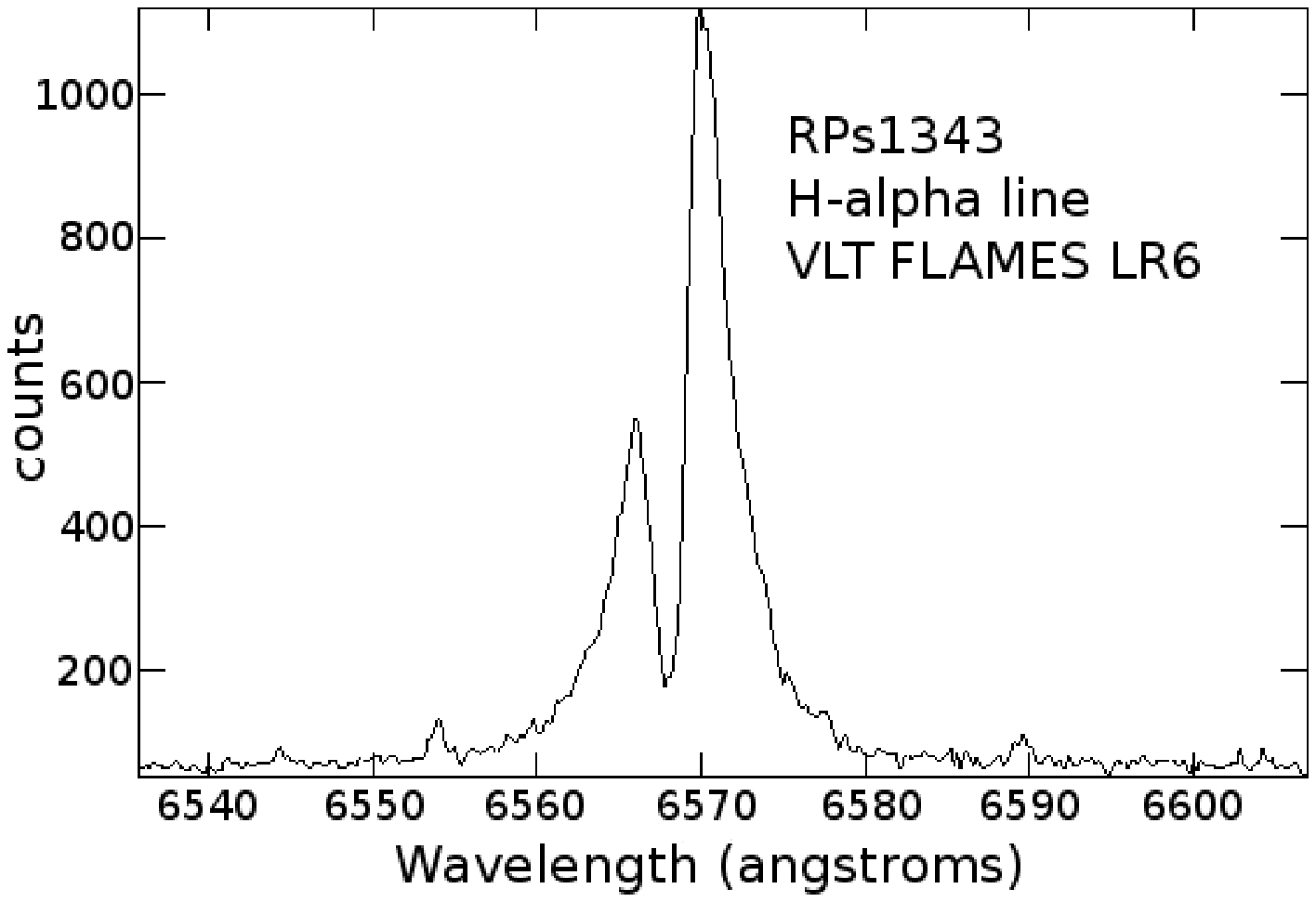}
\end{minipage}
\hspace{0.2cm}
\begin{minipage}[b]{0.46\linewidth}
\centering
\includegraphics[scale=0.295]{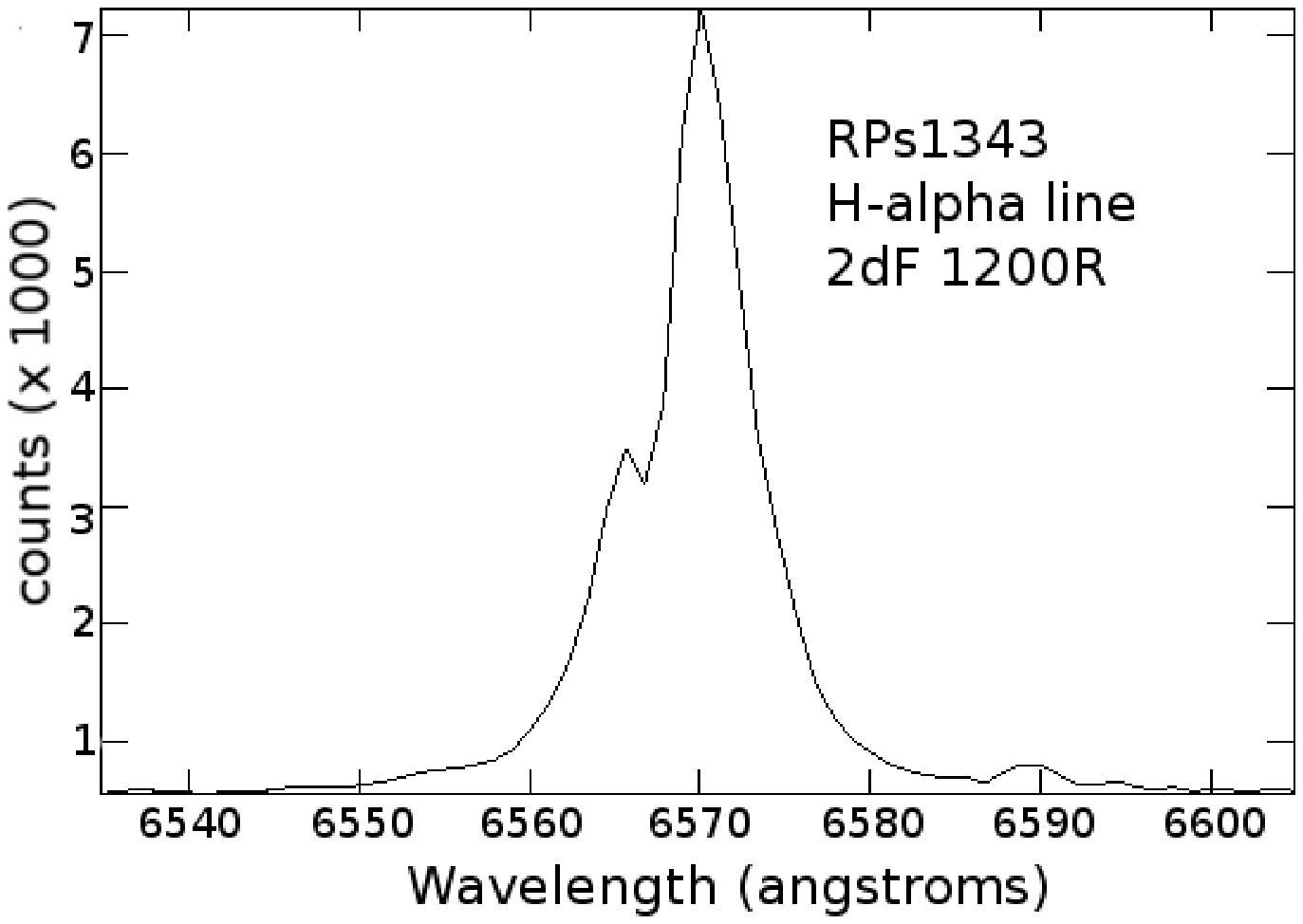}
\end{minipage}
\caption{A comparison of H$\alpha$ Balmer line splitting in emission-line star RPs1343 as seen with the FLAMES LR6 R8600 grating (left) and the 2dF 1200R grating (to the right). The VLT spectrum with its increased detail provides the clear detection of line splitting and some microstructure. The 2dF spectrum is able to detect the presence of line splitting but the amplitude of the same is unable to be measured due to the lower resolution of the 1200R grating. No microstructure can be seen in the 1200R spectrum.
}
\label{Figure16}
\begin{minipage}[b]{0.45\linewidth}
\centering
\includegraphics[scale=0.29]{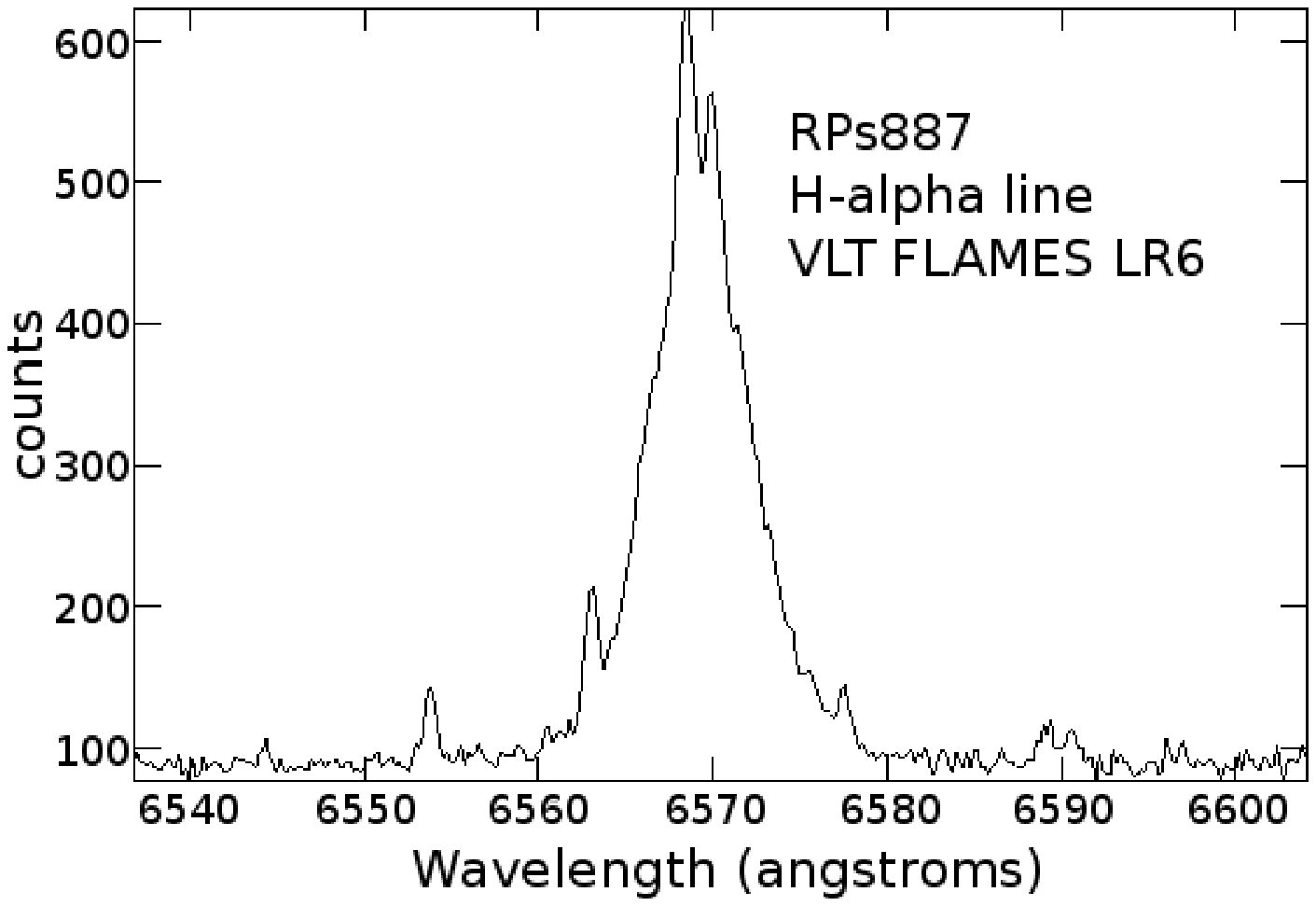}
\end{minipage}
\hspace{0.2cm}
\begin{minipage}[b]{0.45\linewidth}
\centering
\includegraphics[scale=0.29]{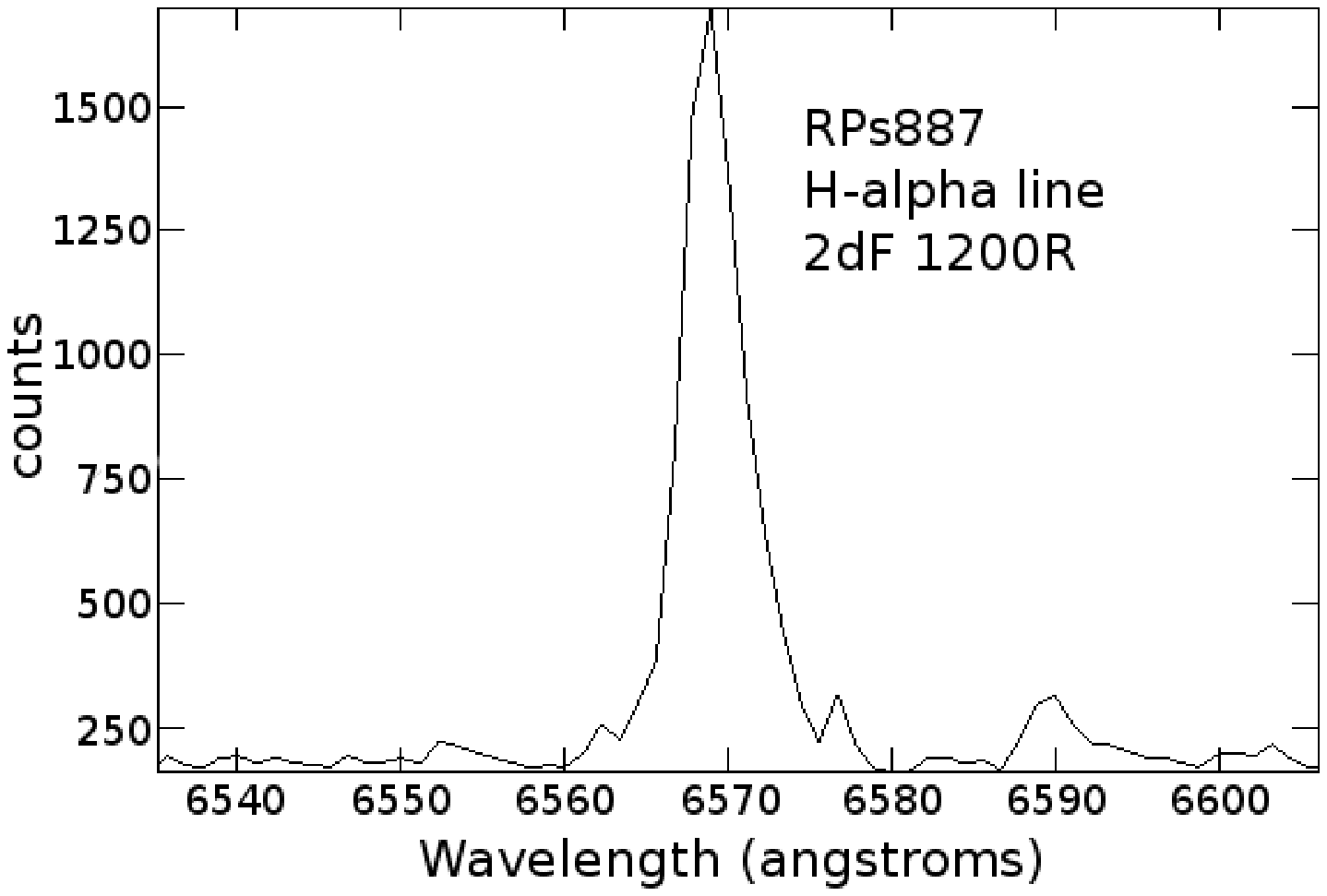}
\end{minipage}
\caption{A comparison of H$\alpha$ Balmer line splitting in emission-line star RPs887 as seen with the FLAMES
LR6 R8600 grating left and the 2dF 1200R grating to the right. The VLT spectrum shows fine line splitting to the
top right but this cannot be seen in the 2dF 1200R spectrum which has a dispersion of 1.105\AA/pixel. In this case
the peaks at the top of the H$\alpha$ line are separated by 1.4\AA, making fine detail impossible to detect at the
lower resolution.}
\label{Figure17}
\end{figure}

\begin{table}
  \caption{H$\alpha$ profile features found in 122 emission-line stars observed with VLT FLAMES using the LR6 grating. Micro features are miniature peaks identified on the sides and at the top of a main peak.}
  \begin{tabular}{|l|c|c|c|c|}
    \hline
    Main feature & Total & \% & Number & Number \\
                  &   number  & of  &  micro  &  bottle   \\
       &          & total    &    features     &     shape  \\
    \hline\hline
    Single peak & 100 & 82 & 11 & 9 \\
    Double peak V=R & 10 & 8 & 2 &  \\
    Double peak R$>$V & 6 & 5 & 3 &  \\
    Double peak V$>$R & 4 & 3 & 3 &  \\
    Shell & 2 & 2 & 2 &  \\
    \hline
  \end{tabular}\label{Table6}
  \end{table}

  A more unusual feature among the Be star emission line profiles is the `wine-bottle' shape, often produced by viewing the star near to the pole. The example shown in Figure~\ref{Figure15}~is possibly broadened by a combination of disk rotation and Thomson scattering.

  In attempting to classify the profiles of H$\alpha$ emission according to the particular features mentioned above, it is prudent to refer only to the higher resolution VLT FLAMES data. Figures~\ref{Figure16}~and~\ref{Figure17}~provide a comparison of the VLT FLAMES LR6 and 2dF 1200R spectra for the one object. In the first comparison (Figure~\ref{Figure16}) the strong absorption feature seen in RPs1343 using LR6 on FLAMES is only detectable to a limited extent in the 2dF spectrum to the right. In the second comparison (Figure~\ref{Figure17}) the absorption feature is too narrow to be detected at a resolution of 1200R.

  Using only the 122 emission-line stars observed on the VLT, the features shown in Table~\ref{Table6} were present. All 122 stars in this table reside within a 3 deg$^{2}$ region on the main optical bar of the LMC. With 100 detections, the single peak profile is the most common. At the time of spectroscopic observation, 11 stars were found to exhibit micro features such as miniature structures on the sides and/or at the peak. In time these may develop into separate peaks or disappear completely. Since emission-line stars are constantly evolving, a table such as this can only provide a snapshot of the percentage of features found at that time.

\label{subsection4.3}

\section{Rotational velocities}

Classical Be stars undergo rapid rotation and possess geometrically thin, circular gaseous disks resulting in hydrogen Balmer emission (Jaschek et al. 1981; Porter \& Rivinius 2003). Typical rotation compared to critical velocity ($\emph{v}_{eq}$/$\emph{v}_{\textrm{crit}}$) has been estimated at $\sim$70\%-80\% (Porter 1996; Porter \& Rivinius 2003). A lower estimate of 40\%-60\% of the critical breakup velocity for such stars was found by Cranmer (2005) but this set of data is not homogeneous. It is likely that both of these estimates may not take all the physical conditions into account. Due to fast rotation it is expected that the star is flattened, causing a variation in temperature and density from pole to equator. This is expected to result in a gravitational darkening of the stellar disk. Based on this theory, Townsend et al. (2004), employing the effects of equatorial gravity darkening, suggest that a degeneracy in the measurement of rotational rates allows Be stars to be rotating at or near their critical breakup velocity. An estimate of rotational velocity for the LMC set of emission-line stars will provide vital information for future studies.

\begin{figure}
\begin{center}
  \includegraphics[width=0.5\textwidth]{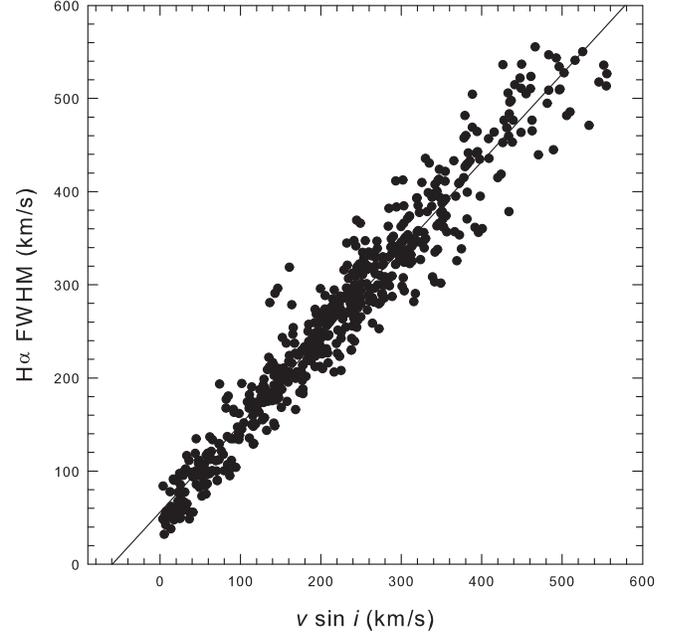}\\
  \caption{The relation between FWHM measured on the H$\alpha$ line and the rotational velocity (\textsl{v} sin \textsl{i}) using equation (3) for all emission-line stars in our sample. The variations away from the line of equality are mainly due to the equivalent width, used to determine \textsl{v} sin \textsl{i}.}
  \label{Figure21}
  \end{center}
 \end{figure}

Although the fine structure across the top of the Be star emission-line profile makes FWHM rather complex to untangle, the strength of the H$\alpha$ line negates any underlying photospheric biases or broadening. This is also true in cases where emission is weak. To derive the projected rotational velocity (v sin\,$\textit{i}$) we used the correlation found by Dachs et al. (1986, Equation (7)) with improvements made by Hanuschik (1989). Their three parameter correlation between FWHM (H$\alpha$), $\textit{v}$ sin $\textit{i}$ and equivalent width (EW) lead them to the relation:

\begin{equation}
\begin{array}{rcl}
\textrm{log}[\textrm{FWHM}(\textrm{H}\alpha) / 1.23~(v~\textrm{sin}~i + 70 \textrm{km} \textrm{s}^{-1})] \\
= -0.08~\textrm{log} EW + 0.14
\end{array}
\end{equation}

which was presented as equation (5) in Hanuschik (1989). We used this equation in the form:
\begin{equation}
v~\textrm{sin}~i = [(\textrm{FWHM}(\textrm{H}\alpha) + 10 ^{0.08~\textrm{log} EW + 0.14}) / 1.23] -70
\end{equation}

to derive $\textit{v}$ sin $\textit{i}$ for all stars in our sample. The resulting relation between FWHM(H$\alpha$) and $\textit{v}$ sin $\textit{i}$ is shown in Figure~\ref{Figure21}. The scatter is mainly due to the equivalent width of the individual line although there will inevitably be a contribution from non-kinematic line broadening due to radiation transfer (Poeckert and Marlborough, 1978), electron scattering, possible turbulence and measurement errors. The median fit to the data in Figure~\ref{Figure21} yields

\begin{equation}
\textrm{FWHM}(\textrm{H}\alpha) = 0.89~v~\textrm{sin}~i + 79 \textrm{km} \textrm{s}^{-1}.
 \end{equation}

 \begin{figure}
\begin{center}
  \includegraphics[width=0.5\textwidth]{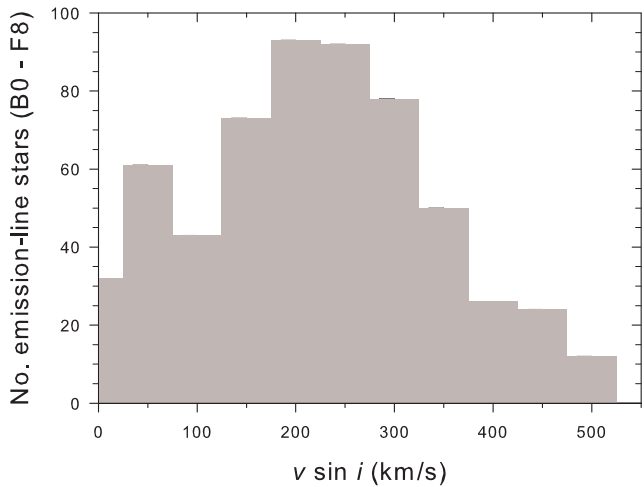}\\
  \caption{The distribution of estimated rotational velocities (\textsl{v} sin \textsl{i}) for all the hot emission-line stars (B0 - F8) in our survey. The large number of stars occupying the bin at 50\,km\,s$^{-1}$ is most likely due to their near pole-on angle to our line of sight.}
  \label{Figure21a}
  \end{center}
 \end{figure}

 \begin{figure}
\begin{center}
  \includegraphics[width=0.5\textwidth]{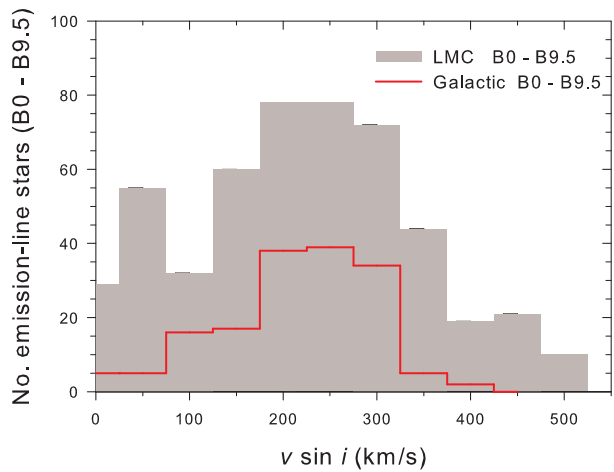}\\
  \caption{The distribution of rotational velocities (\textsl{v} sin \textsl{i}) for all Be stars (B0 - B9.5) in our LMC sample compared to the distribution for Galactic Be stars (B0 - B9.5) found by Slettebak (1982).}
  \label{Figure21b}
  \end{center}
 \end{figure}

 A histogram giving the frequency of \textsl{v} sin \textsl{i} for all hot emission-line stars in our LMC sample is shown in  Figure~\ref{Figure21a}. The distribution covers in excess of 500\,km\,s$^{-1}$ with a maxima at around 200\,km\,s$^{-1}$. With the exception of 30 stars measured using 6dF, all the measurements were taken using the highest resolution 2dF, AAOmega and VLT data. The 30 stars measured using the 6dF red arm 0.62 \AA/pixel data cover a large range from 73$<$\textsl{v} sin \textsl{i}$<$489, indicating that the 6dF data is not introducing any bias to the overall results.

 The number of stars found in the 50\,km\,s$^{-1}$ bin appears to be overstated in relation to the general trend seen in the histogram. This isolated rotational velocity peak probably has little to do with the spectral type or luminosity class, both of which are quite evenly distributed across the histogram. It most likely reflects the number of stars we are viewing close to pole-on (see section~\ref{subsection4.3}), many of which do not display a strong wine-bottle H$\alpha$ emission profile. As with all surveys, the reason may also partly lie in our selection criteria, as we specifically targeted faint stars with relatively strong H$\alpha$ emission.

 The process of estimating rotational velocities using FWHM and EW on H$\alpha$ emission lines which display a strong wine-bottle profile is somewhat complex. Measurement of these parameters using a line such as He$\lambda$4471 is generally considered more accurate because it is less affected by circumstellar rotation. The extra-wide (wine-bottle profile) wings on these particular H$\alpha$ lines substantially increase both measurements, thereby giving these stars a typical rotational velocity between 200$<$$\textsl{v}\,sin\,\textsl{i}$$<$300\,km\,s$^{-1}$.

 We have found that by measuring FWHM and EW on the He$\lambda$4471 absorption line, or by fitting a gaussian curve to the H$\alpha$ wine-bottle profile, the rotational velocity readings drop substantially to levels below 100\,km\,s$^{-1}$. Since both methods of measurement yield similar results (5$<$\textsl{v} sin \textsl{i}$<$40\,km\,s$^{-1}$), we prefer to fit a Gaussian profile to the H$\alpha$ line. This is expected to produce the most accurate measurement of FWHM, EW and \textsl{v} sin \textsl{i} using these peculiar profiles. It not only constrains all FWHM and EW measurements to the H$\alpha$ line for direct comparison across the table (Appendix Table 1) but improves reliability due to the strength of H$\alpha$ compared to the He$\lambda$4471 absorption line, which is weak, difficult to fit and often perturbed by the envelope.

After selecting LMC Be stars between classes B0 and B9.5, we have compared their distribution to the distribution found in the Galaxy for the same Be classes using data from Slettebak (1982). From the histogram in Figure~\ref{Figure21b} we find a correlation coefficient of 0.88 between the LMC and Galactic data sets. Although both data sets are not complete in any sense, the plot indicates that, based on a random selection, the majority of Be stars have a projected rotational velocity between 200$<$$\textsl{v}\,sin\,\textsl{i}$$<$300\,km\,s$^{-1}$.

\label{section6}

\section{Nebula contribution}

Approximately 15 percent of the emission-line stars in our sample are B[e] stars which show evidence of forbidden nebula emission lines such as \FeII\,$\lambda$4244,4287,4415,5273,7155, \NII\,$\lambda$6583,6548, \OI\,$\lambda$6300,6363, \OII\,$\lambda$7320,7330 and \SII\,$\lambda$6716,6731 in their spectrum. Importantly, not all of these lines are necessarily to be found in every B[e] star but the most common are \FeII~and \OI. These stars are often associated with ambient or extended nebula emission which is ionised by the central star with surface temperatures of between 10,000\,K and 33,000\,K. This means that the H$\alpha$ line may be a combination of central star and nebula emission, making them difficult to separate when the RVs of each component are similar and/or when spectral resolution is not sufficient to distinguish the components.

Since the strength of the H$\alpha$ line in each star and the ambient background H$\alpha$ emission affecting the spectrum are constantly changing with location in the LMC, a general sky subtraction may be insufficient in some cases. Each object located in an area of diffuse \HII~emission must therefore be individually assessed for ambient nebula contribution on the basis of H$\alpha$ emission within a 10 arcsec radius of the star, which provides a fair estimate in the crowded regions of the LMC. This can be closely approximated using our deep H$\alpha$ map. The measurement of H$\alpha$ intensities on and off the emission-line star produces a ratio which roughly indicates the percentage of H$\alpha$ spectral flux emitting from the star. If the background emission is deemed to be contributing more than 10\% of the measured flux, a special note about the B[e] status is made in the comments column of Appendix Table 1. If there is no ambient emission surrounding or in the immediate vicinity of the star, we may safely assume that the star is of the B[e] variety, where we expect to find emission-lines of \NII, \SII, \OII~and even \OIII.

\begin{figure}
\begin{minipage}[b]{0.58\linewidth}
\centering
\includegraphics[scale=0.35]{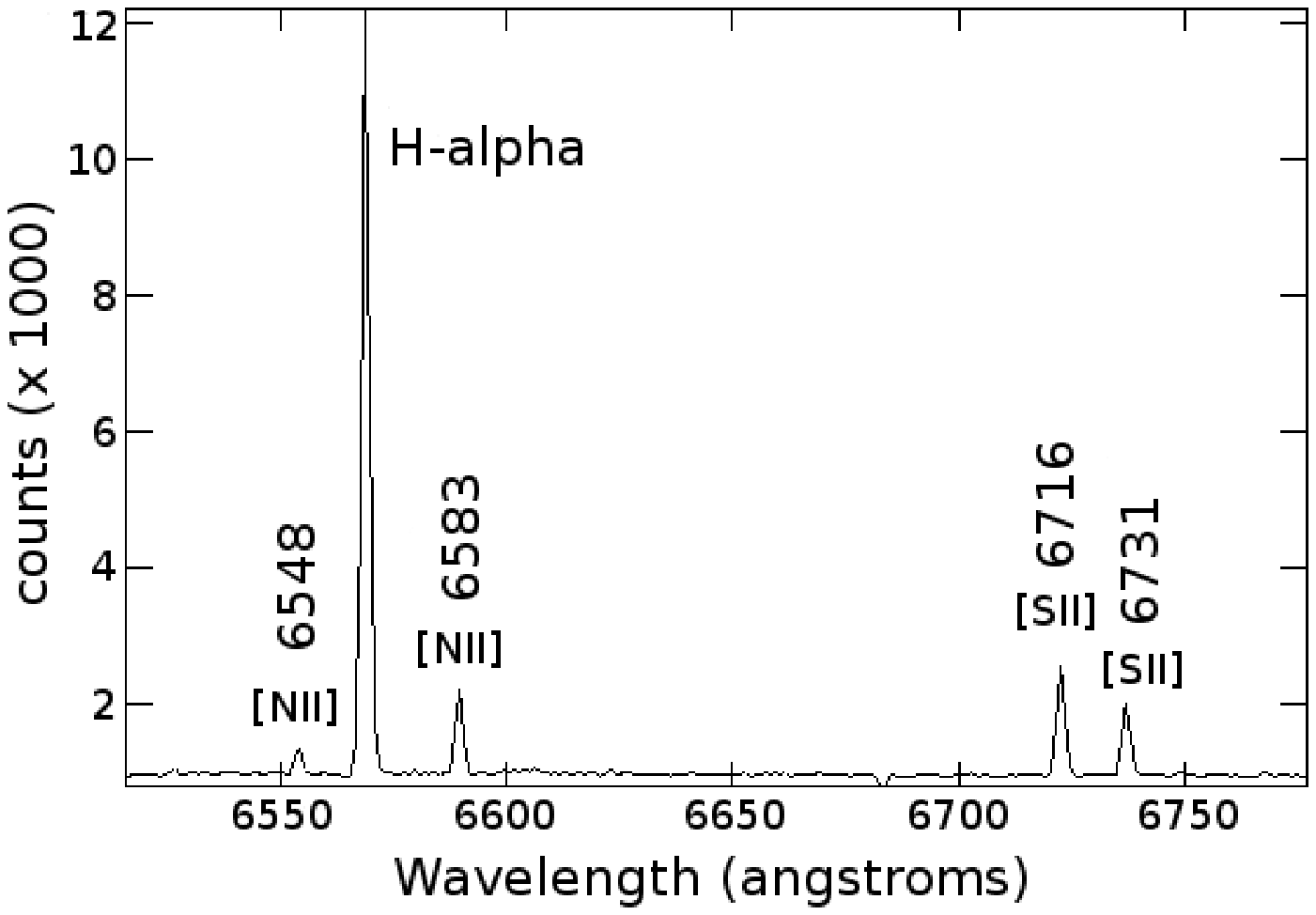}
\end{minipage}
\hspace{0.03cm}
\begin{minipage}[b]{0.32\linewidth}
\centering
\includegraphics[scale=0.182]{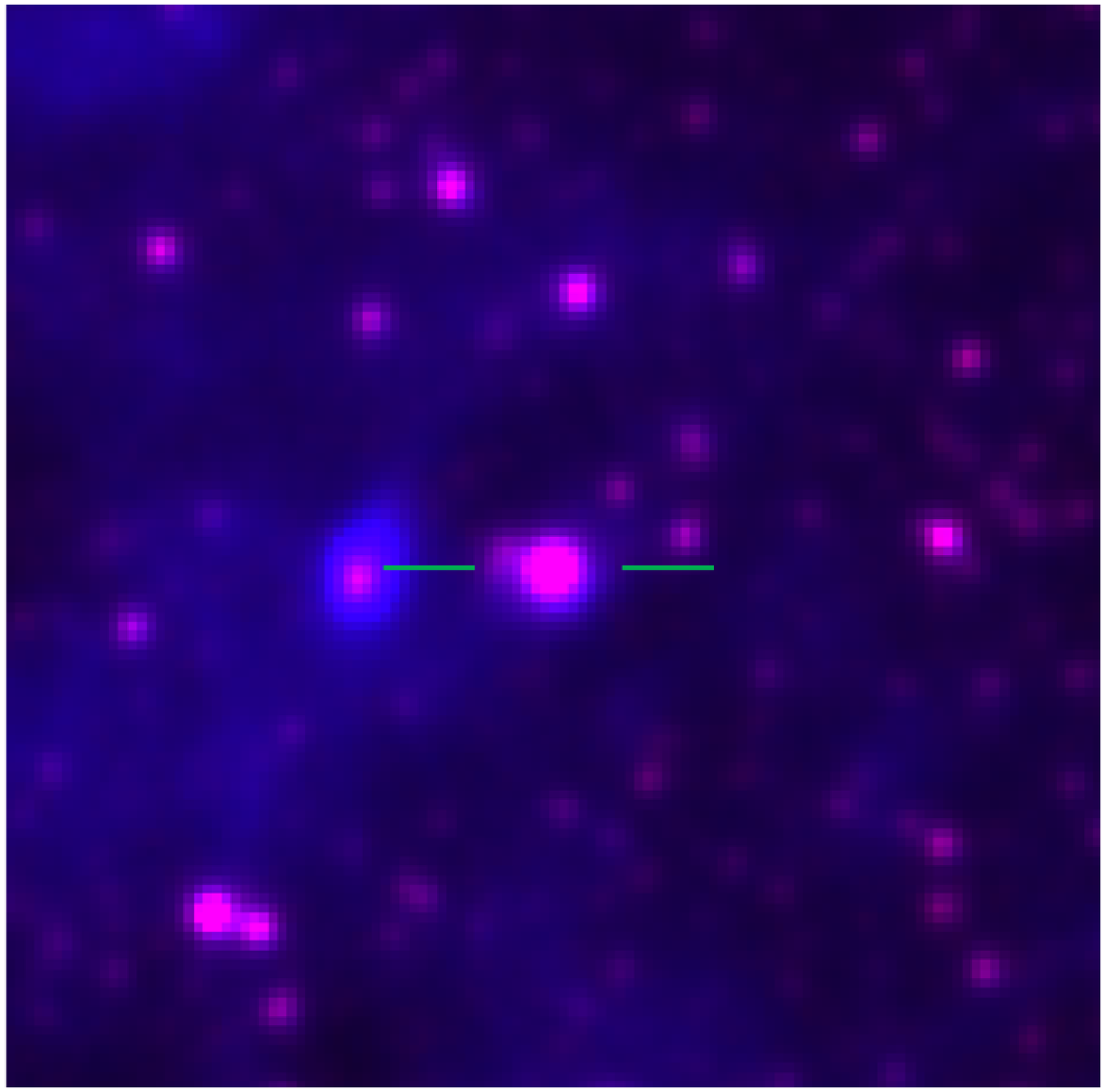}
\end{minipage}
\caption{Left: A 2dF 1200R grating example of the red spectral region around the H$\alpha$ line in B[e] star RPs226 where
there is strong emission from \NII$\lambda$6583,6548 and \SII$\lambda$6716,6731. Right: We show the H$\alpha$ (blue) and short red
(red) combined UKST image, clearly showing the local environment and the contribution from ambient nebulous flux.
Using the H$\alpha$ map alone we can estimate the proportion of nebula contribution by using the Starlink GAIA
task to measure the ambient H$\alpha$ (sky) emission within 10 arcsec ($\sim$2.4pc) of the star and compare the
flux to that at the rim of the star, which represents the H$\alpha$ excess (H$\alpha$--R) emission from the star.
The object, 2 arcsec to the east of RPs226 (immediate
left in the image) is RP225, a compact HII region.}
\label{Figure22}
\begin{minipage}[b]{0.58\linewidth}
\centering
\includegraphics[scale=0.353]{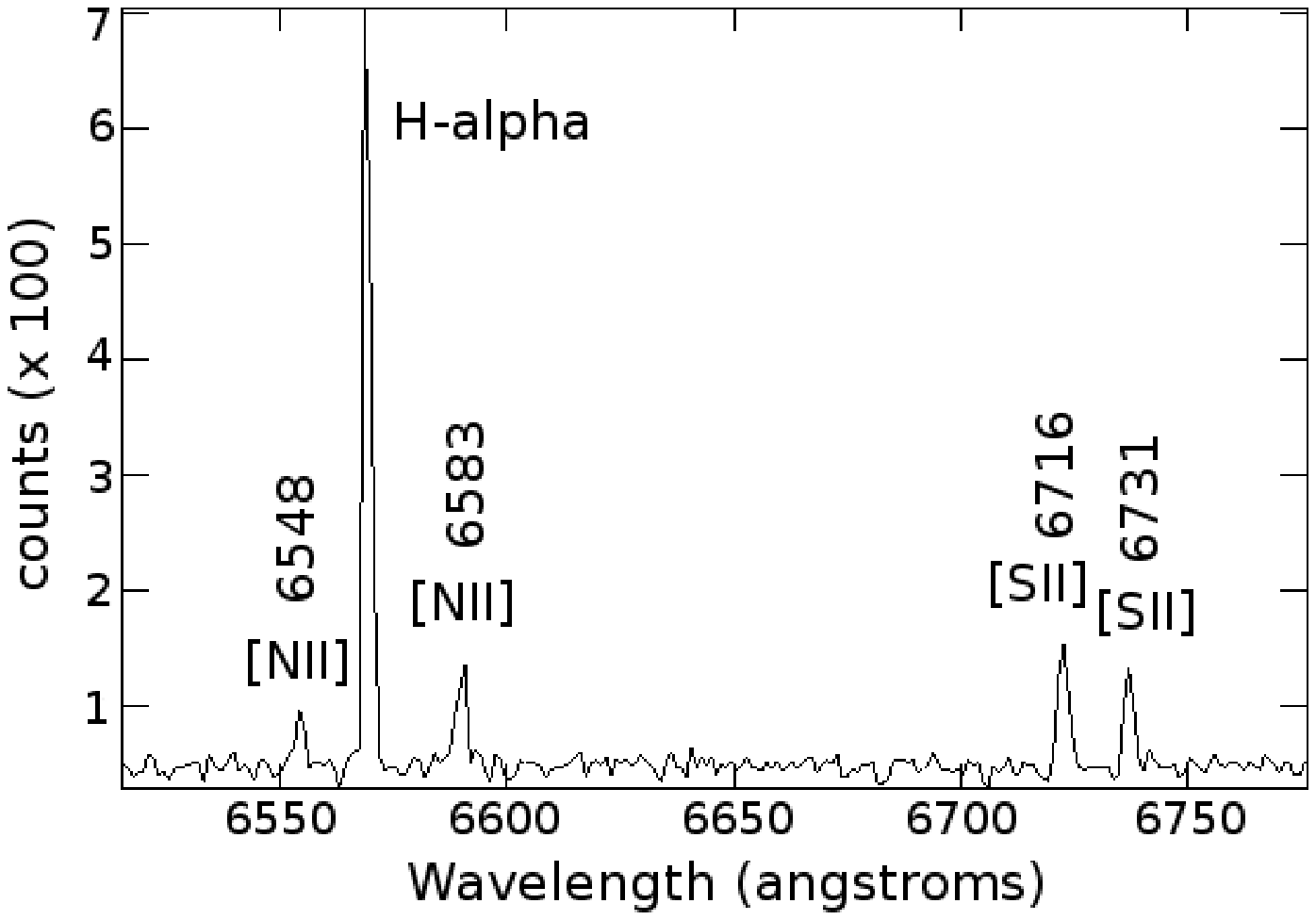}
\end{minipage}
\hspace{0.01cm}
\begin{minipage}[b]{0.32\linewidth}
\centering
\includegraphics[scale=0.40]{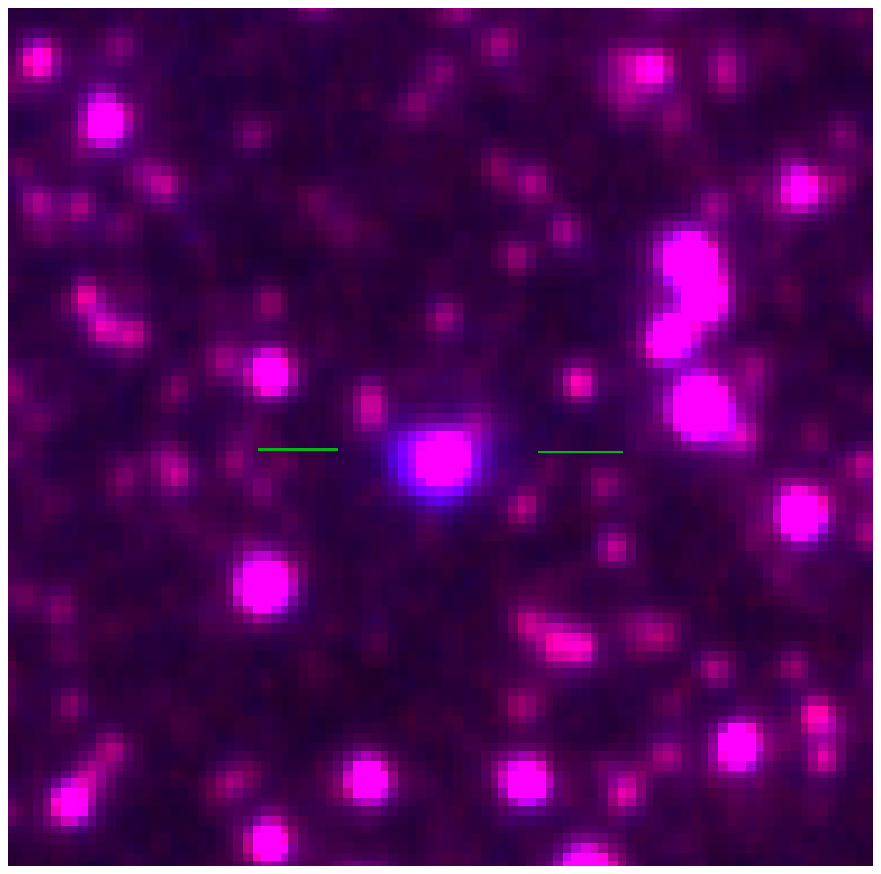}
\end{minipage}
\caption{Left and right images as described in Figure~\ref{Figure22} above. This is an example of a B[e] emission-line star (RPs161) with nebula lines present but no contribution from ambient emission at the stars location. No correction to the H$\alpha$ flux is required. }
\label{Figure23}
\begin{minipage}[b]{0.58\linewidth}
\centering
\includegraphics[scale=0.35]{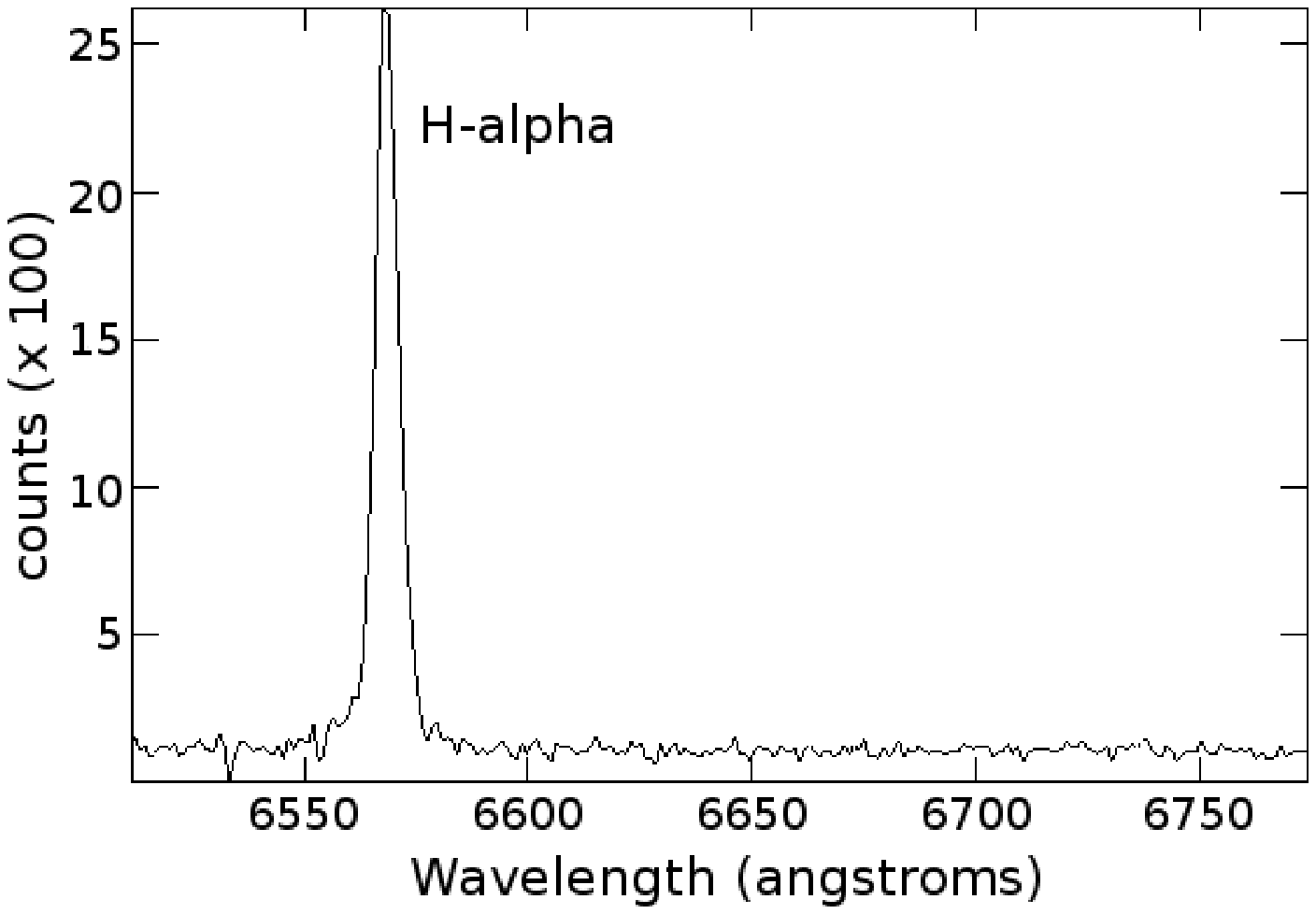}
\end{minipage}
\hspace{0.01cm}
\begin{minipage}[b]{0.32\linewidth}
\centering
\includegraphics[scale=0.183]{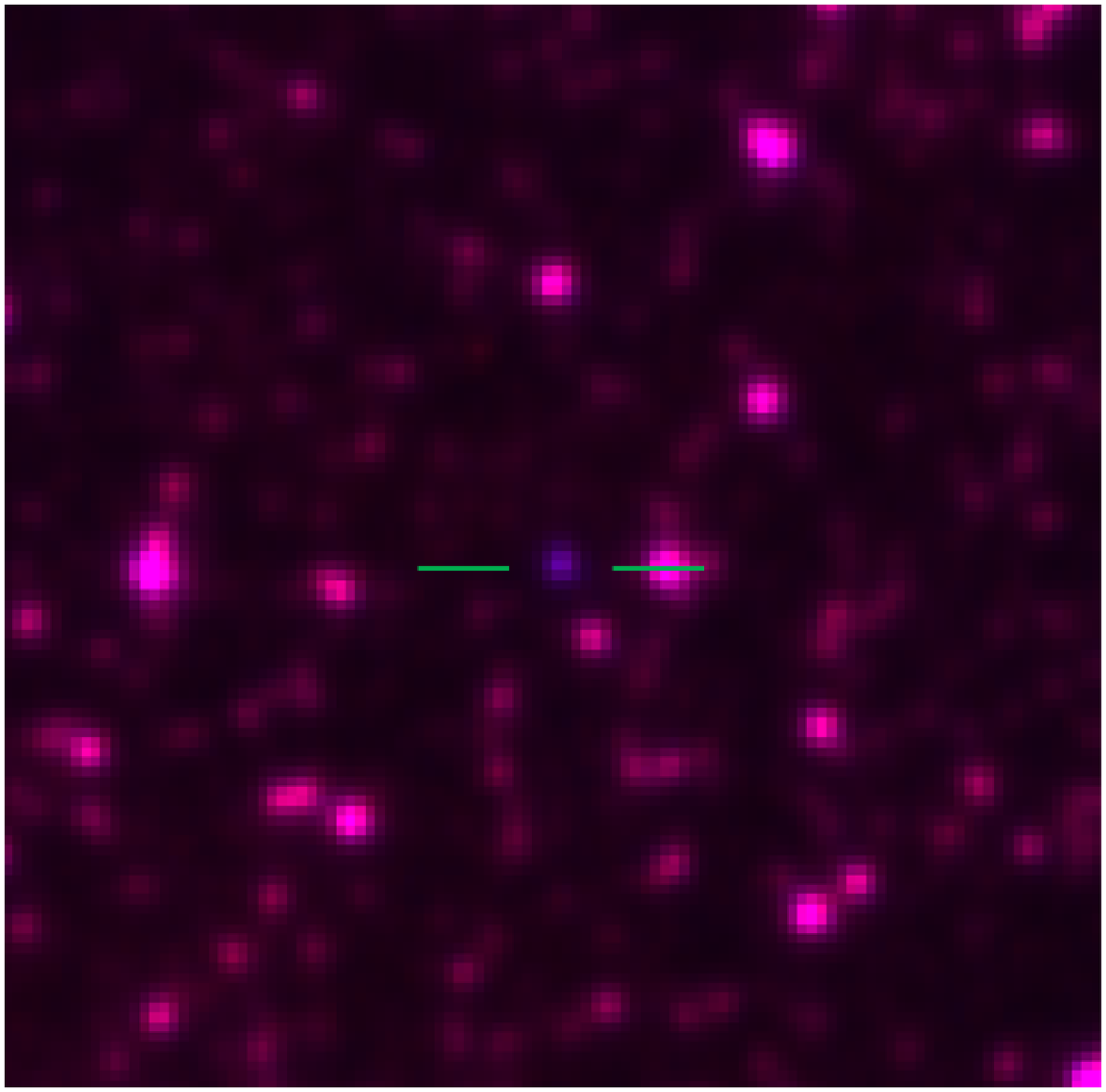}
\end{minipage}
\caption{Left and right images as described in Figure~\ref{Figure22} above. An example of a Be emission-line star (RPs634) not requiring any correction to the measured H$\alpha$ flux. No standard nebula forbidden emission-lines are detected and the environment is free of low level ambient emission.
}
\label{Figure24}
\end{figure}

In Figure~\ref{Figure22} we show an example of a B[e] star with a very significant contribution of emission lines normally associated with nebulae. Where emission lines are this strong we examine the immediate environment for ambient nebula emission. If the deep H$\alpha$ image shows us that much of the emission is environmental, we flag (in the comments column) that the other emission lines in the spectrum may be the result of ambient emission. In Figure~\ref{Figure23} we show a regular B[e] star with some nebula lines present but no apparent environmental contribution requiring correction. Finally in Figure~\ref{Figure24} we show a Be star requiring no nebula subtraction for environmental reasons and no nebula lines. All of these stars are located within a 1 degree radius of each other, emphasising the importance of surveying the immediate location of each star.

\label{section7}

\section{New accurate positions for LMC emission-line stars}

We found that emission-line star positions provided by many earlier surveys (mostly using the
FK4 system) were not sufficiently accurate when converted to the
J2000 equinox and compared to positions across our astronomically accurate survey. As many as 138 previously identified emission-line stars were only published with an accuracy to one decimal fraction of a minute. The majority of these also gave no seconds in DEC. Many true positions were uncertain given the crowded nature of the LMC. In many cases the known emission-line star had to be carefully verified as the object that was previously found. The K-view program in {\small KARMA} initially allowed us to
find the position of peak intensity of any point source within the stacked
H$\alpha$/SR images allowing
accurate positioning to 0.6~arcsec due to our meticulous calibration of our entire map with the SuperCOSMOS world coordinate system.

To improve the positioning and find the most accurate positions for our new emission-line stars, we extracted red image data from the SuperCOSMOS Image Analysis Mode (IAM). The SuperCOSMOS plate measuring machine samples some 1,000 objects within 10 x 10 armin areas in order to define the xy-to-RA/Dec transformation. The resulting coordinates of a given pixel are thought to be accurate to a few tenths of an arcsec. Using both the H$\alpha$/SR discovery images and accurate SuperCOSMOS image positions, we matched each emission-line star to the position provided by the IAM data. This match also allowed us to extract the SuperCOSMOS derived B and R broadband magnitudes for each star, as discussed in section~\ref{section11}.

\label{section8}

\section{Radial Velocities}

Stellar radial velocities for hot B stars are useful for kinematic studies within the LMC. They provide a valuable tool with which to compare young and old populations. Importantly, the radial velocities allow us to verify that our selected emission-line stars reside within the LMC.

Our stellar radial velocities were determined from the medium resolution 2dF 1200R, 6dF 425R and FLAMES spectra as described above. The largest number of velocities (92\%) were measured using the 1200R 2dF grating which has an estimated accuracy of $\pm$4 km s$^{-1}$. Two different methods of
velocity measurement were employed in order to reduce
errors arising as a result of the application of a particular
technique. These techniques have been described in Reid \& Parker (2006b) and will only be repeated briefly here.

Initially, we used the {\scriptsize IRAF EMSAO} technique for measuring multiple,
specified emission lines. Wavelengths for 13 common
emission-lines within the $\lambda$6200-7350\AA~range were
specified to three decimal places. The program found the central position of each available line which was independent of the line width and self-absorption features. It then applied a
weighted gaussian fit to each emission line dependent on its intensity,
derived a weighted average across the spectrum and corrected for the
heliocentric velocity. The {\scriptsize EMSAO}-derived velocity for each star
was displayed and examined.

The second method of velocity determination involved the
cross-correlation technique using {\scriptsize XCSAO} in
{\scriptsize IRAF} (Kurtz \& Mink, 1998). This method requires a
list of template spectra with low internal velocities and
accurately determined published radial velocities against which all
the other spectra may be compared for measurement. Template
emission-line velocities were based on a minimum of four lines, with at
least two of these being fitted by {\scriptsize EMSAO} (Kurtz
\& Mink, 1998). Twenty LMC planetary nebula and emission-line star templates with well established velocities were chosen for the
cross-correlation.

\begin{figure}
\begin{center}
  \includegraphics[width=0.47\textwidth]{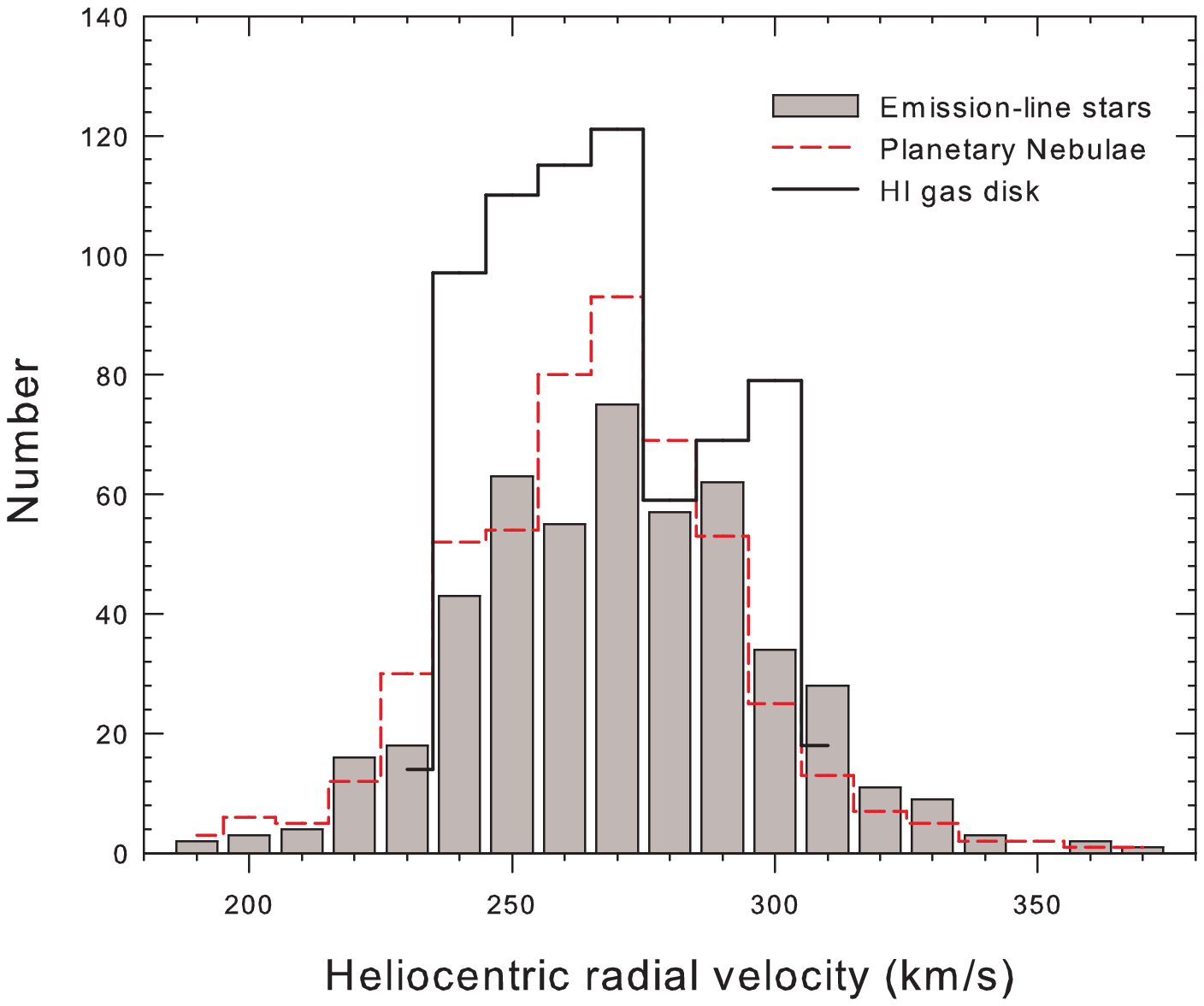}\\
  \caption{The distribution of LMC emission-line star velocities in our survey is shown by number and compared to the distribution for PNe and the HI gas disk. All sources have been placed into 10km s$^{-1}$ heliocentric radial velocity bins. The emission-line stars lie within the range found previously for LMC PNe and \HII~regions which is about 40km s$^{-1}$ wider at each end than the HI distribution. All three distributions share a mean peak number of sources at 270\,km\,s$^{-1}$. The HI has a sudden trough after the peak (280km s$^{-1}$) due to a warp which lies north of the main bar along the line of nodes (see Reid \& Parker 2006b).}
  \label{Figure25}
  \end{center}
\begin{center}
  \includegraphics[width=0.48\textwidth]{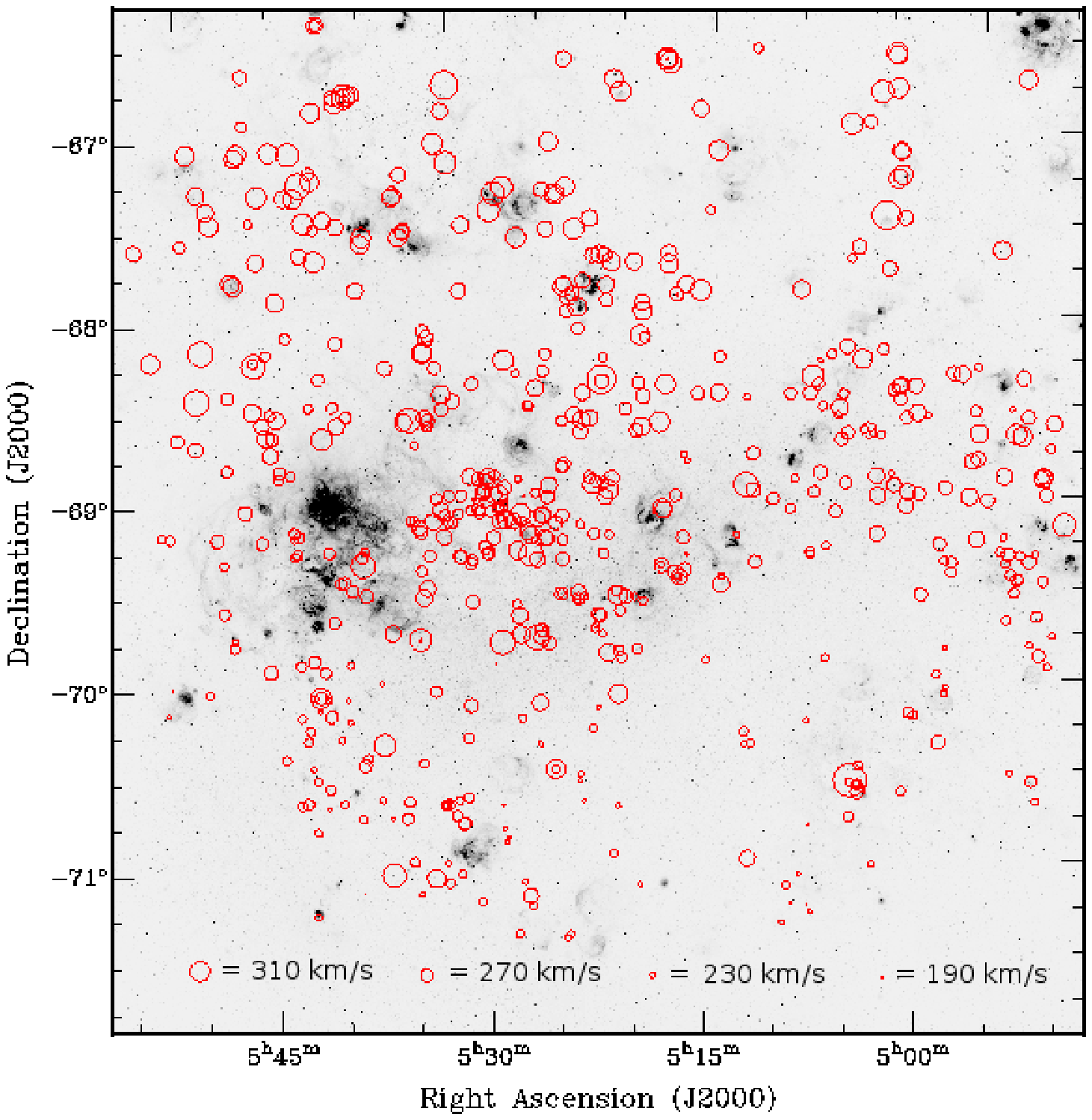}\\
  \caption{An H$\alpha$ map of the central 25deg$^{2}$ region of the LMC showing the distribution of hot emission-line stars as open red circles, the size of which indicates the measured heliocentric radial velocity. The larger the diameter, the higher the velocity. The circle sizes represent a linear scale but are magnified 10x in order to emphasis similarities and contrast the deviations within selected areas. Similar to the HI gas disk, the PNe and HII populations, there is a noticeable gradient from high velocities NE of the main bar to low velocities SW of the main bar. The average velocity of 270\,km s$^{-1}$ for emission-line stars is also common to all three populations on the main bar.}
  \label{Figure26}
  \end{center}
 \end{figure}

To derive a best possible radial velocity from our emission-line and
cross-correlation methods, we examined the error and other
properties (such as the fit and number of lines used) relating directly to each measurement system. From
{\scriptsize EMSAO}, we used weighted velocity results where a large proportion of fitted
lines were used in the compilation and error
values $\leq$13 km s$^{-1}$. These error values sum and weight the difference in emission line velocities for a given object. Errors larger than this value
begin to result from increasingly complex shell velocity
structure. Error values up to 13 km s$^{-1}$ were to be expected using this technique, as velocity ratios between different
lines (eg. H$\alpha$ and \NII$\lambda$6583) can vary depending on shock waves within the surrounding shell and/or in a few cases, ambient emission surrounding the star. In the
cross-correlation technique, we looked for high correlation peaks
and low error values $\leq$2 km s$^{-1}$. In general, we favoured the cross-correlation technique since $>$50\% of the target stars show only Balmer lines in emission and in many cases a weighted result was not possible with {\scriptsize EMSAO}. Results from {\scriptsize EMSAO} were used
where errors from the cross-correlation were above 13km s$^{-1}$.

A small percentage of these radial velocities will combine the true radial velocity with stellar atmospheric effects where the envelope is undergoing a phase of contraction or expansion. The contribution from these motions, unlikely to be much more than 50km s$^{-1}$, will not unduly displace these stars away from their location in the LMC. The Balmer and shell lines used for our radial velocities are formed in the cooler outer atmosphere. The lower order H$\alpha$ line is formed in the outer layers of the atmosphere and is less affected by the large velocity variations which can affect the higher members of the Balmer series which are formed at the deepest layers of the envelope. According the Struve's (1931) model, the mass flux of the star and its excitation steadily decreases towards a distance of several stellar radii where the emission lines are formed.

Our velocities, measured in the envelope, are lower than the escape velocity at the photosphere for stars with high $\textit{v}$ sin $\textit{i}$ found from photospheric HeI lines, which are in turn broad, weak and often perturbed by the envelope. Since each emission-line star is individualistic in terms of its $\textit{v}$ sin $\textit{i}$, shell structure, phase, periodic and non-periodic radial motions and amplitudes, a weighted average and cross-correlation of the emission line in the outer atmosphere is the most efficient and accurate means of applying a radial velocity to each emission-line star in our catalogue.

In figure~\ref{Figure25} we show a histogram of the heliocentrically corrected radial velocities for 501 of the hot emission-line stars in our survey. These are compared with our heliocentric velocities for 515 LMC PNe (Reid \& Parker, 2006b) and 686 HI gas disk pointings from the LMC map of Rohlfs et al. (1984), covering the entire 25deg$^{2}$ survey region. Each pointing averages an $\sim$11.45 arcmin$^{2}$ sub-region, ensuring an unbiased and fully representative distribution and mean can be obtained. The comparison shows that emission-line star velocities lie within the acceptable velocity boundaries and conform well with other LMC population types such as PNe and the \HI~gas (also see Reid \& Parker, 2006b).

LMC emission-line stars and PNe have a very similar distribution but a wider range compared to the HI disk. Although 483 (89\%) of the emission-line stars occupy the HI range 230km s$^{-1}$ to 310km s$^{-1}$, the 52 outliers are to be expected since the HI disk has a narrow vertical velocity dispersion ranging between 17\,km s$^{-1}$ and 2.2\,km s$^{-1}$ with a mean of 6\,km s$^{-1}$ compared to PNe (45\,km s$^{-1}$ to 3\,km s$^{-1}$; mean 22\,km s$^{-1}$) and emission-line stars (43km s$^{-1}$ to 3km s$^{-1}$; mean 23km s$^{-1}$), found by sampling 37 $\times$~37 arcmin sub-regions across the survey area. A few large dispersions in HI can indicate a splitting of the gas disk, which occurs in regions such as the leading arm (see Reid \& Parker 2006b). The central peak of 270km s$^{-1}$ is common to all three distributions and indicates a strong midpoint incorporating both young and old populations. The sudden trough at 280km s$^{-1}$ for the HI gas disk is further proof of a sharp warping of the disk which runs north of the main bar in a SE to NW direction, close to the line of nodes (see Reid \& Parker 2006). This warping produces a large velocity dispersion and fewer velocities at the 280km s$^{-1}$ level. Figure~\ref{Figure26} clearly shows proportionally higher velocities for emission-line stars north-east of the main bar and lower velocities south-west of the main bar. This overall gradient is shared by PNe, HII regions and the HI disk but PNe, HII regions and emission-line stars have a greater vertical dispersion at any point than the HI disk.

The common peak velocity (270\,$\pm$30\,km s$^{-1}$) does not necessarily mean that each population shares the same center of rotation. In fact the rotational centre for PNe and the HI disk have been shown to be located in separate positions (Reid \& Parker, 2006b). What we can see from Figure~\ref{Figure26} is that 270\,$\pm$30\,km s$^{-1}$ is the average velocity for emission-line stars in the main bar region.

\label{section9}

\section{Distribution across the LMC survey area}

\begin{figure*}
\begin{center}
  \includegraphics[angle=270,width=0.8\textwidth]{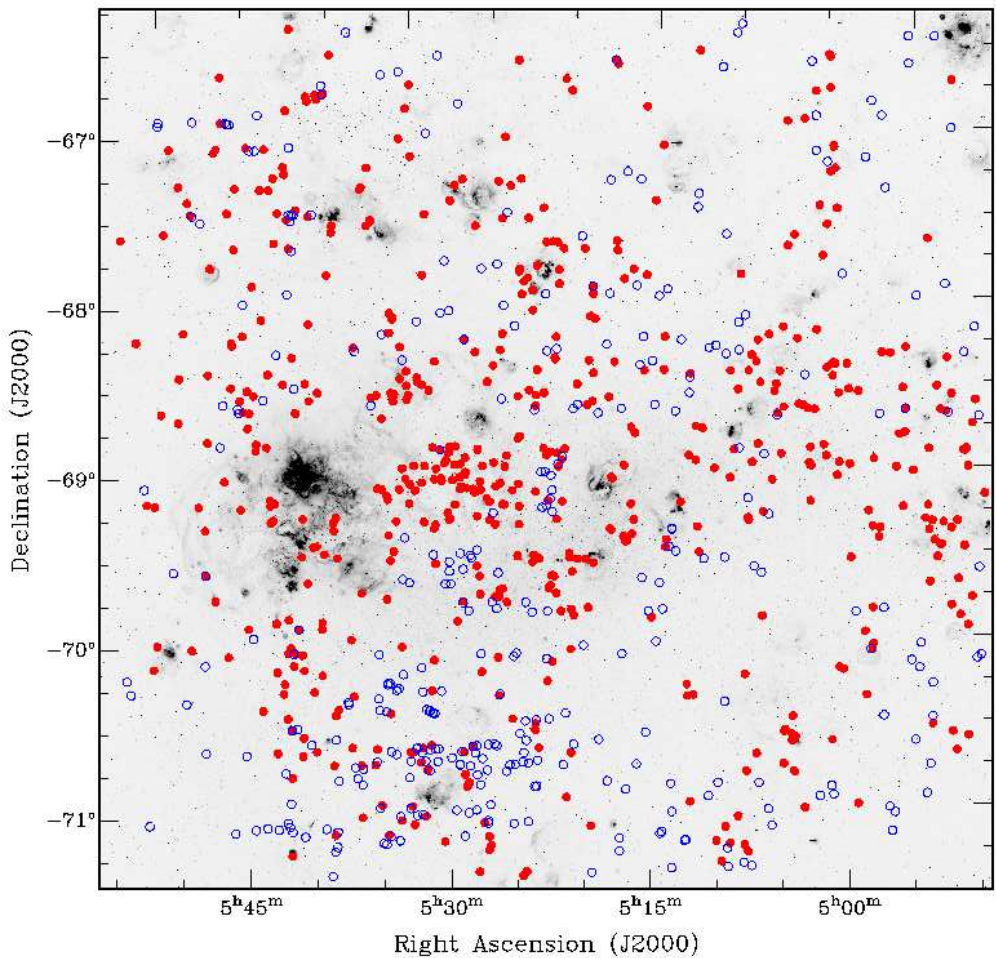}\\
  \caption{An H$\alpha$ map of the central 25deg$^{2}$ region of the LMC showing the distribution of hot emission-line stars as filled red circles. The positions of late-type (mostly M-giant) stars are shown as blue open circles. Stars less than 10 arcsec apart cannot be distinctly separated by the large markers in this image. There is a significant concentration of hot emission-line stars on the immediate north-eastern side of the main bar while M-giants are concentrated to the south.}
  \label{Figure27}
  \end{center}
  \end{figure*}

  In figure~\ref{Figure27} we show the distribution of emission-line stars, superimposed on an H$\alpha$ map of the central 25deg$^{2}$ region of the LMC. The surveyed Be stars in the region are shown as filled red circles while M giants are shown as open blue circles.

Much of the resulting distribution depends on our selection criteria since we were searching for compact emission objects and emission stars fainter than Mag$_{R}$ = 14. For this reason, the most luminous emission-line stars detected in the H$\alpha$ survey were not spectroscopically observed. Objects were selected for spectroscopic follow-up based on the strength of the H$\alpha$ emission relative to the luminosity of the central star. Stars with low ($<$5\% the central star) H$\alpha$ excess were not spectroscopically followed up as their low variability and/or emission excess over the 3 year duration of the survey indicated that they were not strong emission-line star candidates. Where the criteria were met, we extended the selection to the faintest luminosity candidates we could find. Emission-line stars found in clusters and associations were only earmarked where related velocities or previous published work make the association clear.

The densest distribution of B-F emission-line stars occurs across the main bar. From there they form a line northwards, following the line of nodes (see Reid \& Parker 2006b). There is also a large number to be found along the leading arm, south of 30DOR. Being a young population of stars, they trace the more recent star formation regions and \HII~distribution quite well compared to the older population of PNe, which are more randomly distributed within the LMC (see Reid \& Parker 2006b). The somewhat older M population, however, is more evenly distributed across the north and main bar of the LMC. There is a denser distribution of late-type stars along the leading arm which is thought to be a remnant of the LMC's close encounter with the SMC which may have occurred $\sim$2 $\times$ 10$^{8}$ years ago (Murai and Fujimoto, 1980).

\label{section10}

\section{H$\alpha$ luminosity effects as a function of spectral type}

The theory of equatorial darkening suggests that a degeneracy in the rotational rates allows Be stars to be rotating at or near their critical breakup velocity, Townsend et al. (2004). This implies that there will be a maximum mass and hence, luminosity, allowable for a Be star. The question then arises as to whether the intensity of H$\alpha$ emission from these stars relates to the luminosity class for each star. In other words, does the intensity of H$\alpha$ emission increase in hotter stars$?$ To answer this we have constructed the first ever H$\alpha$ luminosity histogram as a function of spectral type for Be stars, using our sample in the LMC. It is based on measuring the total flux emitted in the H$\alpha$ line over and above the continuum level in each star.

In order to do this, measured H$\alpha$ fluxes were converted to the H$\alpha$ magnitude scale by correlating the H$\alpha$ flux from known emission objects with no continuum and no \NII~contamination against well established H$\alpha$ magnitudes for those objects. A zero point scale was then found in order to convert all H$\alpha$ fluxes to H$\alpha$ magnitudes. This allows comparison to other luminosity functions, such as the planetary nebulae luminosity function, which is already extensively used for distance determination.

\begin{figure}
\begin{center}
  \includegraphics[width=0.48\textwidth]{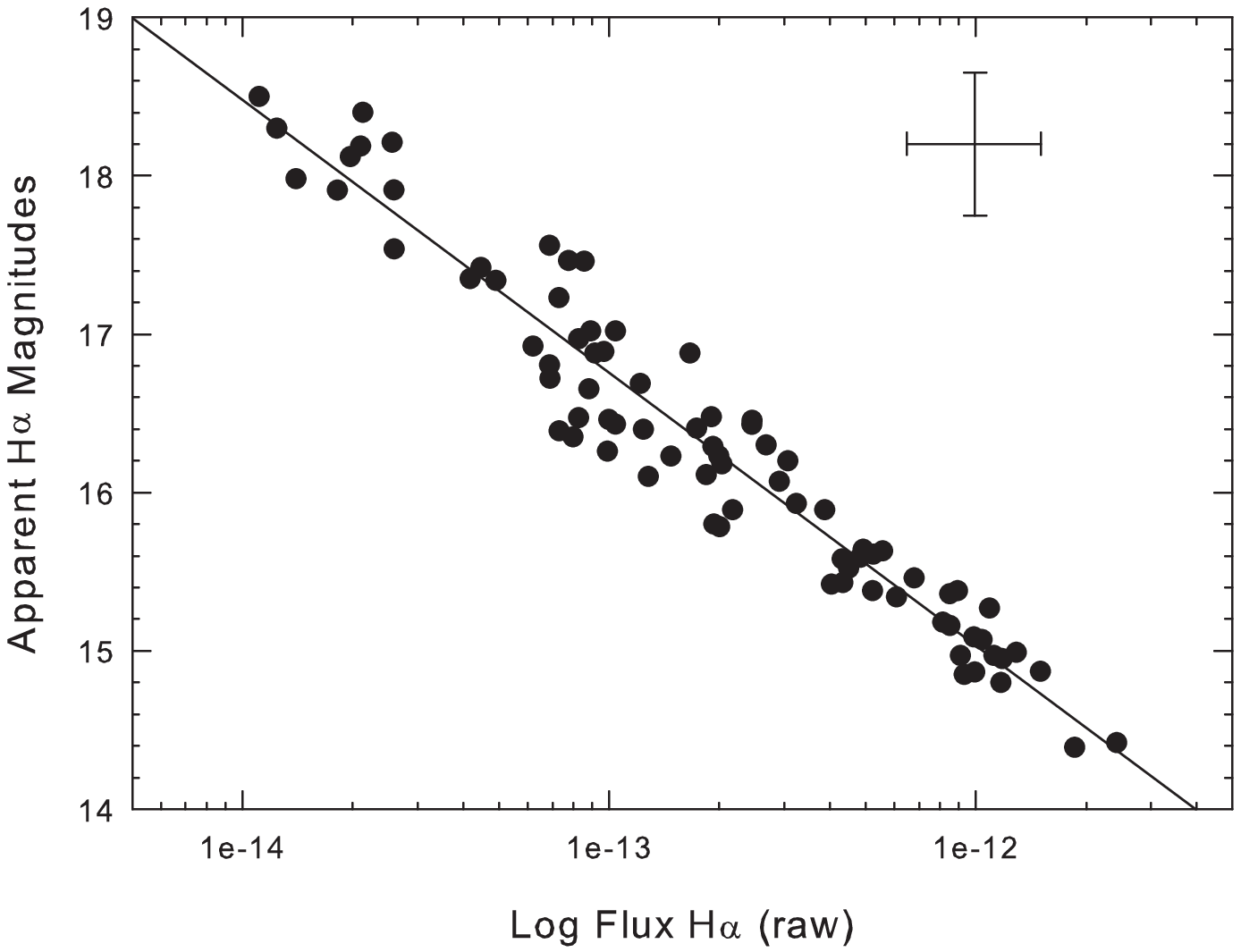}\\
  \caption{The graph, based on our LMC PN sample, used to convert H$\alpha$ flux to H$\alpha$ magnitude. The apparent H$\alpha$ magnitudes for LMC PNe were measured using image photometry on the stacked H$\alpha$ map. Only PNe with extremely low or no \NII 6548, 6583\AA~lines were included in the calibration since the presence of these lines in the wings of the filter can effect the H$\alpha$ measurement. The H$\alpha$ fluxes are from our calibrated 2dF and longslit fluxes for the same PNe. Calibrated this way, there is no need to subtract continuum from the H$\alpha$ line. The scatter is expected due to the limited linear response of the digitised photographic data used to create the H$\alpha$ map. The line of best fit is shown on the graph and the underlying algorithm is provided at equation 4.}
  \label{Figure28}
  \end{center}
  \end{figure}

This inaugural conversion of H$\alpha$ fluxes to magnitudes
was made by choosing PNe with published HST-derived fluxes and 2dF spectra where the PN showed no measurable
\NII\,$\lambda$6548 \& $\lambda$6583 but were bright enough to be seen in broad-band R. PNe were chosen because the continuum level is close to zero, allowing the measurement of H$\alpha$ emission only.
Each PN was located in the H$\alpha$ map data and an
R-band image with an area of 0.1 arcmin radius around each PN was downloaded from
SuperCOSMOS online. Along with the image, the IAM data `tab' file was also
downloaded. This file contains precise object positions and R
magnitudes from the SuperCOSMOS sky survey and the ESO guide star
catalogues. These magnitudes were graphed (see Figure~\ref{Figure28}) against our calibrated LMC PN fluxes (Reid \& Parker 2010) and fluxes from the MCPN catalogue (Stanghellini et al. 2002).

  The fit is sufficient to reveal the position of the line of best fit which will be used to perform the conversion. The scatter, up to half a magnitude, seen on either side of the logarithmic line of best fit is to be expected due to the limited linear
  response of the digitized film and characteristics of the H$\alpha$ filter used on the UKST but does not pose any problem to the calibration of fainter objects since the AB magnitude scale is always fixed at 2.5 log (F$_{H\alpha}$). The logarithmic relation of flux to magnitude means that the slope of the line of best fit will always be fixed. The graph is simply used to attain the magnitude conversion value, which is the final number in the empirical
  relation:

  \begin{equation}
  m_{H\alpha} = -2.5~log F_{H\alpha} - 14.15
  \end{equation}

for the conversion of all the derived H$\alpha$~fluxes (ergs$^{-1}$ cm$^{-2}$ s$^{-1}$) into
H$\alpha$ apparent magnitudes (m$_{H\alpha}$). A mean error estimate
of $\pm$0.27 mag is based on line measurement and flux calibration
errors at a total 7\% plus 0.1 mag for uncertainties in image
photometry.

To check the veracity of this calibration, we used the ESO magnitude-to-flux converter\footnote{ http://archive.eso.org/apps/mag2flux/}~to convert H$\alpha$ fluxes in ergs$^{-1}$ cm$^{-2}$ s$^{-1}$ to H$\alpha$ magnitudes, using a variety of narrow band filters. Using the accepted flux-to-mag conversion of -2.5~$\textrm{log}~\textsl{F}_{5007}$-13.74 for \OIII$\lambda$5007~emission lines (Jacoby 1989), we simulated a variety of narrow band filters and telescopes and found that any given flux value will be between 0.4 and 0.58\,mag brighter in H$\alpha$ than in \OIII$\lambda$5007. With our magnitude correction of -14.15, a given flux will be 0.41\,mag brighter in H$\alpha$ than in \OIII$\lambda$5007, in basic agreement with ESO simulations, giving us added confidence in our new empirical relation.

  \begin{figure}
\begin{center}
  \includegraphics[width=0.50\textwidth]{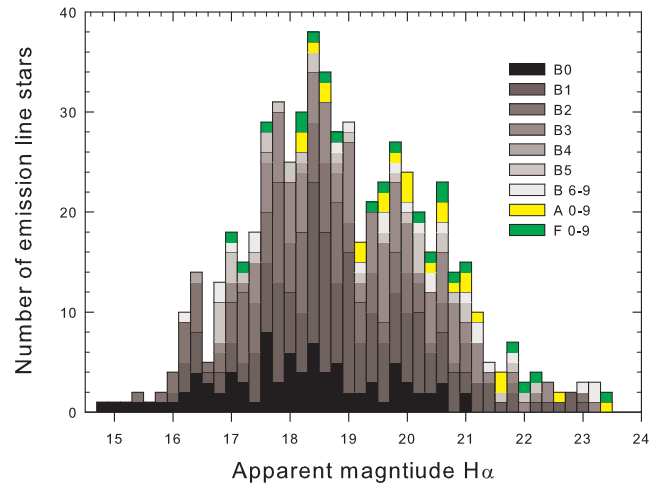}\\
  \caption{The luminosity function of H$\alpha$ emission from hot emission-line stars found in our survey within the central 25deg$^{2}$ region of the LMC. The luminosity bins are 0.2 magnitudes and spectral classifications are indicated in the legend.}
  \label{Figure29}
  \end{center}
  \end{figure}

  Using this conversion we constructed the H$\alpha$ luminosity function, which is a measure of the H$\alpha$ emission above the continuum and is presented in terms of the each stars' spectral classification (see Figure~\ref{Figure29}), a relation which has been unknown to date. The figure shows only a modest increase in H$\alpha$ luminosity with increasing spectral type over a 9 magnitude range. The spectral type or temperature of the star therefore does not correlate strongly with the luminosity of the Balmer emission. As expected, however, classes B0 to B3 dominate the bright end while classes B6 to F9 can mainly be found at the faint end. The bright cutoff occurs at magnitude 15 (an absolute magnitude of -4.5) and the peak in the function (the largest number of stars in any particular bin) occurs at magnitude 18.6. After this peak there is a steady decrease in the distribution over the next 5 magnitudes to the faintest detection at magnitude 23.8. The lone star with a bright H$\alpha$ magnitude of 14.6 is a luminous blue variable (LBV). The shape of the distribution is not unlike that for planetary nebulae in the LMC (see Reid \& Parker 2010) but it is unlikely that this function can be used as an extra-galactic distance scale as the brightest H$\alpha$ emission line is a magnitude fainter than the brightest \OIII$\lambda$5007~line from PNe in the LMC which is traditionally used for the PNLF.

\label{section11}

\section{Photometry}

  We obtained B, V, I magnitudes from OGLE-II photometry for 54 previously known Be and B[e] stars which were found to have strong variability. To this number we add 63 newly discovered Be stars with published OGLE-II photometry (Szyma\'{n}ski, 2005, Udalski et al. 2002), found from the limited OGLE coverage of the main bar only. For other stars not in the OGLE data base we gained I, B and R photometry from SuperCOSMOS. The Starlink {\small GAIA} image detection option was used to detect and isolate sources by placing an ellipse around the assumed centre. For single stars found in relative isolation this works extremely well. For other sources with close companions or within clusters, the de-blending option was employed. The position of each star was manually checked against the downloaded image to ensure accuracy of positioning and non-blending.

  To supplement the OGLE-II V magnitudes we also include GSC2 V magnitudes from ESO. We warn the user to use care in comparing the three photometric sets directly against each other due to intrinsic variability and the change of epoch between the three surveys. Unless specified, we only compare OGLE-II photometry in this section since the published values are an average across the survey period.

  \subsection{V vs (B-V) colour-magnitude diagram}

  B stars congregate at the upper left of the traditional H-R colour-magnitude diagram, close to a 0 B-V colour and where the tracks for main sequence, subgiant and giant stars converge. This area of the H-R diagram is a useful test for our Be stars for two reasons. Firstly, by separating giants from main sequence stars, we can independently test our correlated estimates for luminosity class. Secondly, we can see if the variability of these stars is having any effect on the normal position for these stellar classes.

\begin{figure}
\begin{center}
  \includegraphics[width=0.55\textwidth]{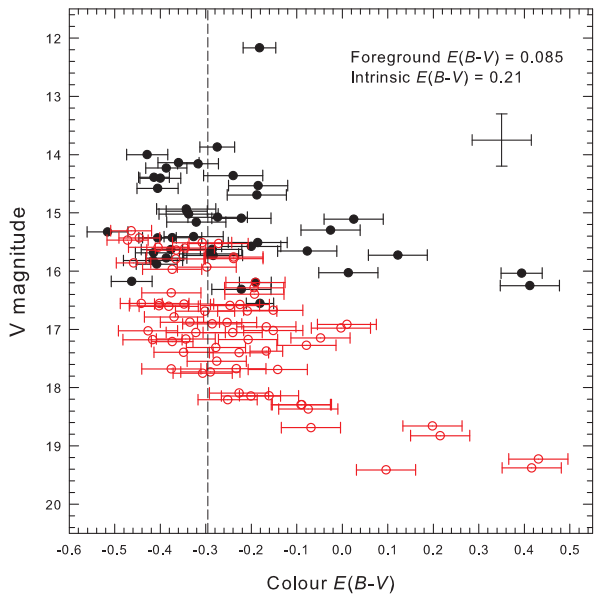}\\
  \caption{The V versus $\textsl{E}$($\textsl{B-V}$) colour-magnitude diagram from OGLE-II photometry for 117 variable hot emission-line stars found on the LMC main bar. Main sequence stars are assigned red open circles while giant stars are represented by filled black circles. We assume a combined foreground and intrinsic reddening of $\textit{E}$($\textit{B}$-$\textit{V}$) = 0.295. Error bars are based on a combination of B and V published error estimates for B stars on or very close to the main bar. Maximum errors on both scales are shown. The observed position of (B-V) = 0 is shown by the broken vertical line. }
  \label{Figure30}
  \end{center}
  \end{figure}

  The V versus (B-V) magnitudes for a sample of 100 of our LMC emission-line stars is shown in Figure~\ref{Figure30}. These magnitudes are derived from OGLE-II photometry where the variability of these stars has been established. Their position on the diagram includes corrections for foreground and intrinsic reddening within the LMC. These reddenings were obtained from the Str\"{o}mgren CCD photometry on LMC fields conducted by Larsen et al (2000) using B stars. Although several of the stars in this small sample exhibit a strong reddening, only 3 of them -  RPs83, RPs1751 and RPs1350 - are visibly surrounded by extended emission halos ($\geq$6 arcsec radius). Rather than applying a small reddening to each individual object, we adopt an uncertainty of $\pm$0.10 for all of these stars since they are close to or on the main bar (Larsen et al. 2000). The stars are separated into main sequence (open red circles) and giants (filled black circles) in Figure~\ref{Figure30}~where the position of the intrinsic (observed) zero point for (B-V) is shown as a broken vertical line.

  The plot indicates that the cross-correlation technique appears to be working very successfully. Main sequence stars appear to be spread across the plot at a broad 45 degree angle from the lower right, following the established track for main sequence stars. Giants on the other hand span across the centre of the plot to the left where they increase in V magnitude.

  \subsection{H$\alpha$ emission}

\begin{figure}
\begin{center}
  \includegraphics[width=0.52\textwidth]{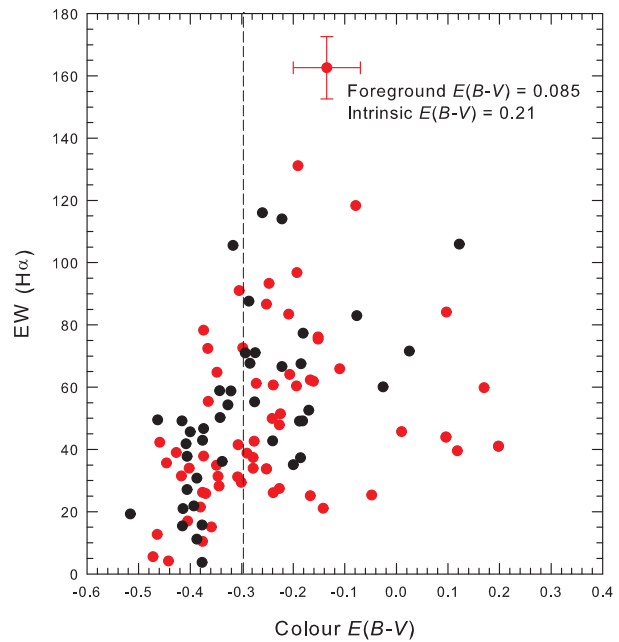}\\
  \caption{The equivalent width (H$\alpha$) versus $\textsl{E}$($\textsl{B-V}$) colour diagram using our measured H$\alpha$ magnitudes and averaged OGLE-II photometry for 100 variable Be stars in the LMC sample. We assume a combined foreground and intrinsic reddening of $\textit{E}$($\textit{B}$-$\textit{V}$) = 0.295. Error bars are based on a $\pm$0.10 estimated error for B stars on or very close to the main bar. The observed position of (B-V) = 0 is shown by the broken vertical line. There is a mild correlation up to $\textsl{E}$($\textsl{B-V}$) of $\sim$-0.3 but as the colour index increases the relation breaks down. Giant stars are again shown in black and main-sequence stars are shown in red. }
  \label{Figure31}
  \end{center}
  \end{figure}

  A correlation has been found between the strength of H$\alpha$ emission and the spatial extent of the emitting region of a Be star (eg. Quirrenbach et al., 1997, Tycner et al. 2005). The H$\alpha$ emission is also thought to be correlated with the observed colour excess (Dachs et al. 1988) where an increase in both H$\alpha$ emission and the colour excess $\textsl{E}$($\textsl{B-V}$) will occur with a larger contribution from bound-free and free-free continuum emission (Beaulieu et al. 2001). For Be stars in our LMC sample, we also find a mild correlation between the Balmer line radiation originating from the stellar envelope as exhibited by H$\alpha$ equivalent width EW(H$\alpha$) and the (B-V) colour indices. This relation is shown in Figure~\ref{Figure31} where the correlation is strongest at low EW(H$\alpha$).

  \begin{figure}
\begin{center}
  \includegraphics[width=0.52\textwidth]{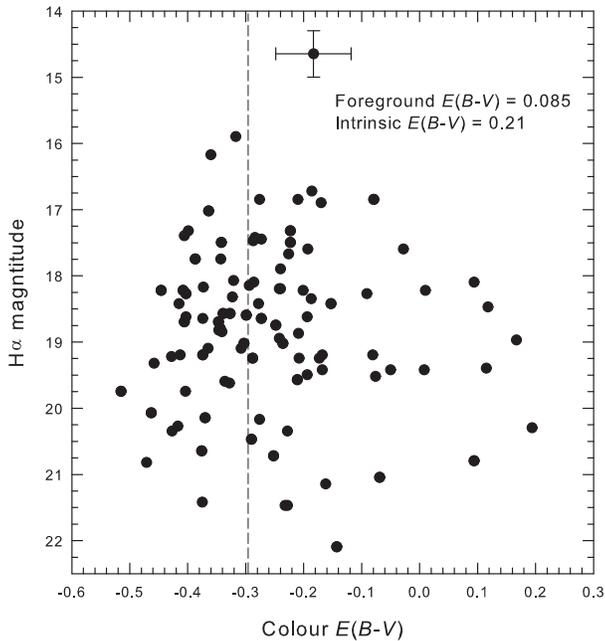}\\
  \caption{The H$\alpha$ emission from stars in the LMC is compared with their $\textsl{E}$($\textsl{B-V}$) colour. No real correlation between the H$\alpha$ magnitude and colour index can be seen. The magnitude of the H$\alpha$ emission is therefore somewhat independent of the $\textsl{E}$($\textsl{B-V}$) colour of the host star. Similar to Figure~\ref{Figure31}, however, the minimum observed $\textsl{E}$($\textsl{B-V}$) colour appears to increase with increasing H$\alpha$ magnitude above magnitude 20.}
  \label{Figure32}
  \end{center}
  \end{figure}

  \begin{figure}
\begin{center}
  \includegraphics[width=0.50\textwidth]{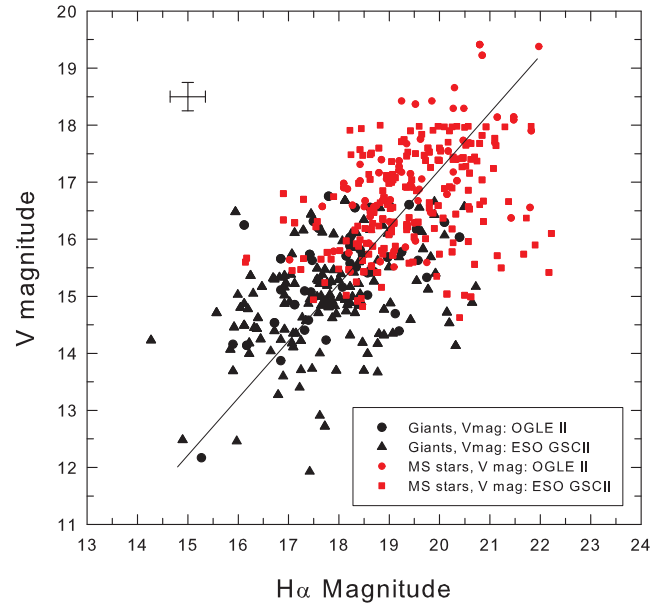}\\
  \caption{The luminosity of H$\alpha$ emission from stars in the LMC is compared with their V magnitude. Giants and main-sequence stars are separated and assigned black and red colours respectively. A mild correlation can be seen down to the faint cut-off of V=18 for the ESO dataset. The linear line of regression indicates that H$\alpha$ magnitudes are generally 2.72 magnitudes fainter than the visual magnitude for these stars. The correlation coefficient is 0.578 and the equation: V$_{mag}$ = H$\alpha$ -- 2.72, may be used as a rough estimate. Maximum expected errors based on published (for V) and measurement/calibration (for H$\alpha$) are shown.}
  \label{Figure33}
  \end{center}
  \end{figure}

  The increasing scatter with increasing EW(H$\alpha$) is partly due to increased reddening from interstellar dust and emission from the circumstellar gas envelope, and partly due to complex variations in the H$\alpha$ emission profiles between the time of our spectroscopic observations and the OGLE-II observations. Since the measured EW(H$\alpha$) is an integrated quantity, it has the tendency to be insensitive to the small-scale variations in the line profile. Effects from OGLE-II photometry, where the LMC was observed repeatedly between 1997 and 2000 will mostly correspond to our 12 H$\alpha$ stacked images, also observed between 1997 and 2000. The average of these photometric variations over 3 year timescales was applied to our spectroscopic observations conducted in 2004 and 2005. Since (B-V) has been averaged out over timescales of years, this ratio is not expected to vary greatly with variation of the star's intrinsic luminosity. For the most variant objects, our spectroscopic measurements of the H$\alpha$ line require slightly larger error margins but remain impossible to estimate without repeated spectroscopic exposures.

  Despite these caveats, a mild correlation is still evident. The decrease in the maximum observed value of (B-V) with increasing EW(H$\alpha$) is one of the main features. It implies that cooler stars will have larger emission shells with a probable maximum size allowable for each spectral class.

  In Figure~\ref{Figure32}~we replace the EW(H$\alpha$) with the H$\alpha$ emission flux above the continuum. There is no correlation evident, however, the range in (B-V) appears to broaden with decreasing flux suggesting that low H$\alpha$~flux can be present in both the brightest and faintest Be stars.

  Figure~\ref{Figure33} shows that the H$\alpha$~magnitude is almost consistently fainter than the V magnitude of the star by a mean of 2.72$\pm$1 mag. The correlation coefficient between the H$\alpha$ and V magnitudes for our set is only 0.578, implying that the V magnitude of any particular emission-line star could be associated with a wide range of H$\alpha$ flux excess. Figure~\ref{Figure33} shows that this could be up to 3 magnitudes either side of the mean correlation, which is V$_{mag}$ = H$\alpha$ -- 2.72.

  The main body of available V magnitudes cease at V=18 due to the limit of the ESO catalogue. OGLE-II magnitudes extend to fainter limits. Stars intrinsically fainter than V=18 include a wide variety of spectral types so it may be that many of them were undergoing a strong emissive phase at the time of our spectroscopic observations. The effect of H$\alpha$ flux variability upon V magnitude is impossible to estimate, however we may assume that a portion of the scatter away from the line of best fit may be due to oscillations.

  The emission-line stars in figure~\ref{Figure33} have been separated into giants and main sequence classifications in order to investigate their positions on the H-R diagram according to luminosity class. As expected, the giants dominate the bright end and the main sequence stars dominate the faint end of the plot. The overlap region of $\sim$2 magnitudes from V=14.7 to V=16.7 contains about 15 main sequence stars with rather low H$\alpha$ emission. There is nothing peculiar that these stars have in common. Their spectral types range from B1Ve to A6IVe. The size separation either side of the overlap region (V$_{mag}$ 16.7-14.7) is very robust, permitting a secure size assessment to be made based on V magnitude alone.

\label{subsection11.1}


\begin{figure}
\begin{minipage}[b]{0.45\linewidth}
\centering
\includegraphics[scale=0.285]{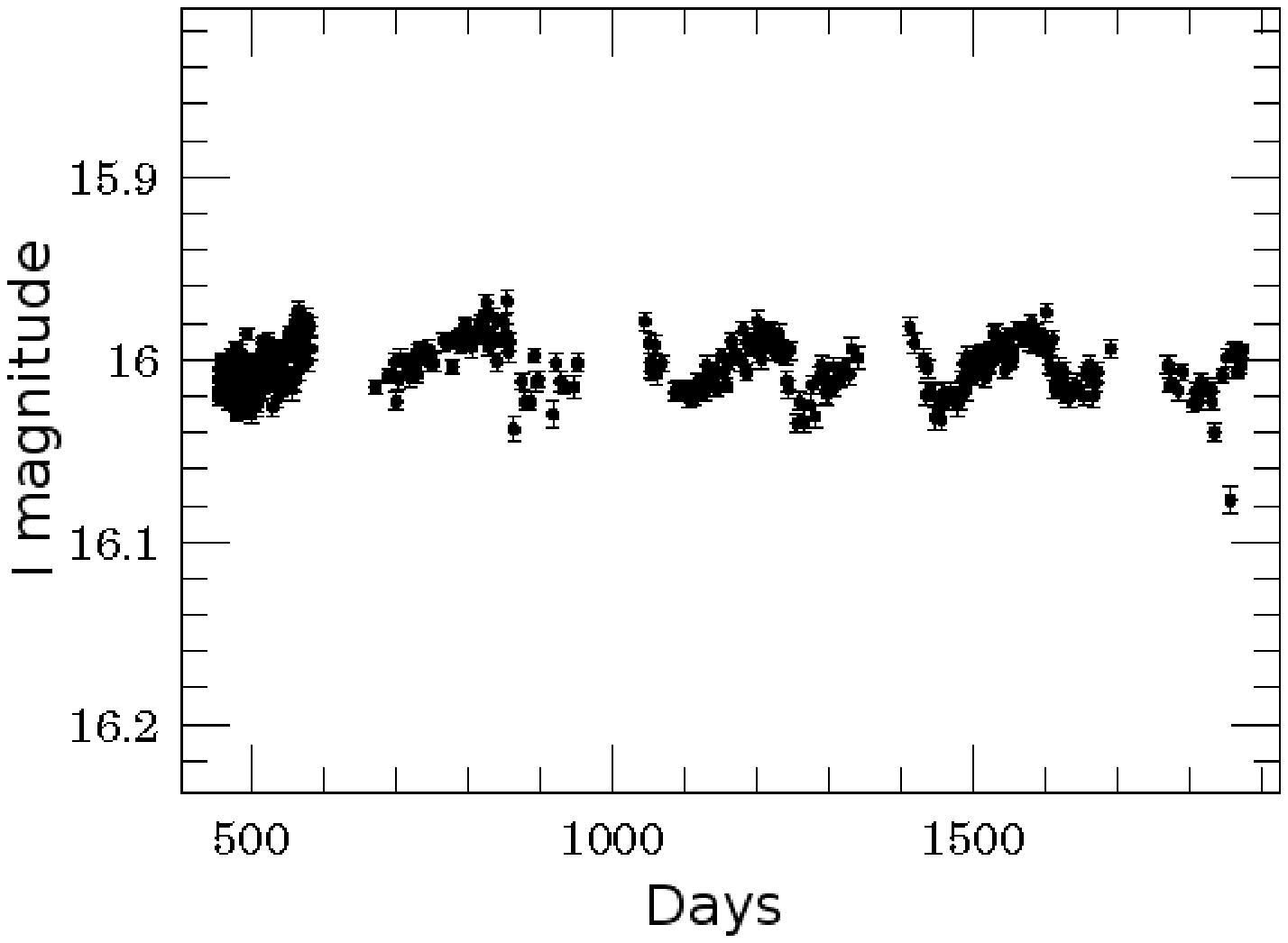}
\end{minipage}
\hspace{0.2cm}
\begin{minipage}[b]{0.45\linewidth}
\centering
\includegraphics[scale=0.285]{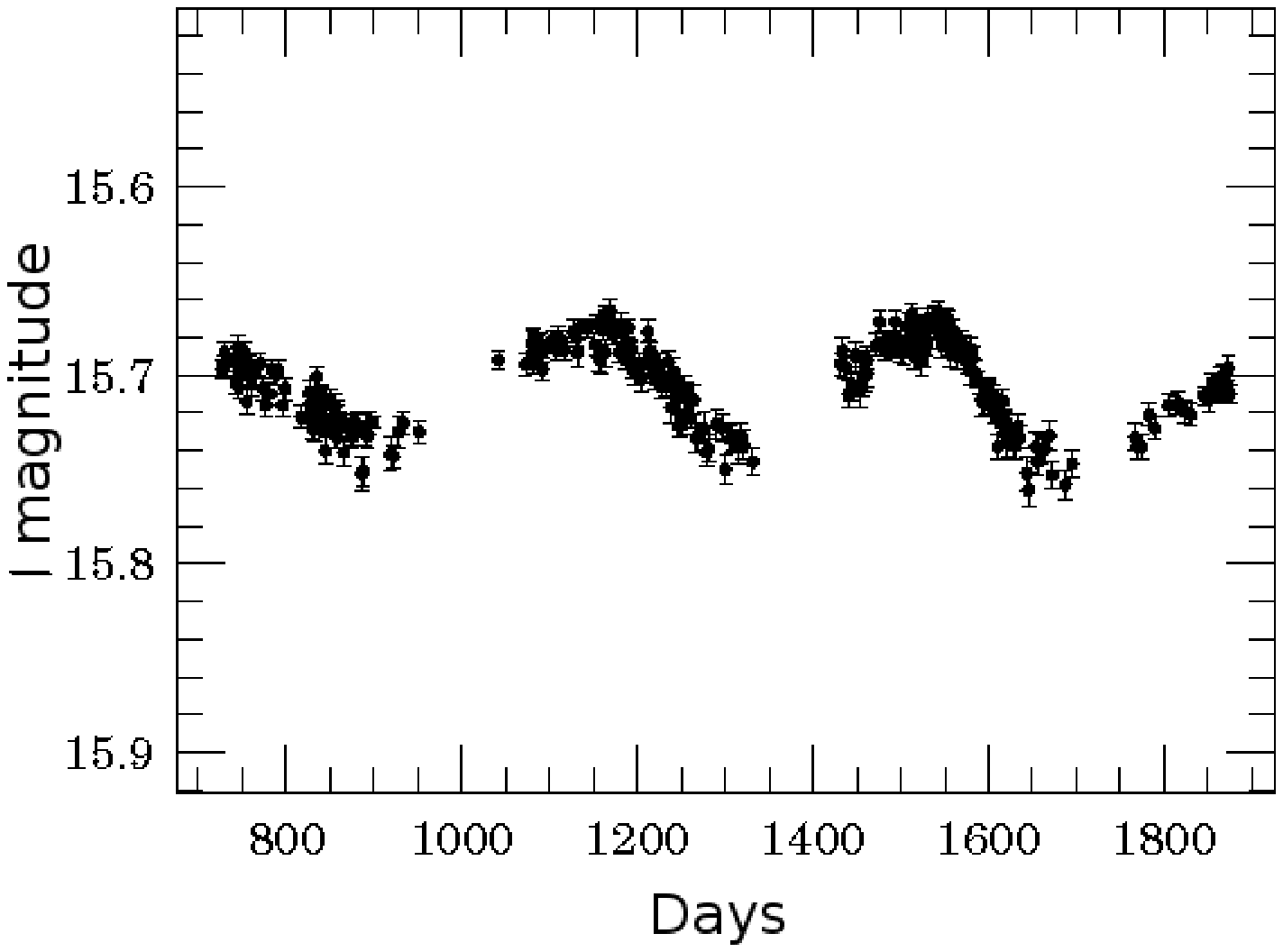}
\end{minipage}
\caption{Light curves for RPs1383 (B0Ve, H$\alpha$: Single Peak, $\textit{v}$ sin $\textit{i}$: 243 $\pm$12\,km\,s$^{-1}$) and RPs1794 (B2V[e], H$\alpha$: Single Peak, $\textit{v}$ sin $\textit{i}$: 185\,$\pm$9\,km\,s$^{-1}$) from the OGLE-II database. These examples show regular variability, with a steady central magnitude.}
\label{Figure35}
\vspace{0.2cm}
\begin{minipage}[b]{0.45\linewidth}
\centering
\includegraphics[scale=0.285]{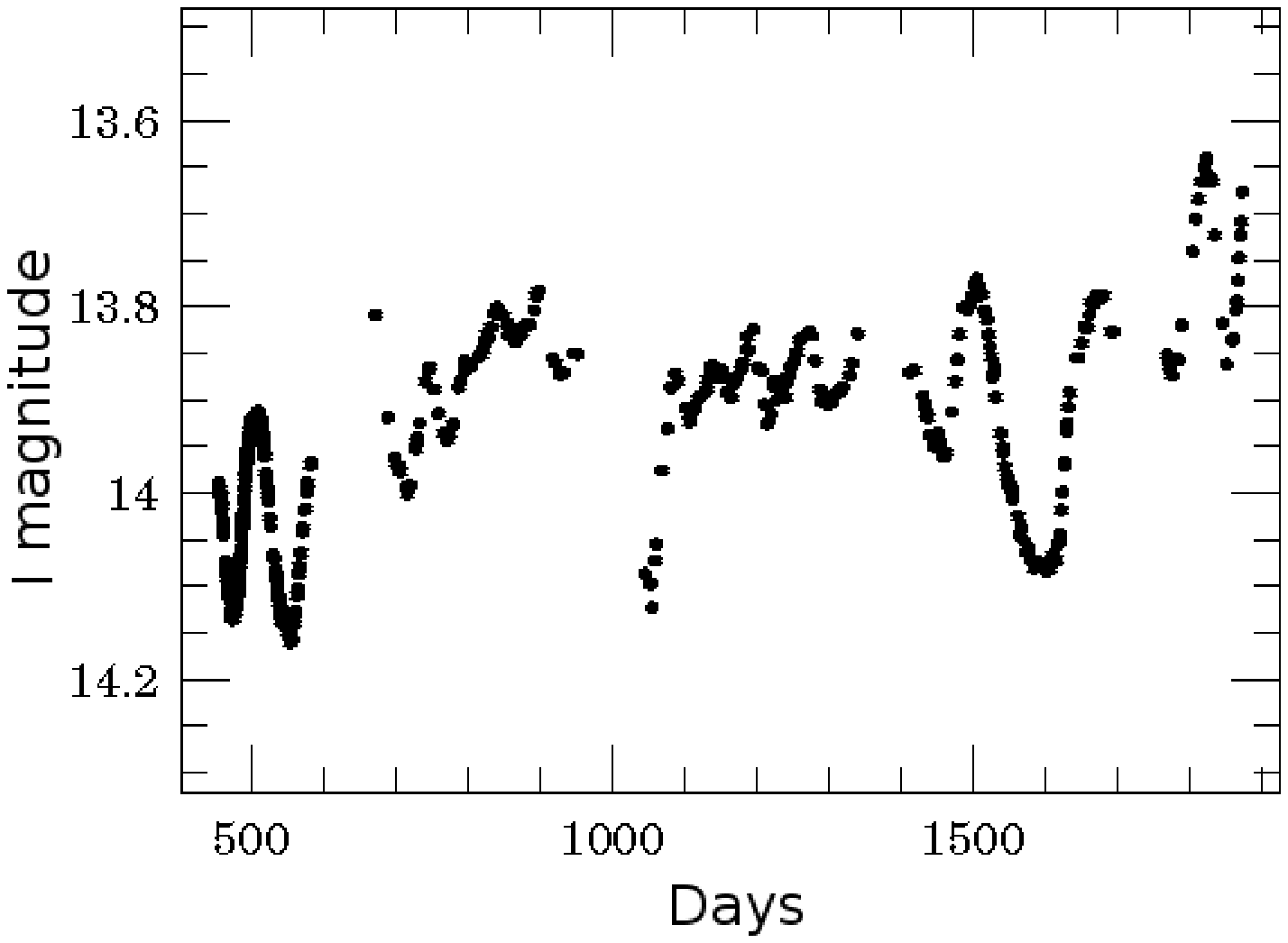}
\end{minipage}
\hspace{0.2cm}
\begin{minipage}[b]{0.45\linewidth}
\centering
\includegraphics[scale=0.285]{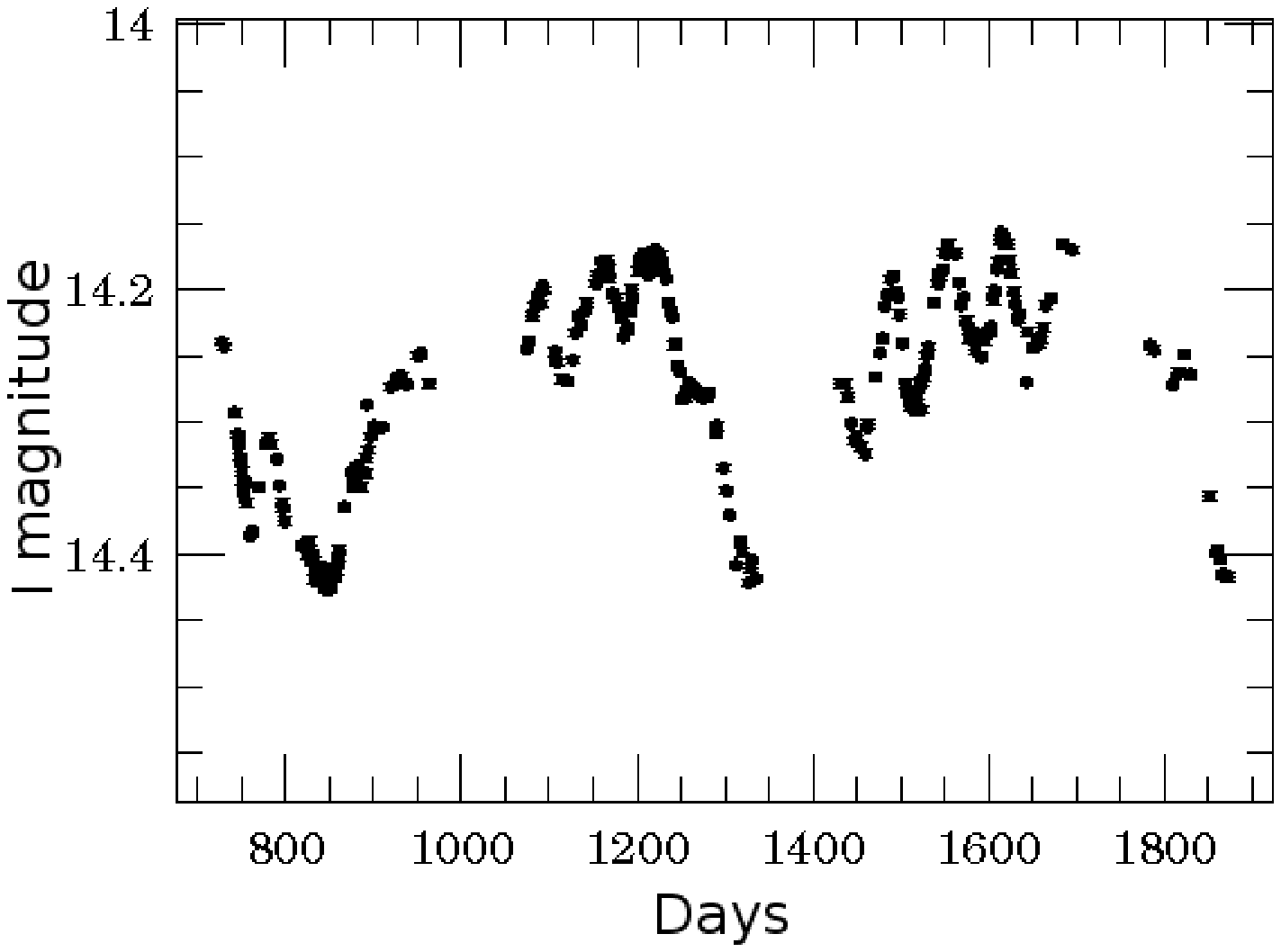}
\end{minipage}
\caption{Light curves for RPs870 (B2IV[e], H$\alpha$: Double Peak R$>$V, $\textit{v}$ sin $\textit{i}$: 245\,$\pm$12\,km\,s$^{-1}$) and RPs2197 (H$\alpha$: Double Peak R$>$V, $\textit{v}$ sin $\textit{i}$: 82\,$\pm$4\,km\,s$^{-1}$) from the OGLE-II database. These examples show semi-regular variability interspersed with minor variations.}
\label{Figure36}
\vspace{0.2cm}
\begin{minipage}[b]{0.45\linewidth}
\centering
\includegraphics[scale=0.285]{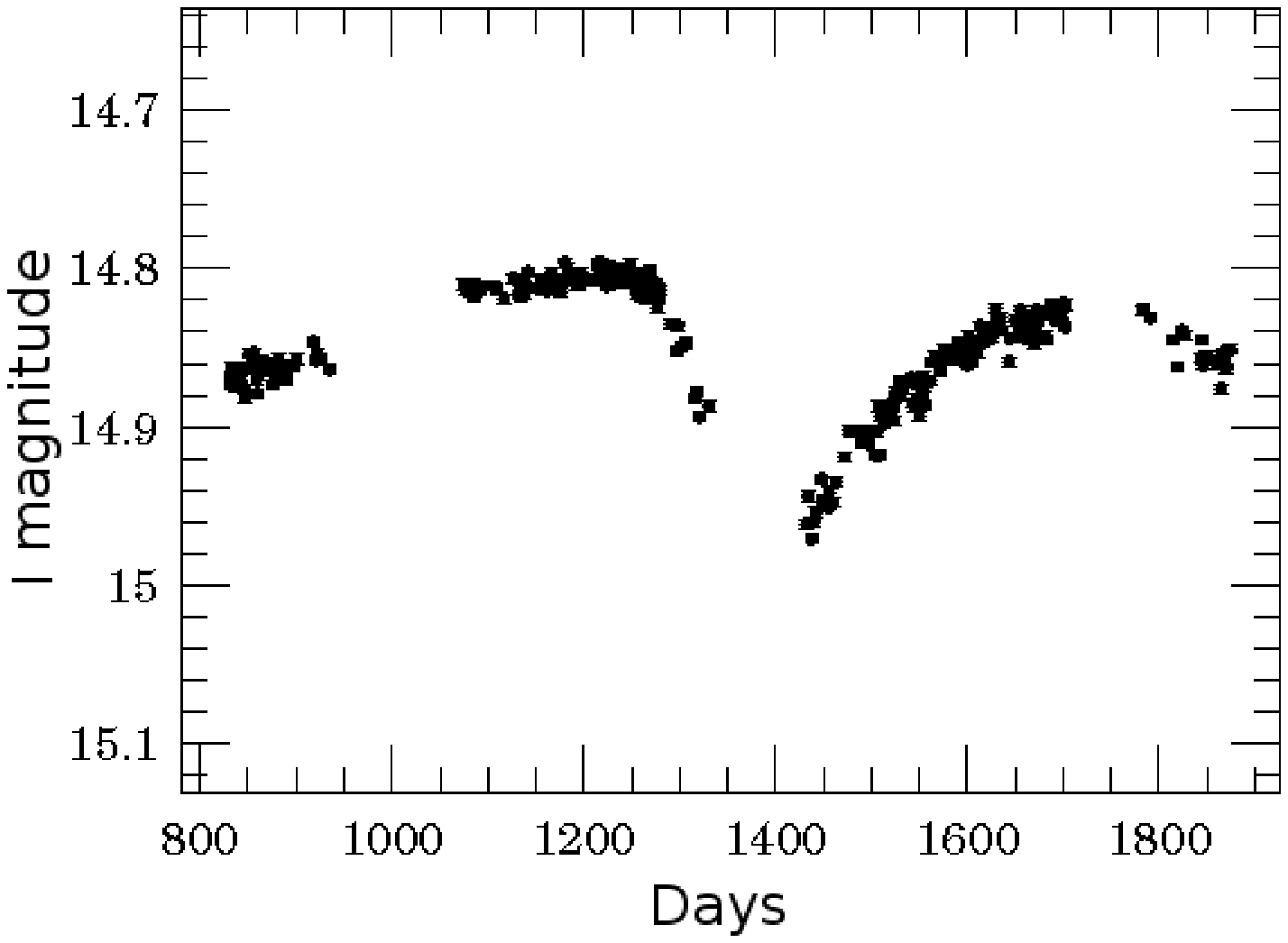}
\end{minipage}
\hspace{0.2cm}
\begin{minipage}[b]{0.45\linewidth}
\centering
\includegraphics[scale=0.285]{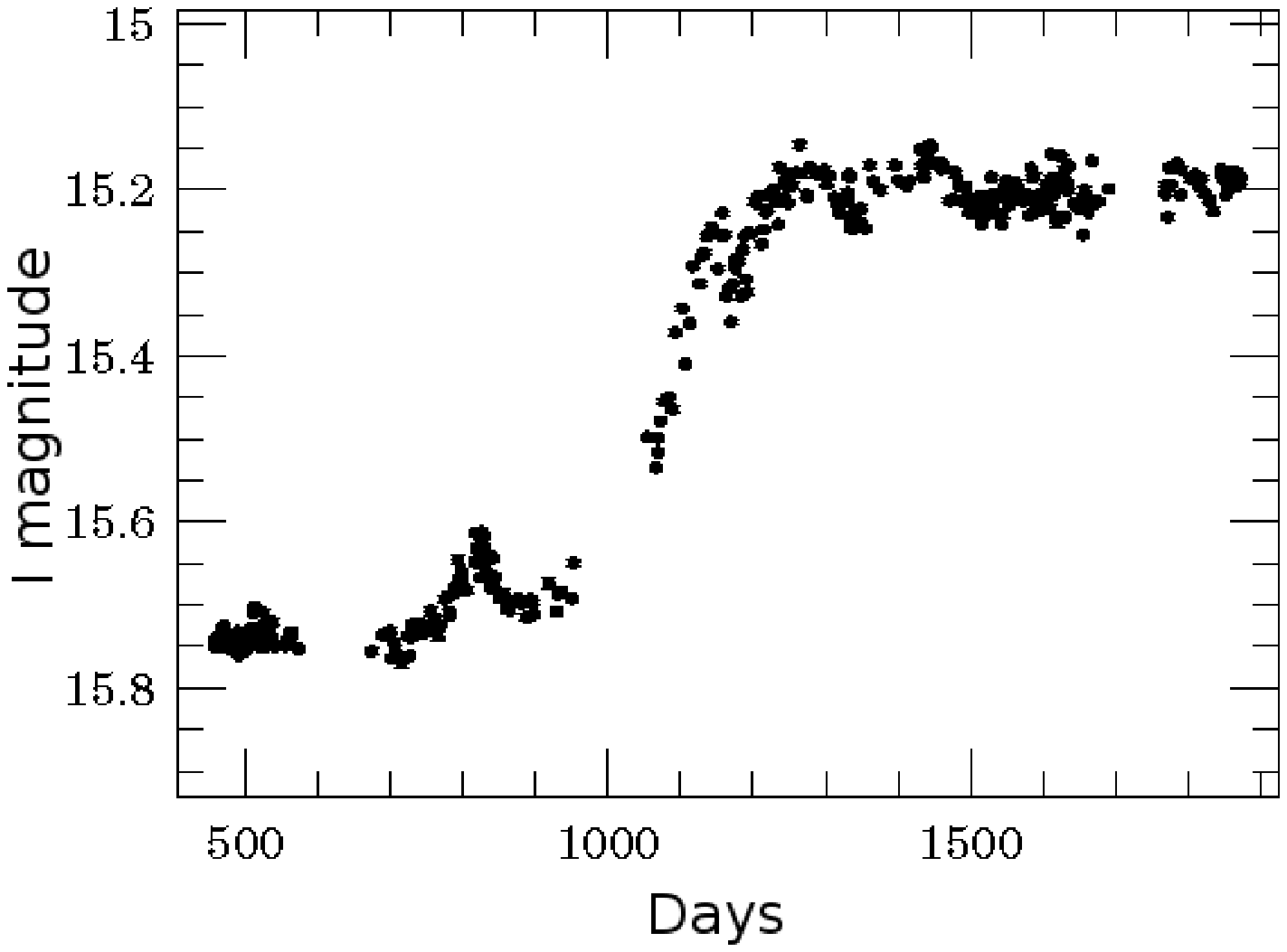}
\end{minipage}
\caption{Light curves for RPs1173 (B1IIIe, H$\alpha$: Single Peak, $\textit{v}$ sin $\textit{i}$: 356\,$\pm$18\,km\,s$^{-1}$) and RPs1382 (B1Ve, H$\alpha$: Double Peak V$>$R, $\textit{v}$ sin $\textit{i}$: 319\,$\pm$16\,km\,s$^{-1}$) from the OGLE-II database. These examples are long period variables which may also include micro features such as that seen at period 1200 days in the right image.}
\label{Figure37}
\vspace{0.2cm}
\begin{minipage}[b]{0.45\linewidth}
\centering
\includegraphics[scale=0.29]{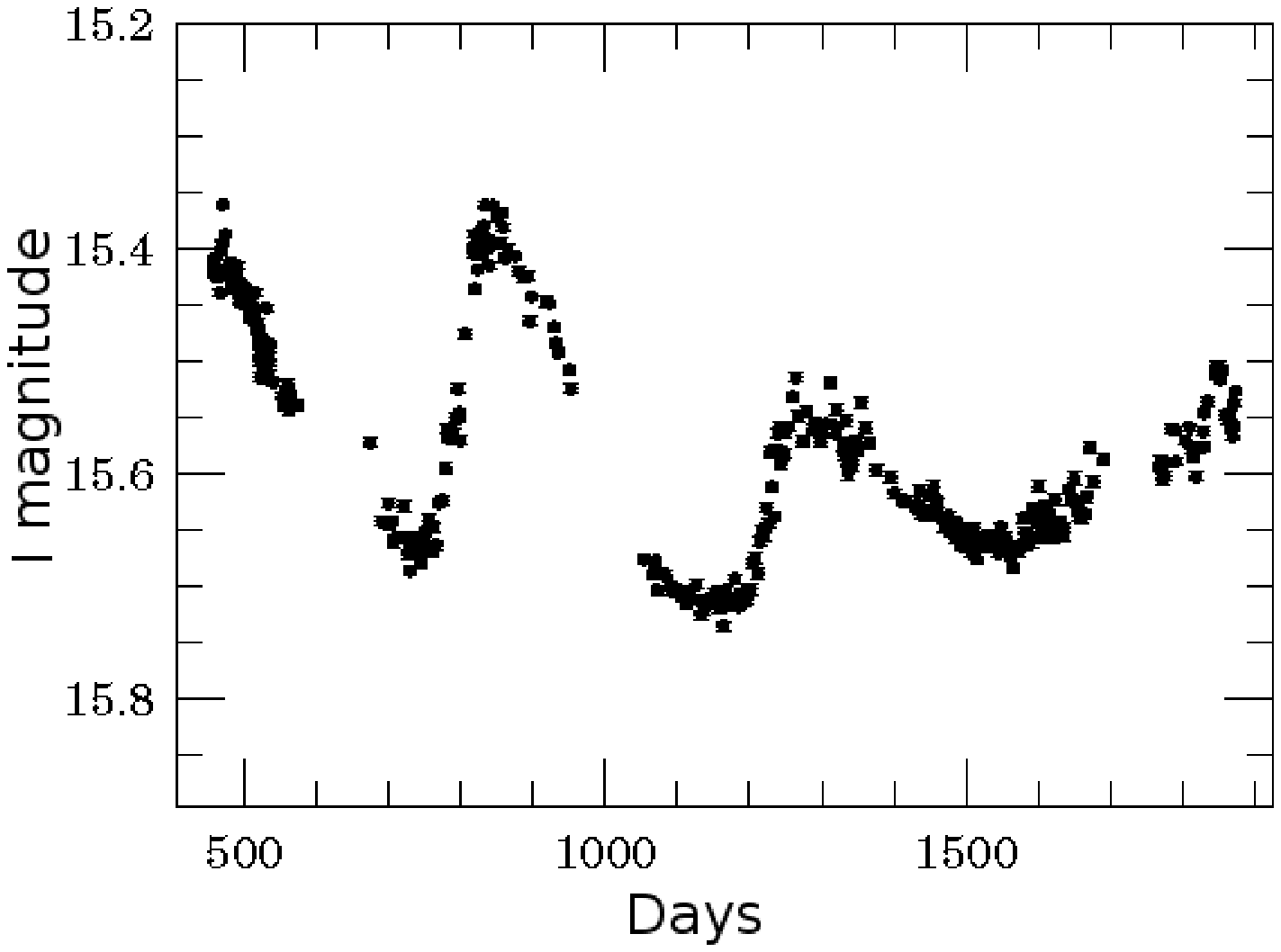}
\end{minipage}
\hspace{0.2cm}
\begin{minipage}[b]{0.45\linewidth}
\centering
\includegraphics[scale=0.29]{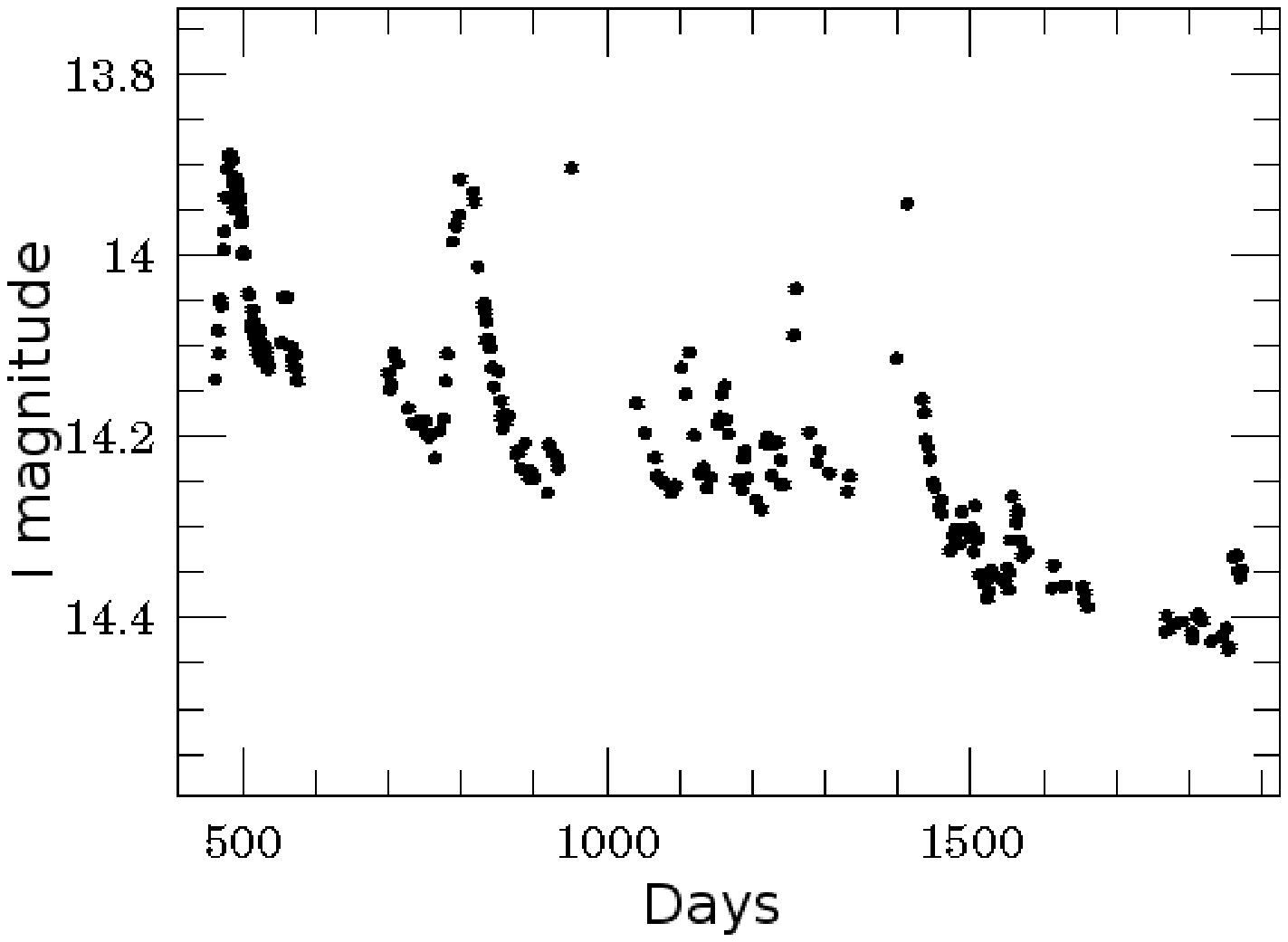}
\end{minipage}
\caption{Light curves for RPs1381 (B1Ve, H$\alpha$: Single Peak, $\textit{v}$ sin $\textit{i}$: 184\,$\pm$9\,km\,s$^{-1}$) and RPs1348 (B1IIIe, H$\alpha$: Single Peak, $\textit{v}$ sin $\textit{i}$: 133\,$\pm$79\,km\,s$^{-1}$) from the OGLE-II database. RPs1381 shows quasi non-regular periods where the brightening can be quite sudden at times. RPs1348 shows semi-regular variability on a broad decline over a long period.}
\label{Figure38}
\end{figure}


 \section{Variability}
The variable nature of Be and B[e] stars is an important feature which relates to the physical stability of the star. As a phenomenon, it has been known for more than a century and may be due to various combinations of physical properties, one or more of which may undergo a transition. Suggested mechanisms are mass loss through stellar winds, rapid rotation and/or non-radial pulsations (see Porter \& Rivinius, 2003). These mechanisms, individually or in combination are usually proposed to explain disc formation and outbursts in Be stars. Sudden brightening and fading episodes are thought to be connected with discrete mass ejection at the surface of Be and B[e] stars. The most notable variations are time dependent variations, known as E/C variations (Hubert-Delplace et al. 1982) where there may be either a change in the emission line intensity or in the continuum level. The latter occasionally cause a veiling effect in the intensity of early-type Be stars.

The jury is still out regarding the mechanism that triggers short-period line profile variability. The possibility of non-radial pulsations has been proposed by several authors (see Rivinius et al. 2003). If the modeling codes (see Townsend, 1997) are observationally confirmed, up to 80\% of early-type Be stars may pulsate in one or more modes. Large numbers of spectroscopic observations to detect multi-periodicity will help to decide this issue. Lastly, and related, are the time variations in the intensities of violet and red components (V/R) as seen in double-peaked emission-line profiles. These probably arise from one-armed density waves in the circumstellar disk.

The classification of Be stars in terms of their light curves was initiated by Mennickent et al. (2002) based on their discovery of 1056 Be star candidates in the Small Magellanic Cloud (SMC) using OGLE II data. Having observed several light curves not seen in Galactic Be stars, they classified four types. Type-1 show outbursts; Type-2 show sudden high and low oscillations; Type-3 show periodic or near periodic variations; Type-4 show the type of light curves seen in Galactic Be stars.

In our data, variability is usually easy to detect by comparing an image taken at a single epoch with our 12 stacked and combined images taken over a 3 year period. While this method strongly indicates variability, the length of the periods can only be measured with repeat exposures over regular intervals.

Although the OGLE photometry currently available does not cover the outer regions of the LMC, there is sufficient data available to find variable stellar candidates within the inner bar region of the LMC. A careful search has found 117 stars with OGLE II data available from our selection of Be stars in the area. Some representative examples of the light curves from these stars are shown in Figures~\ref{Figure35} to~\ref{Figure38}. Periods vary between a few days and over 2,000 days. Magnitude variations range from 0.05 mag (Figure~\ref{Figure35}~left) to 0.6 mag (Figure~\ref{Figure35}~right). Many stars show inconsistent variations in both magnitude and period. For example, Figure~\ref{Figure36}~left shows several 100 day periods followed by a general 0.2 mag increase. The brighter magnitude is accompanied by short bursts which broaden out and increase to 0.3 magnitudes after 900 days. The light curve of RPs1382 (Figure~\ref{Figure37}~right) shows a short 0.15 mag burst after 400 days (from the start of observations)  which precedes and may possibly have pre-cursed a large 0.5 mag increase which was maintained for at least 700 days to the end of the observations. Examples not shown here also include long-term fading episodes such as large decreases over 0.5 in magnitude followed by a wave of short period bursts.

  For our LMC data, the larger magnitude variability ($>$0.2 mag) is mainly confined to early spectral type Be stars (brighter than $\sim$B4). These variations include regular, semi-regular and sporadic outbursts in stars with moderate to high $\textit{v}$ sin $\textit{i}$. Short and mid-term variability on scales of days to weeks can produce amplitudes of up to 0.3 mag. Long-term variability (years to decades) can be accompanied by amplitudes up to 0.8 mag. It is noteworthy that the 2 long period examples with major fading and brightening events shown in Figure~\ref{Figure37} have high rotational velocities which are among the highest 20\% of $\textit{v}$ sin $\textit{i}$ measurements in our sample.

  Figure~\ref{Figure38}~left shows RPs1381, an example of a Be star which features recurrent outbursts (100 days) which slowly decrease over much longer periods, typically about 400 days in this example. Each outburst has a decreasing amplitude which culminates in a gradual brightening after at least 1000 days.

  An interesting phenomenon observed in some variable stars is multiple period oscillations. Low amplitude oscillations are superimposed on much slower and often irregular periods. For example, we find $\sim$100 day luminosity bursts or minor periods occurring as stars experience an overall decrease in magnitude lasting in excess of 2000 days (Figure~\ref{Figure38} right). The variability of emission-line stars is now recognised as one of the main Be star diagnostics; hence further spectroscopic follow-up has been planned.

\label{section12}

\section{The catalogue of emission-line stars}

The full catalogue of emission-line stars uncovered in the stacked H$\alpha$ survey of the central 25deg$^{2}$ of the LMC is provided in the Appendix as Tables 1 and 2. In Table 1 we present our measured and derived data whereas in Table 2 we present a compendium of catalogue identifications and magnitudes where available for each star.

 In the first column of Table 1 we give the Reid \& Parker (RPs) catalogue number where the `s' is added to clearly identify and separate stars from other objects such as PNe, SNRs and \HII~regions which included in the larger RP catalogue. Columns 2 and 3 give the accurate RA and Dec in J2000 world coordinates to 2 decimal places in RA and 1 decimal place in DEC with reference to SuperCOSMOS and verified against 2MASS astrometry. Column 4 provides a published catalogue reference where a star has been previously identified. Please note that this reference does not necessarily indicate that the star was known as an emission-line star. Column 5 shows our estimate of spectral classification and luminosity class as derived from the cross-correlation described in section~\ref{subsection4.1}.

 Column 6 gives the emission flux from H$\alpha$$\lambda$6563 which was measured from our flux-calibrated spectra. All measurements of the H$\alpha$~line including the flux, FWHM (column 7) and the EW (column 8) were measured using the highest quality, medium to high resolution spectra (R$>$1,000) available in our data set. Individual measurements were made using standard IRAF tasks. The heliocentric velocity shown in column 9 represents the measurement which produced the lowest errors, obtained from our high resolution spectra using the {\scriptsize IRAF} ({\scriptsize EMSAO} and {\scriptsize XCSAO}) tasks. Details are provided in section~\ref{section9}. The estimated rotational velocity ($\textit{v}$ sin $\textit{i}$) is shown in column 10 and described in section~\ref{section6}.

 In column 11 we provide comments on each star including a list of which H lines were observed in emission. In the majority of cases, forbidden lines indicate that the star is a B[e] type, however, in several cases, the presence of both low \SII~emission and ambient H$\alpha$ emission (detected in our H$\alpha$ stacked image) suggest that these forbidden lines are not intrinsic to the star itself. In such cases we present the star as a normal Be star but comment on the low \SII~found in the spectrum. Standard abbreviations include `SP' and `DP' indicating an H$\alpha$ line with a single peak or a double peak respectively. The `DP' abbreviation is usually followed by either `V$>$R', `R$>$V', or `centre' indicating the position of the absorption feature (see section~\ref{subsection4.3}).

 The second table presented in the appendix contains a compilation of the B, V, I and R photometry available for each object in this work. In the first three columns we again provide the RP number, RA and DEC of each object to make identification and cross-referencing easier. The fourth column lists the GSC2.2 catalog reference number while column 5 gives the linear distance in arcmins between our position and the GSC2.2 position. Positional errors at the plate epoch are estimated to be in the 200-250 mas range. Column 6 gives the OGLE catalog reference where the matching position has been found in the OGLE II database. Column 7 gives the USNO-A2.0 catalog reference with the linear distance in arcmins between our position and the best matching USNO position shown in column 8. It is estimated that the positional error at plate epoch is near 250 mas. Column9 then provides the position angle (PA) for each object published in the USNO catalog.

 The remaining columns, 10 to 18 provide the magnitudes where available for each star. Since emission-line stars are variable, we should expect a comparison of surveys to reveal minor variations in overall flux estimates. These modulations are confirmed by the data presented here.

 In columns 10 to 12 we present the B magnitudes from OGLE, SuperCosmos (SC) and USNO respectively. Columns 13 and 14 give the V magnitudes from GSC2.2 and OGLE. Columns 15 and 16 give the I magnitudes from SC and OGLE and columns 17 and 18 give the R magnitudes from SC and USNO.

Our online database to be hosted at Macquarie University will contain extra information relating to each star. We will provide optical spectra for each object as well as 1 $\times$ 1 arcmin H$\alpha$/Short Red thumbnail images. At the time of writing, the web site is under construction.

\label{section13}



\section{Summary}

Using our deep H$\alpha$ and SR maps centered on the central 25deg$^{2}$ of the LMC, we have uncovered 1,003 stars which exhibit H$\alpha$ emission. A series of follow-up spectroscopic observations were performed, mostly using 2dF on the AAT during December 2004. The majority of the stars have been assigned a spectral classification using the {\small IRAF} cross-correlation technique and spectral class templates. In addition to the 111 previously known Be stars we have added 468 newly discovered Be, B[e], A and F stars. Most of these stars fall between spectral classes B1 and B3. Analysis of the survey data has also allowed us to identify 315 M (late-type) stars exhibiting chromospheric emission.

For the hot emission-line stars, we provide new, accurate positions, radial and rotational velocities. The distribution of radial velocities has been plotted and compared to the heliocentric distributions for PNe and the \HI~gas disk in the LMC. The good agreement not only indicates that all our emission-line star candidates are located in the LMC, but traces the overall inclination of the LMC's main disk. The distribution of rotational velocities has been plotted and compared to a Galactic sample, revealing a 200-300 km\,s$^{-1}$ agreement in the peak of the distributions. We have also briefly discussed the various H$\alpha$ emission-line profiles identified in our LMC sample.

Emission from the hot B to F stars has been measured and flux calibrated in order to provide the first ever luminosity function for the H$\alpha$ emission from these stars. This included the first ever derived conversion from H$\alpha$ fluxes to magnitudes, with its associated formula, which can be used generally for emission objects in the LMC. The emission has a bright cut-off at magnitude 14.8 (absolute -4.5) and covers a 9 magnitude range. The function shows a steady rise to a distribution peak at 18.5 followed by a decrease over 5 magnitudes to magnitude 23.6. We find a mild correlation between the H$\alpha$ and V magnitudes of the hot stars in our sample. Main sequence stars in our sample are only found at magnitudes below 14.5 in V whereas giants extend to magnitude 12 in V. A compilation of B,V,I and R magnitudes from OGLE II, SuperCOSMOS, ESO GSC 2.2 and USNO are provided in Table 2 of the Appendix.

A plot of the distribution of emission-line stars within the survey area shows that approximately 40\% lie on the main bar. As many as 130 (25\%) of the B-class stars are found to be of the B[e] variety, emitting forbidden lines in \SII, \NII, \OI~and even \OIII. Many of these are located in areas of strong ambient emission or \HII~regions so care is needed to removed these contributions if only emission associated with or emitting from the star is to be measured. These stars with probable contamination have been labeled accordingly within the tables.

\label{section14}

\section{Acknowledgements}
The authors wish to thank the AAO board for observing time on the
AAT and UKST. The authors also thank the European Southern
Observatory for observing time on the VLT and Australian National University along
with their telescope time allocation committees for supporting our
programme of follow-up spectroscopy. We thank the anonymous referee for very helpful comments and suggestions while carefully reviewing the paper. Lastly we would like to thank Suzanne Reid from Kalidus Systems for lending her database design skills to create a repository for the emission-line star data.


\bsp

\label{lastpage}


\begin{thebibliography}{99}

\bibitem[\protect\citeauthoryear{}{}]{}
Acke B., van den Ancker M.E., Dullemond C.P. 2005, A\&A, 436, 209


\bibitem[\protect\citeauthoryear{}{}]{}
Andrews A.D., Lindsay E.M., 1964, IrAJ, 6, 241

\bibitem[\protect\citeauthoryear{}{}]{}
Andrillat Y., \& Houziaux L. 1991, IAUC, 5164, 3

\bibitem[\protect\citeauthoryear{}{}]{}
Beaulieu J-P., et al., 2001, A\&A 380, 168

\bibitem[\protect\citeauthoryear{Bohannan \& Epps}{1974}]{Bohan}
Bohannan B.E., Epps H.W., 1974, A\&AS, 18, 47

\bibitem[\protect\citeauthoryear{}{}]{}
B\"ohm T., \& Catala C. 1994, A\&A, 290, 167

\bibitem[\protect\citeauthoryear{}{}]{}
B\"ohm T., \& Hirth G.A. 1997, A\&A, 324, 177

\bibitem[\protect\citeauthoryear{}{}]{}
Bowyer S., Sasseen T.P., Wu X., Lampton M., 1995, AJ suppl. ser. 96, 461

\bibitem[\protect\citeauthoryear{}{}]{}
Bjorkman J.E., Cassinelli J.P., 1993, ApJ, 409, 429

\bibitem[\protect\citeauthoryear{}{}]{}
Bjorkman K. S., Miroshnichnichenko A.S., McDavid D., Pogrosheva T.M. 2002 ApJ, 573, 812

\bibitem[\protect\citeauthoryear{}{}]{}
Bland-Hawthorn J., Shopbell P.L., Malin D., 1993, AJ, 106, 2154B

\bibitem[\protect\citeauthoryear{}{}]{}
Butler C.J., Wayman P.A., 1974, DunOP, 1, 193


\bibitem[\protect\citeauthoryear{}{}]{}
Cassinelli J.P., Brown J.C., Maheswaran M., Miller N.A., Telfer
D.C., 2002, ApJ, 578, 951

\bibitem[\protect\citeauthoryear{}{}]{}
Cohen M., Kuhi L., 1979, ApJS, 41, 743

\bibitem[\protect\citeauthoryear{}{}]{}
Corcoran M., Ray T.P., 1998 A\&A, 331, 147

\bibitem[\protect\citeauthoryear{}{}]{}
Cranmer S.R., 2005, ApJ, 634, 585


\bibitem[\protect\citeauthoryear{}{}]{}
Dachs J., Hanuschik R.W., Kaiser D., \& Rohe D., 1986, AA., 159, 276

\bibitem[\protect\citeauthoryear{}{}]{}
Dachs J., Kiehling R., Engels D., 1988, A\&A, 194, 167


\bibitem[\protect\citeauthoryear{}{}]{}
Draper H., 1924, Ann, Astron. Obs. Harvard 91-100, 1-6

\bibitem[\protect\citeauthoryear{}{}]{}
Feast M. W.; Thackeray, A. D.; Wesselink, A. J. 1960, MNRAS 121, 337

\bibitem[\protect\citeauthoryear{}{}]{}
Frew D., Parker Q., 2010, PASA, 27, 129

\bibitem[\protect\citeauthoryear{}{}]{}
Gordon K.D., Clayton G.C., Misselt K.A., Landolt A.U., Wolff M.J., 2003, ApJ., 594, 279

\bibitem[\protect\citeauthoryear{}{}]{}
Grebel E.K., 1997, A\&A, 317, 448

\bibitem[\protect\citeauthoryear{}{}]{}
Grebel E.K., Chu Y., 2000, AJ, 199, 787

\bibitem[\protect\citeauthoryear{}{}]{}
Gullbring E., Hartmann L., Briceno C., Calvet N., 1998, ApJ., 492, 323

\bibitem[\protect\citeauthoryear{}{}]{}
Hamann F., 1994, ApJS, 93, 485

\bibitem[\protect\citeauthoryear{}{}]{}
Hambly N. C., et al., 2001, MNRAS, 326, 1279H

\bibitem[\protect\citeauthoryear{}{}]{}
Hanuschik R.W., 1989, Ap\&SS, 161, 61

\bibitem[\protect\citeauthoryear{}{}]{}
Hanuschik R.W., Hummel W., Sutorius E., Dietle O., Thimm G., 1996, A\&AS, 116, 309

\bibitem[\protect\citeauthoryear{}{}]{}
Hartigan P., Edwards S., Ghandour L., 1995, ApJ., 452, 736

\bibitem[\protect\citeauthoryear{}{}]{}
Hartmann L., Hewett R., Calvet N., 1994, ApJ, 426, 669

\bibitem[\protect\citeauthoryear{}{}]{}
Henyey L.G., Lelevier R., Lev\'{e}e R.D., 1955, PASP 67, 154

\bibitem[\protect\citeauthoryear{}{}]{}
Henize K.G., 1956, ApJS, 2, 315

\bibitem[\protect\citeauthoryear{}{}]{}
Hubert-Delplace A. M., Hubert H., Chambon M. T., Jaschek M. 1982 IAUS: Be stars; Proceedings of the Symposium, Munich, West Germany, 1981, Dordrecht, D. Reidel, 98, 125

\bibitem[\protect\citeauthoryear{}{}]{}
Hubert A.M., Floquet M., 1998, A\&A, 335, 565

\bibitem[\protect\citeauthoryear{}{}]{}
Herbig G.H., 1960, ApJS, 4, 337

\bibitem[\protect\citeauthoryear{}{}]{}
Hern\'{a}ndez J., Calvet N., Brice\~{n}o C., Hartmann L., Berlind
P., 2004, AJ127, 1682

\bibitem[\protect\citeauthoryear{}{}]{}
Hillenbrand L.A., Strom S.E., Vrba F.J., Keene J., 1992, ApJ., 397, 613


\bibitem[\protect\citeauthoryear{}{}]{}
Hubrig S., Sch\"{o}ller M., Yudin R.V. 2004, A\&A, 428, L1

\bibitem[\protect\citeauthoryear{}{}]{} 
Hughes S.M.G., 1989, 97, 1634

\bibitem[\protect\citeauthoryear{}{}]{}
Hummel W., Szeifert T., G\"{a}ssler W., Muschielok B., Seifert W., Appenzeller I., Rupprecht G., 1999 A\&A 352, L31

\bibitem[\protect\citeauthoryear{}{}]{}
Hutsem\'ekers D. 1985, A\&AS, 60, 373


\bibitem[\protect\citeauthoryear{}{}]{}
Jaschek M., Jaschek C., 1967, ApJ., 150, 355

\bibitem[\protect\citeauthoryear{}{}]{}
Jacoby G.H., Hunter, D.A., Christian C.A., 1984, ApJS, 56, 257

\bibitem[\protect\citeauthoryear{}{}]{}
Jacoby G.H., 1989. Planetary nebulae as standard candles. I. Evolutionary models. ApJ, 339, 39

\bibitem[\protect\citeauthoryear{}{}]{}
Jaschek M., Jaschek C., 1967, ApJ., 150, 355

\bibitem[\protect\citeauthoryear{}{}]{}
Jaschek M., Slettebak A., Jaschek C., 1981 Be Star Newsl., 4, 9

\bibitem[\protect\citeauthoryear{}{}]{}
Kaler J.B., 1997, Stars and their spectra, Univ. press Cambridge.
ISBN 0 521585708 p 192

\bibitem[\protect\citeauthoryear{}{}]{}
Keller S.C., Wood P.R., Bessell M.S., 1999, A\&AS, 134, 489

\bibitem[\protect\citeauthoryear{}{}]{}
Keller S.C., Bessell M.S., Da Costa G.S., 2000, AJ, 119, 1748

\bibitem[\protect\citeauthoryear{}{}]{}
Keller S.C., Bessell M.S., Cook K. H., Geha M., Syphers D., 2002, AJ, 124, 2039

\bibitem[\protect\citeauthoryear{}{}]{}
K\"{o}nigl A., 1991, ApJ, 370, L39

\bibitem[\protect\citeauthoryear{}{}]{}
Kontizas E., Dapergolas A., Morgan D. H., Kontizas M., 2001, A\&A, 369, 932

\bibitem[\protect\citeauthoryear{}{}]{}
Kurtz M.J., Mink D.J., 1998, PASP, 110, 934



\bibitem[\protect\citeauthoryear{}{}]{}
Lada C.J., Adams F.C., 1992, ApJ., 393, 278

\bibitem[\protect\citeauthoryear{}{}]{}
Larsen S.S., Clausen J.V., Storm J., 2000, A\&A, 364, 455


\bibitem[\protect\citeauthoryear{}{}]{}
Le Borgne J.-F., Bruzual G., Pell\'{o} R., Lan\c{c}on A.,
Rocca-Volmerange B., Sanahuja B., Schaerer D., Soubiran C.,
V\'{i}lchez-G\'{o}mez R., 2003, A\&A 402, 433

\bibitem[\protect\citeauthoryear{}{}]{}
Lewis I.J., et al., 2002, MNRAS, 333, 279

\bibitem[\protect\citeauthoryear{}{}]{}
Lindsay E.M., 1963, IrAJ, 6, 127

\bibitem[\protect\citeauthoryear{}{}]{}
Lindsay E.M., 1974, MNRAS 166, 703

\bibitem[\protect\citeauthoryear{}{}]{}
Lumb D.H., Guainazzi M., Gondoin P., 2001, A\&A, 376, 387


\bibitem[\protect\citeauthoryear{}{}]{}
Melchior A.-L., Hughes S. M. G., Guibert J., 2000, A\&AS, 145, 11

\bibitem[\protect\citeauthoryear{}{}]{}
Mennickent R.E., Pietrzy\'{n}ski G., Gieren W., Szewczyk O, 2002, A\&A, 393, 887

\bibitem[\protect\citeauthoryear{}{}]{}
Miroshnichenko A.S., Fabregat J., Bjorkman K.S., Knauth D.C., Morrison N.D., Tarasov A.E., Reig P., Negueruela I., Blay P., 2001, A\&A, 371, 600

\bibitem[\protect\citeauthoryear{}{}]{}
Morgan W. W., Keenan P.C., Kellman E. 1943, ``An atlas of stellar spectra, with an outline of spectral classification", Chicago, Ill., The University of Chicago press

\bibitem[\protect\citeauthoryear{}{}]{}
Murai T., Fujimoto M., 1980, PASJ, 32, 581

\bibitem[\protect\citeauthoryear{}{}]{}
Muzerolle J., Calvet N., Hartmann L., 1998, ApJ, 492, 743

\bibitem[\protect\citeauthoryear{}{}]{}
Muzerolle J., Calvet N., Hartmann L., 2001, ApJ, 550, 944

\bibitem[\protect\citeauthoryear{}{}]{}
Muzerolle J., D'Alessio P., Calvet N., Hartmann L., 2004, ApJ, 617,
406

\bibitem[\protect\citeauthoryear{}{}]{}
Olsen K.A.G., Kim S., Buss J.F., 2001 AJ, 121, 3075

\bibitem[\protect\citeauthoryear{}{}]{}
Parker Q.A., Bland-Hawthorn J., 1998, PASA, 15, 33p

\bibitem[\protect\citeauthoryear{}{}]{}
Parker Q.A., Malin D., 1999, PASA, 16, 288P


\bibitem[\protect\citeauthoryear{}{}]{}
Parker Q.A., et al., 2005 MNRAS 362, 689

\bibitem[\protect\citeauthoryear{}{}]{}
Pasquini L., et al., 2002, Msngr. 110, 1

\bibitem[\protect\citeauthoryear{}{}]{}
Poeckert R., Marlborough J.M., 1978, A.J.suppl. 38, 229

\bibitem[\protect\citeauthoryear{}{}]{}
Porter J.M., 1996, MNRAS, 280, 31

\bibitem[\protect\citeauthoryear{}{}]{}
Porter J.M., Rivinius T., 2003 PASP, 115, 1153

\bibitem[\protect\citeauthoryear{}{}]{}
Pickles A.J., 1998, PASP, 110, 863

\bibitem[\protect\citeauthoryear{}{}]{}
Quirrenbach A., Bjorkman K.S., Bjorkman J.E., Hummel C.A., Buscher D.F., Armstrong J.T., Mozurkewich D., Elias N.M.II, Babler B.L., 1997, ApJ, 479, 477

\bibitem[\protect\citeauthoryear{}{}]{}
Reid N., Glass I.S., Catchpole R.M., 1988, MNRAS, 232, 53

\bibitem[\protect\citeauthoryear{}{}]{}
Reid W.A., Parker Q.A. 2005, AIPC, 804, 12

\bibitem[\protect\citeauthoryear{}{}]{}
Reid W.A., Parker Q.A. 2006a, MNRAS, 365, 401

\bibitem[\protect\citeauthoryear{}{}]{}
Reid W.A., Parker Q.A. 2006b, MNRAS, 373, 521

\bibitem[\protect\citeauthoryear{}{}]{}
Reid W.A., Parker Q.A. 2010, MNRAS, 405, 1349

\bibitem[\protect\citeauthoryear{}{}]{}
Rivinius Th., Baade D., \v{S}tefl S., Townsend R.H.D., Stahl O., Wolf
B., Kaufer A., 2001, A\&A, 369, 1058

\bibitem[\protect\citeauthoryear{}{}]{}
Rivinius Th., Baade D., \v{S}tefl S., 2003, A\&A, 411, 229

\bibitem[\protect\citeauthoryear{}{}]{}
Rohlfs K., Kreitschmann J., Feitzinger J.V., Siegman B.C., 1984, A\&A, 137, 343


\bibitem[\protect\citeauthoryear{}{}]{}
Sabogal B.E., Mennickent R.E., Pietrzy$\acute{n}$ski G., Gieren W., 2005, MNRAS, 361, 1055


\bibitem[\protect\citeauthoryear{}{}]{}
Schwering P.B.W., 1989, A\&A Suppl. Ser. 79, 105

\bibitem[\protect\citeauthoryear{}{}]{}
Silaj J., Jones C.E., Tycner C., Sigut T.A.A., Smith A.D., 2010, ApJS, 187, 228


\bibitem[\protect\citeauthoryear{}{}]{}
Silva D.R., Cornell M.E., 1992, ApJS, 81, 865

\bibitem[\protect\citeauthoryear{}{}]{}
Skrutskie M.F., et al., 2006, AJ., 131, 1163

\bibitem[\protect\citeauthoryear{}{}]{}
Slettebak A., 1982, AJ. Supp. Series, 50, 55

\bibitem[\protect\citeauthoryear{}{}]{}
Stanghellini L., Shaw R.A., Mutchler M., Palen S., Balick B., Blades J.C., 2002, ApJ, 575, 178

\bibitem[\protect\citeauthoryear{}{}]{}
Struve O., 1931, ApJ, 73, 94


\bibitem[\protect\citeauthoryear{}{}]{}
Szyma\'nski M.K., 2005, Acta Astron, 55, 43

\bibitem[\protect\citeauthoryear{}{}]{}
Telting J.H., 2000 in ASP Conf. Proc. 214, The Be Phenomenon in Early-Type Stars, ed. M.A. Smith \& H.F. Henrichs (San Francisco: ASP), 422

\bibitem[\protect\citeauthoryear{}{}]{}
Tonry J.L., Davis M., 1979, AJ. 43, 393

\bibitem[\protect\citeauthoryear{}{}]{}
Townsend R.H.D., 1997, MNRAS, 284, 839

\bibitem[\protect\citeauthoryear{}{}]{}
Townsend R.H.D., Owocki S.P., Howarth I.D., 2004, MNRAS, 350, 189

\bibitem[\protect\citeauthoryear{}{}]{}
Turnshek D.E., Turnshek D.A., Craine E.R., Boeshaar, P.C., 1985,
$\textit{An atlas of digital spectra of cool stars}$ Astronomy and
Astrophysics Series, v.1, Tucson: Western Research Company, ISBN 0-934525-00-5

\bibitem[\protect\citeauthoryear{}{}]{}
Tycner C., et al., 2005, ApJ, 624, 359

\bibitem[\protect\citeauthoryear{}{}]{}
Uchida Y., Shibata K., 1985, PASJ, 37, 515

\bibitem[\protect\citeauthoryear{}{}]{}
Udalski A., Szyma\'nski M., Kubiak M., Pietrzy\'nski G., Soszy\'nski I., Wo\'zniak P., and Zebru\'n K., 2000, Acta Astron, 50, 307

\bibitem[\protect\citeauthoryear{}{}]{}
Underhill A., Doazan V., 1982,``B STARS with and without emission" NASA SP-456, Washington DC.: NASA


\bibitem[\protect\citeauthoryear{}{}]{}
Van Winckel H., 2003, ARA\&A, 41, 391

\bibitem[\protect\citeauthoryear{}{}]{}
Viera S.L.A., Corradi W.J.B., Alencar S.H.P., Mendes L.T.S., Torres C.A.O., Quast G.R., Guimar\~{a}es M.M., da Silva L. 2003, AJ, 126, 2971

\bibitem[\protect\citeauthoryear{}{}]{}
Waters L.B.F.M., Waelkens C., 1998, ARA\&A, 36, 233

\bibitem[\protect\citeauthoryear{}{}]{}
Wade G.A. et al., 2005, A\&A, 442, L31

\bibitem[\protect\citeauthoryear{}{}]{}
Wisniewski J.P., Bjorkman K.S., 2006 ApJ, 652, 458


\end{thebibliography}
\end{document}



\pagerange{\pageref{firstpage}--\pageref{lastpage}} \pubyear{2011}


\label{firstpage}
\setcounter{table}{1}
\begin{landscape}
\begin{table*}
\begin{minipage}{240mm}
\section{Appendix: Table 1}
{Basic data for all hot emission-line stars in the LMC UKST H$\alpha$ survey. Please refer to section 15 for more detailed information. Column 1 gives the Reid Parker (RPs) number for the star (s). Columns 2 and 3 give the RA and Dec in J2000 coordinates. Column 4 gives the published catalogue and number where a star has been previously identified. Column 5 gives our estimate of spectral classification and luminosity class. Column 6 gives a log of the H$\alpha$ emission flux obtained from our flux calibrated spectra. Column 7 gives the Full Width at Half Maximum (FWHM) in km/s as measured from our data. Column 8 gives the Equivalent Width (EW) of the H$\alpha$ line in Angstroms. Column 9 gives the heliocentric radial velocity and column 10 gives the rotational velocity ($\textit{v}$ sin $\textit{i}$). In column 11 we provide comments where em=emission, A.Em=ambient emission, SP=single peak, DP=double peak, sh=shell. Notes on column 4 are given at the end of the table.}
\centering
\scriptsize
\begin{tabular}{|l|l|l|l|l|c|c|c|c|c|l|}
\hline
RP & RA   &   DEC &  Other &  Spec  &  Log F   &  FWHM   &   EW   &  Vel. (helio)  &  $\textit{v}$ sin $\textit{i}$  &   Comments   \\
Object  &  J2000  &  J2000 & Catalog &  Type &  H$\alpha$ &  H$\alpha$  & H$\alpha$  &   km/s   &   km/s    &                \\
         &         &       &  Reference        &    &   &  km/s   &   \AA    &      &    &            \\
\hline\hline
RPs1741	&	04 52 55.41	&	-70 42 12.3	&		&	B7Ve	&	-13.92	&	464	$\pm$	23	&	70	$\pm$	3	&	326	 $\pm$	6	 &	 449	 $\pm$	 22	&	 H$\alpha$,$\beta$	\\
RPs1647	&	04 53 01.03	&	-70 42 12.3	&		&	B3Ve	&	-13.97	&	342	$\pm$	17	&	18	$\pm$	1	&	241	 $\pm$	1	 &	 243	 $\pm$	 12	&	 H$\alpha$, \SII~from faint, diffuse local emission	\\
RPs1701	&	04 53 02.01	&	-69 47 04.1	&		&	B2Ve	&	-13.73	&	333	$\pm$	17	&	55	$\pm$	2	&	236	 $\pm$	5	 &	 296	 $\pm$	 15	&	 H$\alpha$,$\beta$	\\
RPs1699	&	04 53 06.86	&	-69 57 24.7	&		&	B1.7V[e]	&	-13.33	&	56	$\pm$	3	&	59	$\pm$	2	&	 242	$\pm$	 1	&	 22	 $\pm$	1	&	 H$\alpha$,$\beta$,$\gamma$, \SII, \NII, \OII	\\
RPs1671	&	04 53 25.43	&	-70 35 39.7	&		&	B0III[e]	&	-12.78	&	247	$\pm$	12	&	19	$\pm$	1	&	 262	$\pm$	 1	&	 165	 $\pm$	8	&	 H$\alpha$,$\beta$,$\gamma$, \SII, \NII, \OIII	\\
\hline
\end{tabular}
\end{minipage}
\end{table*}
\end{landscape}
\label{secondpage}
\setcounter{table}{2}
\begin{landscape}
\begin{table*}
\begin{minipage}{240mm}
\centering
\scriptsize
\begin{tabular}{|l|l|l|l|l|c|c|c|c|c|l|}
\hline
RP & RA   &   DEC &  Other &  Spec  &  Log F   &  FWHM   &   EW   &  Vel. (helio)  &  $\textit{v}$ sin $\textit{i}$  &   Comments   \\
Object  &  J2000  &  J2000 & Catalog &  Type &  H$\alpha$ &  H$\alpha$  & H$\alpha$  &   km/s   &   km/s    &                \\
  &         &       &  Reference        &    &   &  km/s   &   \AA    &      &    &            \\
\hline\hline
RPs1656	&	04 59 46.62	&	-70 25 27.7	&		&	B1Ve	&	-13.74	&	509	$\pm$	25	&	60	$\pm$	2	&	278	 $\pm$	9	 &	 483	 $\pm$	 24	&	 H$\alpha$ only, Pcyg on H$\beta$	\\
RPs1698	&	04 59 48.42	&	-69 54 13.0	&	L6370	&	B1Ve	&	-12.99	&	290	$\pm$	14	&	78	$\pm$	3	&	 226	$\pm$	 4	&	 276	 $\pm$	14	&	 H$\alpha$,$\beta$	\\
RPs1856	&	04 59 54.75	&	-68 23 07.4	&		&	B0.5IIIe	&	-13.16	&	410	$\pm$	20	&	50	$\pm$	2	&	 289	$\pm$	 6	&	 373	 $\pm$	19	&	 H$\alpha$,$\beta$,$\gamma$	\\
RPs1777	&	05 00 13.12	&	-70 02 56.5	&	SVDV 284	&	B1.5Ve	&	-13.56	&	291	$\pm$	15	&	79	$\pm$	3	&	 246	 $\pm$	4	&	 277	 $\pm$	14	&	 H$\alpha$,$\beta$	\\
RPs1722	&	05 00 14.06	&	-69 25 13.9	&	SAB2338	&	B2Ve	&	-13.01	&	309	$\pm$	15	&	61	$\pm$	2	&	 250	$\pm$	 4	&	 280	 $\pm$	14	&	 H$\alpha$,$\beta$	\\
\hline
\end{tabular}
\end{minipage}
\end{table*}
\end{landscape}
\label{secondpage}
\setcounter{table}{2}
\begin{landscape}
\begin{table*}
\begin{minipage}{240mm}
\centering
\scriptsize
\begin{tabular}{|l|l|l|l|l|c|c|c|c|c|l|}
\hline
RP & RA   &   DEC &  Other &  Spec  &  Log F   &  FWHM   &   EW   &  Vel. (helio)  &  $\textit{v}$ sin $\textit{i}$  &   Comments   \\
Object  &  J2000  &  J2000 & Catalog &  Type &  H$\alpha$ &  H$\alpha$  & H$\alpha$  &   km/s   &   km/s    &                \\
  &         &       &  Reference        &    &   &  km/s   &   \AA    &      &    &            \\
\hline\hline
RPs1822	&	05 06 38.01	&	-68 44 41.5	&		&	B1.7V[e]	&	-13.71	&	167	$\pm$	8	&	27	$\pm$	1	&	 272	 $\pm$	1	 &	 111	 $\pm$	6	&	 H$\alpha$,$\beta$,$\gamma$, \SII, \NII, \OII, \OIII. Dense halo surrounding 	 \\
	&		&		&		&		&		&				&				&				&				&	star + diffuse emission 10 arcsec radius surrounding assoc. of stars.	\\
RPs1822	&	05 06 38.01	&	-68 44 41.5	&		&	B1.7V[e]	&	-13.71	&	167	$\pm$	8	&	27	$\pm$	1	&	 272	$\pm$	 1	&	 111	 $\pm$	6	&	 H$\alpha$,$\beta$,$\gamma$, \SII, \NII, \OII, \OIII. Dense halo surrounding 	 \\
	&		&		&		&		&		&				&				&				&				&	star + diffuse emission 10 arcsec radius surrounding assoc. of stars.	\\
RPs1754	&	05 06 39.31	&	-69 01 22.5	&		&	B1Ve	&	-13.54	&	457	$\pm$	23	&	48	$\pm$	2	&	268	 $\pm$	3	 &	 408	 $\pm$	 20	&	 H$\alpha$,$\beta$, \SII~only 3 sigma noise	\\
\hline
\end{tabular}
\end{minipage}
\end{table*}
\end{landscape}
\label{secondpage}
\setcounter{table}{2}
\begin{landscape}
\begin{table*}
\begin{minipage}{240mm}
\centering
\scriptsize
\begin{tabular}{|l|l|l|l|l|c|c|c|c|c|l|}
\hline
RP & RA   &   DEC &  Other &  Spec  &  Log F   &  FWHM   &   EW   &  Vel. (helio)  &  $\textit{v}$ sin $\textit{i}$  &   Comments   \\
Object  &  J2000  &  J2000 & Catalog &  Type &  H$\alpha$ &  H$\alpha$  & H$\alpha$  &   km/s   &   km/s    &                \\
  &         &       &  Reference        &    &   &  km/s   &   \AA    &      &    &            \\
\hline\hline
RPs1260	&	05 15 10.00	&	-70 01 21.6	&		&	B3III[e]	&	-14.07	&	52	$\pm$	3	&	61	$\pm$	2	&	 241	$\pm$	 1	&	 18	 $\pm$	1	&	 H$\alpha$,$\beta$, $\gamma$, \SII, \NII, \OII	\\
RPs1524	&	05 15 41.41	&	-67 58 52.4	&	BE74231	&	B4III[e]	&	-12.29	&	412	$\pm$	21	&	18	$\pm$	1	&	 316	 $\pm$	1	 &	302	 $\pm$	15	&	 H$\alpha$,$\beta$,$\gamma$, \SII, \NII, \OII, dense \HII~halo surrounding, SP	 \\
RPs1582	&	05 15 45.61	&	-66 58 42.3	&		&	B1Ve	&	-14.99	&	42	$\pm$	2	&	55	$\pm$	2	&	291	 $\pm$	1	 &	 7	$\pm$	 0	&	H$\alpha$, p cyg on H$\beta$	\\
RPs1469	&	05 15 48.74	&	-68 33 04.0	&		&	B1.7Ve	&	-13.92	&	98	$\pm$	5	&	27	$\pm$	1	&	272	 $\pm$	3	 &	 49	$\pm$	 2	&	H$\alpha$ only	\\
RPs1414	&	05 16 24.57	&	-68 55 27.6	&		&	B3V[e]	&	-12.97	&	302	$\pm$	15	&	186	$\pm$	7	&	232	 $\pm$	1	 &	 349	 $\pm$	 18	&	 H$\alpha$,$\beta$,$\gamma$, \SII, \NII, \OII, Poor spectral fit	\\
\hline
\end{tabular}
\end{minipage}
\end{table*}
\end{landscape}
\label{secondpage}
\setcounter{table}{2}
\begin{landscape}
\begin{table*}
\begin{minipage}{240mm}
\centering
\scriptsize
\begin{tabular}{|l|l|l|l|l|c|c|c|c|c|l|}
\hline
RP & RA   &   DEC &  Other &  Spec  &  Log F   &  FWHM   &   EW   &  Vel. (helio)  &  $\textit{v}$ sin $\textit{i}$  &   Comments   \\
Object  &  J2000  &  J2000 & Catalog &  Type &  H$\alpha$ &  H$\alpha$  & H$\alpha$  &   km/s   &   km/s    &                \\
  &         &       &  Reference        &    &   &  km/s   &   \AA    &      &    &            \\
\hline\hline
RPs1360	&	05 21 17.30	&	-69 19 53.9	&	BE74257	&	B5IIIe	&	-12.73	&	262	$\pm$	13	&	38	$\pm$	2	&	 251	$\pm$	 4	&	 213	 $\pm$	11	&	 H$\alpha$,$\beta$	\\
RPs1542	&	05 21 20.43	&	-67 47 06.8	&	IRAS-P.C05214	&	B2IV[e]	&	-13.46	&	49	$\pm$	2	&	233	$\pm$	9	 &	 279	 $\pm$	1	 &	36	$\pm$	2	 &	H$\alpha$,$\beta$, \NII, \SII, \OII, SP	\\
	&		&		&	-6749	&		&		&				&				&				&				&		 \\
RPs1266	&	05 21 21.48	&	-69 59 00.2	&		&	A2Ve	&	-14.02	&	320	$\pm$	16	&	30	$\pm$	1	&	300	 $\pm$	1	 &	 246	 $\pm$	 12	&	 H$\alpha$,$\beta$, similar size star 3 arcsec to SW.	\\
RPs1362	&	05 21 29.45	&	-69 07 25.4	&		&	B1Ve	&	-14.04	&	256	$\pm$	13	&	41	$\pm$	2	&	285	 $\pm$	1	 &	 207	 $\pm$	 10	&	 H$\alpha$,$\beta$	\\
\hline
\end{tabular}
\end{minipage}
\end{table*}
\end{landscape}
\label{secondpage}
\setcounter{table}{2}
\begin{landscape}
\begin{table*}
\begin{minipage}{240mm}
\centering
\scriptsize
\begin{tabular}{|l|l|l|l|l|c|c|c|c|c|l|}
\hline
RP & RA   &   DEC &  Other &  Spec  &  Log F   &  FWHM   &   EW   &  Vel. (helio)  &  $\textit{v}$ sin $\textit{i}$  &   Comments   \\
Object  &  J2000  &  J2000 & Catalog &  Type &  H$\alpha$ &  H$\alpha$  & H$\alpha$  &   km/s   &   km/s    &                \\
  &         &       &  Reference        &    &   &  km/s   &   \AA    &      &    &            \\
\hline\hline
RPs778	&	05 24 14.14	&	-69 27 40.3	&		&	B1III[e]	&	-13.41	&	101	$\pm$	5	&	9	$\pm$	0	&	 277	$\pm$	 1	&	 31	 $\pm$	2	&	 H$\alpha$,$\beta$, \SII \& some associated \NII emission, SP	\\
RPs601	&	05 24 17.73	&	-71 31 50.2	&	BE74580	&	B3V[e]	&	-12.45	&	96	$\pm$	5	&	43	$\pm$	2	&	 238	$\pm$	 1	&	 55	 $\pm$	3	&	 H$\alpha$,$\beta$,$\gamma$,$\delta$, \OII~3727, \NII, \SII,	\\
	&		&		&		&		&		&				&				&				&				&	 relatively low continuum levels suggest active \HII~region	\\
RPs873	&	05 24 23.07	&	-69 39 08.9	&	BE74579	&	B3V[e]	&	-13.04	&	88	$\pm$	4	&	62	$\pm$	2	&	 272	$\pm$	 1	&	 55	 $\pm$	3	&	 H$\alpha$,$\beta$, \SII, \OII, \NII	\\
RPs874	&	05 24 23.70	&	-69 38 53.6	&	BE74579	&	B3V[e]	&	-13.31	&	338	$\pm$	17	&	50	$\pm$	2	&	 246	$\pm$	 25	&	 301	 $\pm$	15	&	 H$\alpha$,$\beta$, \SII, \OII	\\
\hline
\end{tabular}
\end{minipage}
\end{table*}
\end{landscape}
\label{secondpage}
\setcounter{table}{2}
\begin{landscape}
\begin{table*}
\begin{minipage}{240mm}
\centering
\scriptsize
\begin{tabular}{|l|l|l|l|l|c|c|c|c|c|l|}
\hline
RP & RA   &   DEC &  Other &  Spec  &  Log F   &  FWHM   &   EW   &  Vel. (helio)  &  $\textit{v}$ sin $\textit{i}$  &   Comments   \\
Object  &  J2000  &  J2000 & Catalog &  Type &  H$\alpha$ &  H$\alpha$  & H$\alpha$  &   km/s   &   km/s    &                \\
  &         &       &  Reference        &    &   &  km/s   &   \AA    &      &    &            \\
\hline\hline
RPs605	&	05 27 49.37	&	-71 31 19.8	&		&	A1V[e]	&	-13.37	&	251	$\pm$	13	&	60	$\pm$	2	&	247	 $\pm$	2	 &	 221	 $\pm$	 11	&	 H$\alpha$,$\beta$,$\gamma$,$\delta$, \OII~3727, \NII, \SII, \OIII~5007	\\
RPs1057	&	05 28 02.72	&	-67 31 16.5	&	AL 207	&	B2IIIe	&	-12.18	&	256	$\pm$	13	&	72	$\pm$	3	&	 324	 $\pm$	4	 &	 239	 $\pm$	12	&	 H$\alpha$,$\beta$ some diffuse emission in area.	\\
RPs840	&	05 28 03.99	&	-69 09 57.2	&		&	B4IVe	&	-12.96	&	453	$\pm$	23	&	70	$\pm$	3	&	248	 $\pm$	17	 &	 438	 $\pm$	 22	&	 H$\alpha$,$\beta$, SP	\\
RPs2185	&	05 28 04.23	&	-69 05 26.8	&	SVHV 2565	&	B0Ve	&	-13.89	&	182	$\pm$	9	&	19	$\pm$	1	&	 294	 $\pm$	4	&	 111	 $\pm$	6	&	 H$\alpha$,$\beta$, DP centre	\\
RPs823	&	05 28 04.41	&	-69 13 33.0	&		&	B1.7Ve	&	-13.17	&	430	$\pm$	22	&	47	$\pm$	2	&	241	 $\pm$	11	 &	 382	 $\pm$	 19	&	 H$\alpha$,$\beta$, SP	\\
\hline
\end{tabular}
\end{minipage}
\end{table*}
\end{landscape}
\label{secondpage}
\setcounter{table}{2}
\begin{landscape}
\begin{table*}
\begin{minipage}{240mm}
\centering
\scriptsize
\begin{tabular}{|l|l|l|l|l|c|c|c|c|c|l|}
\hline
RP & RA   &   DEC &  Other &  Spec  &  Log F   &  FWHM   &   EW   &  Vel. (helio)  &  $\textit{v}$ sin $\textit{i}$  &   Comments   \\
Object  &  J2000  &  J2000 & Catalog &  Type &  H$\alpha$ &  H$\alpha$  & H$\alpha$  &   km/s   &   km/s    &                \\
  &         &       &  Reference        &    &   &  km/s   &   \AA    &      &    &            \\
\hline\hline
RPs992	&	05 31 21.06	&	-68 31 34.3	&		&	B2V[e]	&	-12.13	&	32	$\pm$	2	&	108	$\pm$	4	&	302	 $\pm$	1	 &	5	 $\pm$	 0	&	 H$\alpha$,$\beta$,$\gamma$, \SII, \NII, \OII, \OIII~within compact dense \HII~cloud. SP	\\
RPs915	&	05 31 22.11	&	-68 36 42.0	&		&	B2V[e]	&	-13.24	&	82	$\pm$	4	&	57	$\pm$	2	&	275	 $\pm$	2	 &	 49	$\pm$	 3	&	 H$\alpha$,$\beta$, \SII, \OIII, \OII~in cocoon, SP	\\
RPs2208	&	05 31 24.25	&	-68 41 33.6	&	BE74334	&		&	-11.81	&	374	$\pm$	19	&	34	$\pm$	1	&	278	 $\pm$	19	 &	 310	 $\pm$	 16	&	 H$\alpha$,$\beta$, Late-type star characteristics	\\
RPs832	&	05 31 30.75	&	-69 11 52.0	&	MACHO82.8284.23	&	B3IIIe	&	-12.84	&	258	$\pm$	13	&	31	$\pm$	1	 &	 265	 $\pm$	4	 &	197	$\pm$	10	 &	H$\alpha$,$\beta$ \OIII~from ambient emission	\\
RPs432	&	05 31 33.11	&	-70 46 17.9	&	MACHO7.8260	&	B3Ve	&	-13.49	&	383	$\pm$	19	&	26	$\pm$	1	&	 246	 $\pm$	0	&	 294	 $\pm$	15	&	 H$\alpha$, H$\beta$	\\
\hline
\end{tabular}
\end{minipage}
\end{table*}
\end{landscape}
\label{secondpage}
\setcounter{table}{2}
\begin{landscape}
\begin{table*}
\begin{minipage}{240mm}
\centering
\scriptsize
\begin{tabular}{|l|l|l|l|l|c|c|c|c|c|l|}
\hline
RP & RA   &   DEC &  Other &  Spec  &  Log F   &  FWHM   &   EW   &  Vel. (helio)  &  $\textit{v}$ sin $\textit{i}$  &   Comments   \\
Object  &  J2000  &  J2000 & Catalog &  Type &  H$\alpha$ &  H$\alpha$  & H$\alpha$  &   km/s   &   km/s    &                \\
  &         &       &  Reference        &    &   &  km/s   &   \AA    &      &    &            \\
\hline\hline
RPs364	&	05 35 49.19	&	-66 51 26.0	&		&	B1.5Ve	&	-13.86	&	329	$\pm$	16	&	28	$\pm$	1	&	302	 $\pm$	5	 &	 254	 $\pm$	 13	&	 H$\alpha$	 \\
RPs365	&	05 35 53.94	&	-66 52 32.3	&		&	F8Ve	&	-13.58	&	399	$\pm$	20	&	65	$\pm$	3	&	262	 $\pm$	6	 &	 382	 $\pm$	 19	&	 H$\alpha$,$\beta$, some local faint \HII~emission	\\
RPs368	&	05 35 55.73	&	-66 51 08.6	&		&	B2IVe	&	-13.33	&	271	$\pm$	14	&	28	$\pm$	1	&	335	 $\pm$	4	 &	 203	 $\pm$	 10	&	 H$\alpha$,$\beta$	\\
RPs369	&	05 35 57.84	&	-66 50 53.7	&	AL350	&	B7IIIe	&	-13.56	&	281	$\pm$	14	&	30	$\pm$	1	&	 282	$\pm$	 4	&	 217	 $\pm$	11	&	 H$\alpha$,$\beta$	\\
RPs1033	&	05 36 04.27	&	-67 55 37.0	&	BE74385	&	B2IVe	&	-11.05	&	263	$\pm$	13	&	54	$\pm$	2	&	 289	$\pm$	 4	&	 227	 $\pm$	11	&	 H$\alpha$,$\beta$, LBV poss	\\
\hline
\end{tabular}
\end{minipage}
\end{table*}
\end{landscape}
\label{secondpage}
\setcounter{table}{2}
\begin{landscape}
\begin{table*}
\begin{minipage}{240mm}
\centering
\scriptsize
\begin{tabular}{|l|l|l|l|l|c|c|c|c|c|l|}
\hline
RP & RA   &   DEC &  Other &  Spec  &  Log F   &  FWHM   &   EW   &  Vel. (helio)  &  $\textit{v}$ sin $\textit{i}$  &   Comments   \\
Object  &  J2000  &  J2000 & Catalog &  Type &  H$\alpha$ &  H$\alpha$  & H$\alpha$  &   km/s   &   km/s    &                \\
  &         &       &  Reference        &    &   &  km/s   &   \AA    &      &    &            \\
\hline\hline
RPs83	&	05 40 04.53	&	-70 39 44.9	&	SAB1449	&	B0IIIe	&	-13.38	&	108	$\pm$	5	&	16	$\pm$	1	&	 255	$\pm$	 2	&	 48	 $\pm$	2	&	 H$\alpha$,$\beta$, Forbidden lines due to surrounding thick \HII~disk 16arcsrc radius	\\
RPs138	&	05 40 10.00	&	-70 09 20.1	&		&	B3Ve	&	-13.50	&	62	$\pm$	3	&	22	$\pm$	1	&	270	 $\pm$	1	 &	 14	$\pm$	 1	&	H$\alpha$ only	\\
RPs139	&	05 40 12.00	&	-70 09 15.1	&		&	F0III[e]	&	-13.84	&	411	$\pm$	21	&	43	$\pm$	2	&	 314	$\pm$	 5	&	 354	 $\pm$	18	&	 H$\alpha$, \SII, SP	\\
RPs625	&	05 40 15.33	&	-70 46 13.2	&		&	B1Ve	&	-13.51	&	543	$\pm$	27	&	44	$\pm$	2	&	221	 $\pm$	9	 &	 493	 $\pm$	 25	&	 H$\alpha$,$\beta$ 	\\
RPs288	&	05 40 19.60	&	-68 29 19.9	&		&		&	-13.42	&	515	$\pm$	26	&	37	$\pm$	1	&				 &	 441	 $\pm$	 22	 &	 H$\alpha$,$\beta$	\\
\hline
\end{tabular}
\end{minipage}
\end{table*}
\end{landscape}
\label{secondpage}
\setcounter{table}{2}
\begin{landscape}
\begin{table*}
\begin{minipage}{240mm}
\centering
\scriptsize
\begin{tabular}{|l|l|l|l|l|c|c|c|c|c|l|}
\hline
RP & RA   &   DEC &  Other &  Spec  &  Log F   &  FWHM   &   EW   &  Vel. (helio)  &  $\textit{v}$ sin $\textit{i}$  &   Comments   \\
Object  &  J2000  &  J2000 & Catalog &  Type &  H$\alpha$ &  H$\alpha$  & H$\alpha$  &   km/s   &   km/s    &                \\
  &         &       &  Reference        &    &   &  km/s   &   \AA    &      &    &            \\
\hline\hline
RPs274	&	05 44 41.56	&	-68 51 29.3	&		&	B1Ve	&	-12.63	&	527	$\pm$	26	&	61	$\pm$	2	&	259	 $\pm$	6	 &	 502	 $\pm$	 25	&	 H$\alpha$,$\beta$	\\
RPs344	&	05 44 47.54	&	-67 19 33.8	&	IRAS - P.C05448-6720	&	B1.5V[e]	&	-13.27	&	95	$\pm$	5	&	 160	$\pm$	 6	&	 294	 $\pm$	2	&	87	 $\pm$	4	&	H$\alpha$,$\beta$, \SII, \NII, \OII	\\
RPs131	&	05 44 49.65	&	-70 08 26.1	&		&		&	-13.49	&	212	$\pm$	11	&	19	$\pm$	1	&	262	$\pm$	 11	&	 136	 $\pm$	 7	&	weak continuum, H$\alpha$,$\beta$, \OIII	\\
RPs383	&	05 45 08.11	&	-67 05 58.1	&		&	B2Ve	&	-13.25	&	293	$\pm$	15	&	45	$\pm$	2	&	306	 $\pm$	4	 &	 250	 $\pm$	 13	&	 H$\alpha$,$\beta$	\\
RPs244	&	05 45 19.02	&	-69 48 02.4	&		&	B3Ve	&	-13.65	&	330	$\pm$	16	&	43	$\pm$	2	&	243	 $\pm$	24	 &	 277	 $\pm$	 14	&	 H$\alpha$,$\beta$, (\SII~ambient)	\\
\hline
\end{tabular}
{notes:
AL: Andrews A.D., Lindsay E.M., (1964);
AGPRS: Melchior A.-L., Hughes S.M.G., Guibert J., (2000);
BE74: Bohannan B., Epps H. W., (1974);
FAUST: Bowyer S., Sasseen T.P., Wu X., Lampton M., (1995);
GRV: Reid N., Glass I.S., Catchpole R.M., (1988);
HD: Draper H., (1924);
IRAS: http://irsa.ipac.caltech.edu/IRASdocs/iras.html;
KDM: Kontizas E., Dapergolas A., Morgan D. H., Kontizas M., (2001);
L63: Lindsay E.M., (1963);
LI: Schwering P.B.W., (1989);
MACHO: Keller S.C., Bessell M.S., Cook K. H., Geha M., Syphers D., (2002);
S: Henize K.G., (1956);
SAB: Sabogal B.E., Mennickent R.E., Pietrzynski G., Gieren W., (2005);
SV (DV, HV): Butler C.J., Wayman P.A., (1974);
SHV: Hughes S.M.G., (1989);
XMMU: Lumb D.H., Guainazzi M., Gondoin P., (2001);
2MASS: Skrutskie M.F. et al. (2006)}
\end{minipage}
\end{table*}
\end{landscape}

\newpage
\begin{landscape}
\begin{table*}
\begin{minipage}{240mm}
\section{Appendix: Table 2}
{Compiled magnitude data for all hot emission-line stars in the LMC UKST H$\alpha$ survey. Please refer to section 15 for more detailed information. Column 1 gives the Reid Parker (RPs) number for the star (s). Columns 2 and 3 give the RA and Dec in J2000 coordinates. Column 4 gives the published GSC2.2 catalogue number where a star has been previously identified. Column 5 gives the distance between our position and the GSC2.2 position. Column 6 gives Ogle Catalog reference. Column 7 gives the reference number from the USNO catalogue provided by ESO. Column 8 gives the distance between our position and that provided by USNO. Column 9 gives the PA as provided by USNO. Columns 10-12 give the B magnitudes from Ogle, Super Cosmos (SC) and USNO respectively. Columns 13 and 14 give the V magnitudes from Ogle and SC respectively. Columns 15 and 16 give the I magnitudes from OGLE and SC respectively and columns 17 and 18 give the R magnitudes from SC and USNO respectively.}
\centering
\scriptsize
\begin{tabular}{|l|l|l|l|c|c|c|c|c|l|l|l|l|l|l|l|l|l}
\hline
RP & RA   &   DEC &    GSC2.2  &  d   &    OGLE   &   USNO      &     d    &  PA   &   B    &  B  &  B  &    V  & V & I &  I  & R  &  R \\
Object  &  J2000  &  J2000 &   Catalog &   mins  &   Catalog    &   Catalog  &  mins  & deg.  & Mag  & Mag  &  Mag & Mag & Mag & Mag & Mag  & Mag & Mag\\
         &        &       &   Reference & GSC2.2 &  Reference  &   Reference &  USNO   &    &  Ogle &  SC  &  USNO & GSC2.2 & OGLE & SC &  OGLE &  SC  & USNO\\
\hline\hline
RPs1741	&	04 52 55.41	&	-69 10 02.30	&	S013202239559	&	0.02	&		&		&		&		&		&	18.279	&		 &		 &		&	16.274	&		&	17.318	&		\\
RPs1647	&	04 53 01.03	&	-70 42 12.33	&	S0132020114	&	0.01	&		&		&		&		&		&	14.561	&		 &	16.53	 &		&		&		&	13.55	&		\\
RPs1701	&	04 53 02.01	&	-69 47 04.14	&	S013202222732	&	0.01	&		&	U0150-02229476	&	0.006	&	289	&		 &	17.306	 &	18.1	&	17.38	&		&	16.003	&		&	16.159	&	16.9	\\
RPs1699	&	04 53 06.86	&	-69 57 24.74	&	S013202215240	&	0.01	&		&	U0150-02231025	&	0.01	&	333	&		 &	16.067	 &	16.1	&	16.98	&		&	15.711	&		&	15.424	&	15.5	\\
RPs1671	&	04 53 25.43	&	-70 35 39.75	&	S013202078	&	0.01	&		&		&		&		&		&	12.896	&		 &	12.91	 &		&	12.939	&		&	11.915	&		\\
\hline
\end{tabular}
\end{minipage}
\end{table*}
\end{landscape}
\begin{landscape}
\begin{table*}
\begin{minipage}{240mm}
\centering
\scriptsize
\begin{tabular}{|l|l|l|l|c|c|c|c|c|l|l|l|l|l|l|l|l|l}
\hline
RP & RA   &   DEC &    GSC2.2  &  d   &    OGLE  &   USNO      &     d    &  PA   &   B    &  B  &  B  &    V  & V & I &  I  & R  &  R \\
Object  &  J2000  &  J2000 &   Catalog &   mins  &   Catalog    &   Catalog  &  mins  & deg.  & Mag  & Mag  &  Mag & Mag & Mag & Mag & Mag  & Mag & Mag\\
         &        &       &   Reference & GSC2.2 &  Reference  &   Reference &  USNO   &    &  Ogle &  SC  &  USNO & GSC2.2 & OGLE & SC &  OGLE &  SC  & USNO\\
\hline\hline
RPs1722	&	05 00 14.06	&	-69 25 13.98	&		&		&	SC15 5854	&	U0150-02382843	&	0.016	&	312	&	15.815	&		 &	15.8	 &		&	15.759	&		&	15.498	&	14.639	&	15.4	\\
RPs1855	&	05 00 28.46	&	-68 23 04.41	&	S013201139687	&	0	&		&	U0150-02388956	&	0.007	&	161	&		&	 16.75	&	 17.6	&	16.3	&		&	15.408	&		&	15.656	&	16.9	\\
RPs1755	&	05 00 28.77	&	-69 01 06.37	&	S013203017314	&	0.01	&	SC15 31595	&		&		&		&	16.637	&		 &		 &	16.44	&	16.589	&		&	16.363	&	14.215	&		\\
RPs1729	&	05 00 35.39	&	-69 19 53.36	&	S013202266654	&	0.02	&	SC15 9370	&	U0150-02391673	&	0.027	&	 307	&	 15.377	&		&	16.8	&	14.71	&	15.409	&		&	15.236	&	14.067	&	14.6	\\
RPs1662	&	05 01 32.53	&	-70 16 56.11	&	S013202335246	&	0.01	&		&	U0150-02416226	&	0.001	&	0	&		 &	16.796	 &	17.7	&	16.35	&		&	15.38	&		&	15.583	&	16.8	\\
\hline
\end{tabular}
\end{minipage}
\end{table*}
\end{landscape}
\begin{landscape}
\begin{table*}
\begin{minipage}{240mm}
\centering
\scriptsize
\begin{tabular}{|l|l|l|l|c|c|c|c|c|l|l|l|l|l|l|l|l|l}
\hline
RP & RA   &   DEC &    GSC2.2  &  d   &    OGLE  &   USNO      &     d    &  PA   &   B    &  B  &  B  &    V  & V & I &  I  & R  &  R \\
Object  &  J2000  &  J2000 &   Catalog &   mins  &   Catalog    &   Catalog  &  mins  & deg.  & Mag  & Mag  &  Mag & Mag & Mag & Mag & Mag  & Mag & Mag\\
         &        &       &   Reference & GSC2.2 &  Reference  &   Reference &  USNO   &    &  Ogle &  SC  &  USNO & GSC2.2 & OGLE & SC &  OGLE &  SC  & USNO\\
\hline\hline
RPs1837	&	05 07 11.00	&	-68 36 31.84	&	S013203045734	&	0	&		&	U0150-02574919	&	0.007	&	118	&		&	 16.031	&	 17.9	&	15.46	&		&	14.917	&		&	14.893	&	15.9	\\
RPs1744	&	05 07 11.44	&	-69 10 50.04	&	S01320308726	&	0.01	&		&		&		&		&		&	14.615	&		 &	 15.42	&		&		&	15.346	&	15.08	&		\\
RPs1914	&	05 07 11.54	&	-67 02 23.09	&	S013201330984	&	0.01	&		&	U0225-01605095	&	0.005	&	106	&		 &	14.877	 &	14.6	&	14.92	&		&	14.784	&		&	14.502	&	15.4	\\
RPs1836	&	05 07 16.78	&	-68 39 06.64	&	S013203043527	&	0	&		&		&		&		&		&	16.763	&		 &	15.6	 &		&	15.039	&		&	14.954	&		\\
RPs1262	&	05 07 26.42	&	-69 59 41.56	&	S013202345622	&	0.02	&	SC12 163287	&	U0150-02582713	&	0.004	&	 262	&	 15.55	&		&	16.3	&	15.25	&	15.527	&		&	15.346	&	14.376	&	15	\\
\hline
\end{tabular}
\end{minipage}
\end{table*}
\end{landscape}
\begin{landscape}
\begin{table*}
\begin{minipage}{240mm}
\centering
\scriptsize
\begin{tabular}{|l|l|l|l|c|c|c|c|c|l|l|l|l|l|l|l|l|l}
\hline
RP & RA   &   DEC &    GSC2.2  &  d   &    OGLE  &   USNO      &     d    &  PA   &   B    &  B  &  B  &    V  & V & I &  I  & R  &  R \\
Object  &  J2000  &  J2000 &   Catalog &   mins  &   Catalog    &   Catalog  &  mins  & deg.  & Mag  & Mag  &  Mag & Mag & Mag & Mag & Mag  & Mag & Mag\\
         &        &       &   Reference & GSC2.2 &  Reference  &   Reference &  USNO   &    &  Ogle &  SC  &  USNO & GSC2.2 & OGLE & SC &  OGLE &  SC  & USNO\\
\hline\hline
RPs1381	&	05 17 01.02	&	-69 34 10.57	&	S013203121696	&	0	&	SC8 291817	&	U0150-02842280	&	0.012	&	69	&	 15.57	&		 &	20	&	15.21	&	15.64	&		&	15.574	&		&	15.9	\\
RPs1525	&	05 17 06.93	&	-68 00 27.87	&	S013203267977	&	0	&		&	U0150-02844993	&	0.009	&	130	&		&	 16.88	&	 17	&	17.68	&		&	16.546	&		&	16.591	&	17.2	\\
RPs1526	&	05 17 07.33	&	-68 00 18.54	&	S013203268101	&	0.01	&		&	U0150-02845215	&	0.013	&	59	&		 &	16.376	 &	16.8	&	16.5	&		&	15.502	&		&	15.686	&	16.3	\\
RPs1331	&	05 17 08.13	&	-69 07 02.17	&	S013203161741	&	0.01	&	SC8 343463	&		&		&		&	17.164	&		 &		 &	17.16	&	17.021	&		&	16.886	&		&		\\
RPs2054	&	05 17 08.93	&	-69 32 21.18	&	S013203124892	&	0	&		&		&		&		&		&		&		&	 13.19	&		 &		&		&		&		\\
\hline
\end{tabular}
\end{minipage}
\end{table*}
\end{landscape}
\begin{landscape}
\begin{table*}
\begin{minipage}{240mm}
\centering
\scriptsize
\begin{tabular}{|l|l|l|l|c|c|c|c|c|l|l|l|l|l|l|l|l|l}
\hline
RP & RA   &   DEC &    GSC2.2  &  d   &    OGLE  &   USNO      &     d    &  PA   &   B    &  B  &  B  &    V  & V & I &  I  & R  &  R \\
Object  &  J2000  &  J2000 &   Catalog &   mins  &   Catalog    &   Catalog  &  mins  & deg.  & Mag  & Mag  &  Mag & Mag & Mag & Mag & Mag  & Mag & Mag\\
         &        &       &   Reference & GSC2.2 &  Reference  &   Reference &  USNO   &    &  Ogle &  SC  &  USNO & GSC2.2 & OGLE & SC &  OGLE &  SC  & USNO\\
\hline\hline
RPs2173	&	05 21 54.96	&	-69 46 35.99	&	S01320315582	&	0.03	&	SC6 273912	&	U0150-02970166	&	0.012	&	 355	&	 15.671	&		&	20.8	&	15.42	&	15.576	&		&	15.263	&	16.07	&	16.7	\\
RPs852	&	05 22 02.11	&	-69 02 05.82	&	S013203165603	&	0.01	&		&		&		&		&		&	17.39	&		 &		 &		&	16.937	&		&	15.911	&		\\
RPs2174	&	05 22 02.73	&	-69 46 14.48	&	S01320315848	&	0	&		&	U0150-02973617	&	0.007	&	31	&		&	 19.393	&	 20.6	&	15.17	&		&	15.57	&		&	15.788	&	16.5	\\
RPs985	&	05 22 03.16	&	-67 47 04.28	&	S0132032136219	&	0.02	&		&	U0150-02973891	&	0.032	&	45	&		 &	17.785	 &	11.1	&	17.08	&		&		&		&	16.28	&	11.5	\\
RPs606	&	05 22 07.95	&	-70 17 01.10	&	S010101233265	&	0.02	&		&	U0150-02975873	&	0.011	&	16	&		 &	16.942	 &	18.8	&	15.63	&		&	15.094	&		&	15.404	&	16.3	\\
\hline
\end{tabular}
\end{minipage}
\end{table*}
\end{landscape}
\begin{landscape}
\begin{table*}
\begin{minipage}{240mm}
\centering
\scriptsize
\begin{tabular}{|l|l|l|l|c|c|c|c|c|l|l|l|l|l|l|l|l|l}
\hline
RP & RA   &   DEC &    GSC2.2  &  d   &    OGLE  &   USNO      &     d    &  PA   &   B    &  B  &  B  &    V  & V & I &  I  & R  &  R \\
Object  &  J2000  &  J2000 &   Catalog &   mins  &   Catalog    &   Catalog  &  mins  & deg.  & Mag  & Mag  &  Mag & Mag & Mag & Mag & Mag  & Mag & Mag\\
         &        &       &   Reference & GSC2.2 &  Reference  &   Reference &  USNO   &    &  Ogle &  SC  &  USNO & GSC2.2 & OGLE & SC &  OGLE &  SC  & USNO\\
\hline\hline
RPs2149	&	05 25 20.02	&	-69 20 24.34	&	S013203142564	&	0	&		&	U0150-03057554	&	0.007	&	87	&		&	 16.931	&	 18.6	&	14.88	&		&	14.289	&		&	14.722	&	14.9	\\
RPs869	&	05 25 25.72	&	-69 12 55.70	&	S013203153169	&	0.01	&		&	U0150-03059808	&	0.016	&	179	&		 &	15.61	 &	17.6	&	15.24	&		&	15.238	&		&	14.827	&	14.9	\\
RPs724	&	05 25 29.68	&	-69 50 30.40	&	S0132031501	&	0.02	&	SC4 53506	&		&		&		&	15.33	&		 &		&	 12.46	&	15.081	&		&	14.715	&	15.118	&		\\
RPs725	&	05 25 39.77	&	-69 50 59.75	&	S013203178612	&	0.02	&	SC4 176812	&		&		&		&	17.564	&		 &		 &	17.52	&	17.545	&		&	17.518	&	17.35	&		\\
RPs942	&	05 25 40.09	&	-68 30 46.16	&	S013203337006	&	0.01	&		&		&		&		&		&	14.978	&		 &	 14.87	&		&	13.521	&		&	13.911	&		\\
\hline
\end{tabular}
\end{minipage}
\end{table*}
\end{landscape}
\begin{landscape}
\begin{table*}
\begin{minipage}{240mm}
\centering
\scriptsize
\begin{tabular}{|l|l|l|l|c|c|c|c|c|l|l|l|l|l|l|l|l|l}
\hline
RP & RA   &   DEC &    GSC2.2  &  d   &    OGLE  &   USNO      &     d    &  PA   &   B    &  B  &  B  &    V  & V & I &  I  & R  &  R \\
Object  &  J2000  &  J2000 &   Catalog &   mins  &   Catalog    &   Catalog  &  mins  & deg.  & Mag  & Mag  &  Mag & Mag & Mag & Mag & Mag  & Mag & Mag\\
         &        &       &   Reference & GSC2.2 &  Reference  &   Reference &  USNO   &    &  Ogle &  SC  &  USNO & GSC2.2 & OGLE & SC &  OGLE &  SC  & USNO\\
\hline\hline
RPs784	&	05 28 53.10	&	-69 24 51.39	&	S0132002149010	&	0	&	SC3 328853	&	U0150-03136671	&	0.006	&	252	&	 16.894	&		 &	19.1	&	17.32	&	17.026	&		&	16.978	&	16.063	&	17.3	\\
RPs783	&	05 28 57.69	&	-69 25 09.57	&	S0132002148868	&	0	&	SC3 328843	&	U0150-03138364	&	0.008	&	81	&	 17.055	&		 &	19.4	&	17.3	&	17.177	&		&	17.094	&	17.594	&	17.7	\\
RPs2182	&	05 28 58.82	&	-69 03 55.04	&	S0132002123411	&	0	&		&	U0150-03138765	&	0.001	&	0	&		&	 15.75	&	 15	&	15.94	&		&	15.287	&		&	14.172	&	15.8	\\
RPs794	&	05 29 01.97	&	-69 22 50.66	&	S0132002149891	&	0.01	&	SC3 333494	&	U0150-03139883	&	0.02	&	 187	&	 14.074	&		&	16.8	&	14.01	&	14.139	&		&	13.924	&	14.779	&	14.8	\\
RPs2183	&	05 29 02.92	&	-69 04 29.12	&	S0132002181	&	0	&		&	U0150-03140235	&	0.011	&	342	&		&	 18.514	&	12.5	 &	12.94	&		&	14.883	&		&	15.492	&	10.9	\\
\hline
\end{tabular}
\end{minipage}
\end{table*}
\end{landscape}
\begin{landscape}
\begin{table*}
\begin{minipage}{240mm}
\centering
\scriptsize
\begin{tabular}{|l|l|l|l|c|c|c|c|c|l|l|l|l|l|l|l|l|l}
\hline
RP & RA   &   DEC &    GSC2.2  &  d   &    OGLE  &   USNO      &     d    &  PA   &   B    &  B  &  B  &    V  & V & I &  I  & R  &  R \\
Object  &  J2000  &  J2000 &   Catalog &   mins  &   Catalog    &   Catalog  &  mins  & deg.  & Mag  & Mag  &  Mag & Mag & Mag & Mag & Mag  & Mag & Mag\\
         &        &       &   Reference & GSC2.2 &  Reference  &   Reference &  USNO   &    &  Ogle &  SC  &  USNO & GSC2.2 & OGLE & SC &  OGLE &  SC  & USNO\\
\hline\hline
RPs911	&	05 32 26.47	&	-68 39 03.91	&	S013200285951	&	0.01	&		&		&		&		&		&	17.57	&		 &	 18.79	&		&	16.469	&		&	16.621	&		\\
RPs1014	&	05 32 26.98	&	-70 47 44.35	&	S0101032159496	&	0	&		&	U0150-03225440	&	0.014	&	93	&		&	 18.045	&	 18.1	&	17.98	&		&	16.846	&		&	17.051	&	17.8	\\
RPs447	&	05 32 27.03	&	-70 47 44.05	&	S0101032159496	&	0.01	&		&	U0150-03225440	&	0.012	&	120	&		 &	18.045	 &	18.1	&	17.98	&		&	16.846	&		&	17.051	&	17.8	\\
RPs449	&	05 32 28.35	&	-70 47 28.58	&		&		&		&		&		&		&		&	17.737	&		&		&		 &	 16.48	&		&	17.222	&		\\
RPs532	&	05 32 30.10	&	-71 13 30.24	&	S0101033150569	&	0.02	&		&	U0150-03226663	&	0.012	&	292	&		 &	15.85	 &	15.3	&	16.7	&		&	15.322	&		&	15.132	&	15.8	\\
\hline
\end{tabular}
\end{minipage}
\end{table*}
\end{landscape}
\begin{landscape}
\begin{table*}
\begin{minipage}{240mm}
\centering
\scriptsize
\begin{tabular}{|l|l|l|l|c|c|c|c|c|l|l|l|l|l|l|l|l|l}
\hline
RP & RA   &   DEC &    GSC2.2  &  d   &    OGLE  &   USNO      &     d    &  PA   &   B    &  B  &  B  &    V  & V & I &  I  & R  &  R \\
Object  &  J2000  &  J2000 &   Catalog &   mins  &   Catalog    &   Catalog  &  mins  & deg.  & Mag  & Mag  &  Mag & Mag & Mag & Mag & Mag  & Mag & Mag\\
         &        &       &   Reference & GSC2.2 &  Reference  &   Reference &  USNO   &    &  Ogle &  SC  &  USNO & GSC2.2 & OGLE & SC &  OGLE &  SC  & USNO\\
\hline\hline
RPs298	&	05 37 28.44	&	-68 12 34.72	&	S0132003185437	&	0.02	&		&	U0150-03382920	&	0.015	&	95	&		 &	17.026	 &	17.1	&	16.98	&		&	16.041	&		&	16.342	&	16.7	\\
RPs257	&	05 37 35.37	&	-69 34 59.56	&	S0101001153570	&	0.01	&		&	U0150-03386570	&	0.007	&	4	&		 &	16.633	 &	18	&	16.48	&		&	15.297	&		&	15.591	&	15.7	\\
RPs29	&	05 37 38.98	&	-70 32 46.34	&	S0101033169023	&	0.01	&	SC16  211454	&	U0150-03388603	&	0.009	&	 84	&	 17.395	&		&	17.8	&	17.74	&	17.148	&		&	16.938	&	16.782	&	16.9	\\
RPs1043	&	05 37 40.59	&	-67 31 37.14	&	S0132001226952	&	0.01	&		&	U0150-03389408	&	0.009	&	312	&		 &	15.412	 &	15.5	&	15.56	&		&	14.785	&		&	16.118	&	15.6	\\
RPs283	&	05 37 48.30	&	-68 39 54.66	&	S0132002170098	&	0.01	&		&	U0150-03393904	&	0.012	&	174	&		 &		&	 16	&	16.58	&		&		&		&		&	16.2	\\
\hline
\end{tabular}
\end{minipage}
\end{table*}
\end{landscape}
\begin{landscape}
\begin{table*}
\begin{minipage}{240mm}
\centering
\scriptsize
\begin{tabular}{|l|l|l|l|c|c|c|c|c|l|l|l|l|l|l|l|l|l}
\hline
RP & RA   &   DEC &    GSC2.2  &  d   &    OGLE  &   USNO      &     d    &  PA   &   B    &  B  &  B  &    V  & V & I &  I  & R  &  R \\
Object  &  J2000  &  J2000 &   Catalog &   mins  &   Catalog    &   Catalog  &  mins  & deg.  & Mag  & Mag  &  Mag & Mag & Mag & Mag & Mag  & Mag & Mag\\
         &        &       &   Reference & GSC2.2 &  Reference  &   Reference &  USNO   &    &  Ogle &  SC  &  USNO & GSC2.2 & OGLE & SC &  OGLE &  SC  & USNO\\
\hline\hline
RPs159	&	05 41 15.29	&	-69 58 10.34	&		&		&	SC18 107209	&		&		&		&	15.565	&		&		&		 &	 15.685	&		&	15.791	&	13.784	&		\\
RPs343	&	05 41 20.97	&	-67 22 02.98	&	S013200123035	&	0.01	&		&	U0225-02299544	&	0.007	&	199	&		 &	17.787	 &	17.6	&	17.44	&		&	16.799	&		&	16.566	&	17.3	\\
RPs1022	&	05 41 22.94	&	-68 36 24.99	&	S0132003175066	&	0	&		&	U0150-03517086	&	0.003	&	339	&		&	 15.671	&	 17.1	&	16.44	&		&	16.367	&		&	15.831	&	15.9	\\
RPs1036	&	05 41 30.41	&	-68 53 44.06	&	S0132002161489	&	0	&		&	U0150-03521288	&	0.012	&	339	&		&	 15.661	&	 16.3	&	14.78	&		&	14.5	&		&	14.109	&	14.7	\\
RPs20	&	05 41 31.63	&	-70 15 31.42	&	S0101032177542	&	0	&		&	U0150-03521919	&	0.014	&	295	&		&	 16.35	&	 17.3	&	16.81	&		&		&	15.895	&	15.872	&	16.7	\\
\hline
\end{tabular}
\end{minipage}
\end{table*}
\end{landscape}

\bsp

\label{lastpage}